\documentclass[12pt,preprint,showpacs,preprintnumbers]{revtex4}
\usepackage{amsfonts}
\usepackage{amsmath}
\usepackage{amssymb}
\usepackage{hyperref}
\usepackage{color}
\usepackage{graphicx}
\usepackage{amssymb}
\usepackage{amsmath}
\usepackage{graphicx}
\usepackage{dcolumn}
\usepackage{bm}
\usepackage{epsfig}
\usepackage[T1]{fontenc}
\usepackage{ae,aecompl}

\begin{document}

\title{Discrete Boltzmann model of compressible flows with spherical or
cylindrical symmetry}
\author{ Aiguo Xu$^{1,2,3}$\thanks{%
Corresponding author: Xu\_Aiguo@iapcm.ac.cn}, Guangcai Zhang$^{1,3}$,
Chuandong Lin$^{4}$ }
\affiliation{$^1$ National Key Laboratory of Computational Physics, Institute of Applied
Physics and Computational Mathematics, P. O. Box 8009-26, Beijing 100088,
P.R.China \\
$^2$ Center for Applied Physics and Technology, MOE Key Center for High
Energy Density Physics Simulations, College of Engineering, Peking
University, Beijing 100871, China \\
$^3$ State Key Laboratory of Theoretical Physics, Institute of Theoretical
Physics, Chinese Academy of Sciences,Beijing 100190, China\\
$^4$ State Key Laboratory for GeoMechanics and Deep Underground Engineering,
China University of Mining and Technology, Beijing 100083, P.R.China}
\date{\today }

\begin{abstract}
To study simultaneously the hydrodynamic and thermodynamic behaviors in
compressible flow systems with spherical or cylindrical symmetry, we present
a theoretical framework for constructing Discrete Boltzmann Model(DBM) with
spherical or cylindrical symmetry in spherical or cylindrical coordinates.
To this aim, a key technique is to use \emph{local} Cartesian coordinates to
describe the particle velocity in the kinetic model. Thus, the geometric
effects, like the divergence and convergence, are described as a
\textquotedblleft force term\textquotedblright . Even though the
hydrodynamic models are one- or two-dimensional, the DBM needs a Discrete
Velocity Model(DVM) with 3 dimensions. We use a DVM with 26 velocities to
formulate the DBM which recovers the Navier-Stokes equations with spherical
or cylindrical symmetry in the hydrodynamic limit. For the system with \emph{%
global} cylindrical symmetry, we formulated also a DBM based on a DVM with 2
dimensions and 16 velocities. In terms of the nonconserved moments, we
define two sets of measures for the deviations of the system from its
thermodynamic equilibrium state. The extension of current model to the
multiple-relaxation-time version is straightforward.
\end{abstract}

\maketitle


\section{Introduction}

During recent decades the lattice Boltzmann (LB) modeling and simulation
have achieved great success in various complex flows \cite%
{Succi-Book,XuReview2014}. According to the discretization scheme, the LB
methods can be roughly classified as the standard LB\cite{Succi-Book},
Finite-Difference(FD) LB\cite{HeLuoPRE1997,Cao1997,VictorJCP2003},
Finite-Volume(FV) LB\cite{Nannelli1992} and Finite-Element(FE) LB\cite%
{LiPRE2004,LiPRE2005}, etc. According to the purpose and/or function, the LB
methods can be roughly classified as Partial Differential Equations(PDE)
solvers, mesoscopic kinetic models, etc. The LB methods working as
mesoscopic kinetic models are generally designed to access some
nonequilibrium behaviors of the complex system which can not be or are not
convenient to be described by the traditional hydrodynamic models. When
constructing LB mesoscopic kinetic models, one has to ensure that the main
behaviors under consideration to be included\cite{Succi-Book,XuReview2014}.
The formulating of a LB PDE solver can be more flexible. To distinguish from
the PDE solvers, the LB mesoscopic kinetic models are referred to as the
Discrete Boltzmann Model(DBM) in this work. The appropriately designed DBM
should inherit some functions of the Boltzmann equation\cite%
{XuReview2014,XuReview2012}.

Given the great importance of shock waves in many fields of physics and
engineering, constructing LB models for high speed compressible flows has
attracted considerable interest since the early days of LB research \cite%
{Succi-Book}. The LB model for high speed compressible flows has seen
significant progress in recent years\cite{XuReview2012}. In 1992 Alexander
et al \cite{Alexander1992} formulated a compressible LB model for flows at
high Mach number via introducing a flexible sound speed. This model works
only for nearly isothermal compressible systems. In 1999 Yan et al \cite%
{Yan1999} proposed a LB scheme for compressible Euler equations. In the
years of 1998 and 2003 Sun and his coworker \cite{Sun1998,Sun2003} presented
an adaptive LB scheme for the two- and three-dimensional systems,
respectively. In this model the particle velocities vary with the Mach
number and internal energy, so that the particle velocities are no longer
constrained to fixed values. All of those models belong to the standard LB
solvers of PDE.

In 1997 Cao et al \cite{Cao1997} proposed to use the FD scheme to calculate
the spatial and temporal derivatives in the LB equation and apply nonuniform
grids so that the numerical stability can be improved a little. In the past
decade, Tsutahara, Watari and Kataoka \cite%
{Kataoka2004a,Kataoka2004b,Watari2003,Watari2004,Watari2007} proposed
several nice FD LB models for the Euler and Navier-Stokes equations. In 2005
Xu \cite{Xu2005PRE,Xu2005EPL} extended the idea to handle binary fluids.
However, similar to the case of standard LB models, these FD LB schemes
still belong to LB solvers of corresponding PDE since no function beyond the
traditional hydrodynamic models are pointed out in corresponding
publications. These FD LB schemes work also only for subsonic flows. To
simulate high speed compressible flows, especially those with shocks, many
attempts and considerable progress have been achieved in recent years \cite%
{XuPan2007,XuGan2008CTP,XuGan2008PhysA,XuGan2011CTP,LB2KHI2011,XuChen2009,
MRT2010EPL,XuChen2010,MRT2011CTP,MRT2011PLA,XuChen2011,MRT2011TAML,Review2012}%
.

Up to now, most of LB models for compressible fluids are in Cartesian
coordinates. In many cases the flows show divergent, convergent, and/or
rotational behaviors, for example, in cylindrical or spherical devices. For
such flow systems, LB models in polar, spherical, cylindrical or rotational
coordinates are more convenient and are less exposed to numerical errors.
There have been a number of LB methods in curvilinear coordinates or
axisymmetric cylindrical coordinates. Early in 1992, Nannelli and Succi \cite%
{Nannelli1992} presented a general framework to extend the LB equation to
arbitrary lattice geometries. In this work a FV LB was formulated. Then some
other versions of FV LB were proposed for irregular meshes\cite%
{Succi1995,Amati1997,Peng1998,Peng1999,Ubertini2003}. In 1997 He and Doolen
\cite{He1997} extended the LB to general curvilinear coordinate systems. In
the following year Mei and Shyy \cite{Mei1998} developed a FD LB in
body-fitted curvilinear coordinates with non-uniform grids. Later, Halliday
et al \cite{Halliday2001} proposed a Polar Coordinate Lattice
Boltzmann(PCLB) for hydrodynamics. In 2005 Premnath and Abraham \cite%
{Premnath2005} presented a LB for axisymmetric multiphase flows. In this
work source terms were added to a two-dimensional standard LB equation for
multiphase flows such that the emergent dynamics can be transformed into the
axisymmetric cylindrical coordinate system. But all those LB methods in
non-Cartesian coordinates work only for isothermal and nearly incompressible
flows. In 2010 Asinari et al \cite{Asinari2010} formulated a LB to analyze
the radiative heat transfer problems in a participation medium, but did not
take into account the effects of fluid flow. In 2011 Watari \cite{Watari2011}
formulated a FD PCLB to investigate the rotational flow problems in coaxial
cylinders. This work presents valuable information on the LB application to
the cylindrical system. However, this model works also only for subsonic
flow systems. Based on the same discrete velocity model, Lin et al\cite%
{XuLin2014PRE} formulated a PCLB for high-speed compressible flows. Within
this model\cite{XuLin2014PRE}, a hybrid scheme being similar to, but
different from, the operator-splitting is proposed. The temporal evolution
is calculated analytically and the convection term is solved via a Modified
Warming-Beam (MWB) scheme. Within the MWB scheme a suitable switch function
is introduced.

Besides recovering the macroscopic hydrodynamic equations, designing DBM to
access the nonequilibrium behaviors is attracting more attention with time%
\cite{XuReview2014,XuReview2012}. The idea of DBM has been further specified
and applied in various compressible flow systems via several models\cite%
{XuYan2013,XuGan2013,XuLin2014PRE,XuLin2014arXiv,XuChen2014FoP}. Examples of
LBGK for compressible flow systems are referred to Refs. \cite%
{XuGan2013,XuLin2014PRE} where preliminary studies on shocking behaviors are
shown. An example of MRT-LB for compressible flow systems is referred to
Ref. \cite{XuChen2014FoP}. In Refs.\cite{XuYan2013,XuLin2014arXiv} the
Thermodynamic NonEquilibrium (TNE) behaviors in combustion systems are
initially investigated via LBGK models. The kinetic nature of DBM for the
non-ideal gas systems, particularly, the liquid-vapor system or
single-component two phase flows was considered in Ref.\cite{XuZhangGan2014}.

In traditional modeling the implosion and explosion processes,
one-dimensional or two-dimensional hydrodynamic equations are frequently
used. The one-dimensional hydrodynamic equations are generally used to
describe system with spherical symmetry and systems with cylindrical
symmetry with translational symmetry. The two-dimensional equations are
generally used to describe systems with rotational or cylindrical symmetry.
Since the DBM has been proved to be more convenient for measuring the
macroscopic behaviors of the system due to its deviating from thermodynamic
equilibrium state, in this work we aim to construct the DBM for compressible
flow systems with spherical or cylindrical symmetry.

This paper is organized as below. In section II we briefly review the
kinetic and hydrodynamic models of the fluid system. In terms of their
correlations, we formulate two set of measures for the deviation of the
system from its thermodynamic equilibrium. The discrete Boltzmann models are
formulated in section III. Section IV presents the conclusion and
discussions.

\section{Brief review of fluid models}

\subsection{Kinetic model}

The Boltzmann BGK model\cite{BGK1954} reads%
\begin{equation}
\partial _{t}f+\mathbf{v}\cdot \nabla f=-\frac{1}{\tau }\left(
f-f^{eq}\right) \text{,}  \label{e1}
\end{equation}%
where $f=f\left( \mathbf{R}\text{, }\mathbf{v}\text{, }t\right) =f\left( x%
\text{,}y\text{,}z\text{,}v_{x}\text{,}v_{y}\text{,}v_{z}\text{,}t\right) $,
$\mathbf{R=}x\mathbf{\hat{x}}+y\mathbf{\hat{y}}+z\mathbf{\hat{z}}$ and $%
\mathbf{v=}v_{x}\mathbf{\hat{x}+}v_{y}\mathbf{\hat{y}+}v_{z}\mathbf{\hat{z}}$
in Cartesian coordinates.

\subsubsection{In cylindrical coordinates}

In cylindrical coordinates, the position $\mathbf{R=}r\mathbf{\hat{r}+}z%
\mathbf{\hat{z}}$, where $r\mathbf{\hat{r}}$ is the projection of the
position $\mathbf{R}$ in the $\left( x\text{, }y\right) $ plane. The
included angle between the $x$-axis and the vector $r\mathbf{\hat{r}}$ is $%
\theta $. The unit vectors $\mathbf{\hat{r}}$,$\mathbf{\hat{\theta}}$ and $%
\mathbf{\hat{z}}$ are the changing directions of the position $\mathbf{R}$
along the three parameters, $r$, $\theta $ and $z$, i.e.,%
\begin{equation*}
d\mathbf{r}=\mathbf{\hat{r}}dr+\mathbf{\hat{\theta}}rd\theta +\mathbf{\hat{z}%
}dz\text{.}
\end{equation*}%
Obviously, $\mathbf{\hat{r}}$,$\mathbf{\hat{\theta}}$ and $\mathbf{\hat{z}}$
are orthogonal to each other and satisfy the following relationships,%
\begin{eqnarray*}
\mathbf{\hat{z}}\mathbf{=\hat{r}\times \hat{\theta}} &&\text{,} \\
\mathbf{\hat{\theta}}\mathbf{=\hat{z}\times \hat{r}} &&\text{,} \\
\mathbf{\hat{r}}\mathbf{=\hat{\theta}\times \hat{z}} &&\text{\textbf{.}}
\end{eqnarray*}%
It is easy to find that%
\begin{eqnarray*}
d\mathbf{\hat{r}} &=&\left( \mathbf{\hat{z}}d\theta \right) \times \mathbf{%
\hat{r}=\hat{\theta}}d\theta , \\
d\mathbf{\hat{\theta}} &=&\left( \mathbf{\hat{z}}d\theta \right) \times
\mathbf{\hat{\theta}=-\hat{r}}d\theta , \\
d\mathbf{\hat{z}} &=&0\text{.}
\end{eqnarray*}

We consider two kinds of cylindrical symmetries here. The system is referred
to have \emph{local} cylindrical symmetry if the behavior of the system is
independent of the azimuthal coordinate $\theta $ but maybe dependent of the
hight $z$, and is referred to have \emph{global} cylindrical symmetry if the
behavior of the system is independent of both the azimuthal coordinate $%
\theta $ and the hight $z$.

The description of the particle velocity $\mathbf{v}$ can use the global
Cartesian coodinates $\left( \mathbf{\hat{x}}\text{, }\mathbf{\hat{y}}\text{%
, }\mathbf{\hat{z}}\right) $,

\begin{equation}
\mathbf{v}\mathbf{=}v_{x}\mathbf{\hat{x}+}v_{y}\mathbf{\hat{y}+}v_{z}\mathbf{%
\hat{z}}\text{.}  \label{v1}
\end{equation}%
It can also use the local Cartesian coordinates $\left( \mathbf{\hat{r}}%
\text{, }\mathbf{\hat{\theta}}\text{, }\mathbf{\hat{z}}\right) $,
\begin{eqnarray}
\mathbf{v} &=&v_{r}\mathbf{\hat{r}+}v_{\theta }\mathbf{\hat{\theta}+}v_{z}%
\mathbf{\hat{z}}  \notag \\
&=&\mathbf{v}\cdot \mathbf{\hat{r}\hat{r}+\mathbf{v}\cdot \hat{\theta}\hat{%
\theta}+\mathbf{v}\cdot \hat{z}\hat{z}}\text{.}  \label{e4}
\end{eqnarray}%
Consequently, the distribution function $f$ can also be described in two
different forms,
\begin{equation}
f=f\left( r\text{, }\theta \text{, }z\text{, }v_{x}\text{, }v_{y}\text{, }%
v_{z}\text{, }t\right)  \label{e6}
\end{equation}%
or
\begin{equation}
f=f\left( r\text{, }\theta \text{, }z\text{, }v_{r}\text{, }v_{\theta }\text{%
, }v_{z}\text{, }t\right) \text{.}  \label{e7}
\end{equation}%
In this work we will use the latter form, Eq. (\ref{e7}). This is a key
technique in this work for constructing the DBM. Under the definition (\ref%
{e7}), it should be stressed that the particle velocity $\mathbf{v}$ is
\emph{locally} fixed, while its two components, $v_{r}$, $v_{\theta }$,
varies with the position $\mathbf{R}$, when calculating the spatial
derivative, $\nabla f$. We use the symbol \textquotedblleft $\nabla |_{%
\mathbf{v}}$\textquotedblright\ to replace \textquotedblleft $\nabla $%
\textquotedblright\ in Eq. (\ref{e1}) to stress that $\mathbf{v}$ is fixed
when calculating the spatial derivatives. Thus, Eq. (\ref{e1}) is rewritten
as

\begin{equation}
\partial _{t}f+\mathbf{v}\cdot \nabla |_{\mathbf{v}}f=-\frac{1}{\tau }\left(
f-f^{eq}\right) .  \label{e8}
\end{equation}%
According to the definition (\ref{e7}),

\begin{equation}
\nabla |_{\mathbf{v}}f=\nabla |_{v_{r},v_{\theta }}\text{ }f+\nabla v_{r}%
\frac{\partial f}{\partial v_{r}}+\nabla v_{\theta }\frac{\partial f}{%
\partial v_{\theta }}\text{,}  \label{e9}
\end{equation}%
where
\begin{eqnarray}
\nabla v_{r} &=&\nabla \left( \mathbf{\hat{r}\cdot v}\right)  \notag \\
&=&\left( \nabla \mathbf{\hat{r}}\right) \mathbf{\cdot v}  \notag \\
&=&\left( \mathbf{\hat{r}}\frac{\partial \mathbf{\hat{r}}}{\partial r}%
\mathbf{+\hat{\theta}}\frac{1}{r}\frac{\partial \mathbf{\hat{r}}}{\partial
\theta }+\mathbf{\hat{z}}\frac{\partial \mathbf{\hat{r}}}{\partial z}\right)
\mathbf{\cdot v}  \notag \\
&=&\mathbf{\hat{\theta}}\frac{1}{r}\mathbf{\hat{\theta}\cdot v}  \notag \\
&=&\frac{v_{\theta }}{r}\mathbf{\hat{\theta}}\text{,}  \label{e10}
\end{eqnarray}%
similarly,
\begin{equation}
\nabla v_{\theta }=-\frac{v_{r}}{r}\mathbf{\hat{\theta}}\text{.}  \label{e11}
\end{equation}%
Substituting Eqs. (\ref{e9})-(\ref{e11}) into Eq. (\ref{e8}) gives the
Boltzmann equation in cylindrical coordinates,

\begin{equation}
\partial _{t}f+\mathbf{v}\cdot \nabla |_{v_{r},v_{\theta }}\text{ }f+\frac{%
v_{\theta }^{2}}{r}\frac{\partial f}{\partial v_{r}}-\frac{v_{r}v_{\theta }}{%
r}\frac{\partial f}{\partial v_{\theta }}=-\frac{1}{\tau }\left(
f-f^{eq}\right) .  \label{e12}
\end{equation}%
For macroscopic system with cylindrical symmetry, the distribution function $%
f$ does not depend explicitly on the angle $\theta $. i.e.,
\begin{equation}
f=f\left( r\text{, }z\text{, }v_{r}\text{, }v_{\theta }\text{, }v_{z}\text{,
}t\right) \text{.}  \label{e13}
\end{equation}%
So, the Boltzmann equation becomes%
\begin{equation}
\partial _{t}f+\left( v_{r}\frac{\partial f}{\partial r}+v_{z}\frac{\partial
f}{\partial z}\right) +\left( \frac{v_{\theta }^{2}}{r}\frac{\partial f}{%
\partial v_{r}}-\frac{v_{r}v_{\theta }}{r}\frac{\partial f}{\partial
v_{\theta }}\right) =-\frac{1}{\tau }\left( f-f^{eq}\right) .  \label{e14}
\end{equation}%
It is clear that the term%
\begin{equation*}
\left( \frac{v_{\theta }^{2}}{r}\frac{\partial f}{\partial v_{r}}-\frac{%
v_{r}v_{\theta }}{r}\frac{\partial f}{\partial v_{\theta }}\right)
\end{equation*}%
plays the role of the \emph{force term} in the Boltzmann equation in the
Cartesian coordinates. It creates the divergence or convergence effects in
the flow system. When the system is not far from its thermodynamic
equilibrium, we can use the approximation, $f=f^{eq}$, when calculating the
force term. If further use the macroscopic condition, $u_{\theta }=0$, the
final Boltzmann model for the flow system with cylindrical becomes,%
\begin{equation}
\partial _{t}f+\left( v_{r}\frac{\partial f}{\partial r}+v_{z}\frac{\partial
f}{\partial z}\right) +\left[ \frac{v_{r}v_{\theta }^{2}}{rT}-\frac{%
v_{\theta }^{2}\left( v_{r}-u_{r}\right) }{rT}\right] f^{eq}=-\frac{1}{\tau }%
\left( f-f^{eq}\right) .  \label{e18}
\end{equation}

\subsubsection{In spherical coordinates}

In spherical coordinates, the position $\mathbf{R=}r\mathbf{\hat{r}}$. The
three parameters $r$, $\theta $ and $\varphi $ are the radial, azimuth and
zenith angle, respectively. The unit vectors, $\mathbf{\hat{r}}$,$\mathbf{%
\hat{\theta}}$ and $\mathbf{\hat{\varphi}}$, are the changing directions of
the position vector $\mathbf{R}$ along the three parameters, $r$, $\theta $
and $\varphi $, respectively, i.e.,
\begin{equation*}
d\mathbf{r}=\mathbf{\hat{r}}dr+r\mathbf{\hat{\theta}}d\theta +r\sin \theta
\text{ }\mathbf{\hat{\varphi}}d\varphi \text{.}
\end{equation*}%
Obviously, $\mathbf{\hat{r}}$,$\mathbf{\hat{\theta}}$ and $\mathbf{\hat{%
\varphi}}$ are orthogonal to each other and satisfy the following
relationships,
\begin{eqnarray*}
\mathbf{\hat{\varphi}} &\mathbf{=\hat{r}\times \hat{\theta}}& \mathtt{,} \\
\mathbf{\hat{\theta}} &\mathbf{=\hat{\varphi}\times \hat{r}}& \mathtt{,} \\
\mathbf{\hat{r}} &\mathbf{=\hat{\theta}\times \hat{\varphi}}& \mathtt{.}
\end{eqnarray*}%
It is easy to find that%
\begin{eqnarray*}
d\mathbf{\hat{r}} &=&\left( \boldsymbol{\hat{\varphi}}d\theta +\mathbf{\hat{z%
}}d\varphi \right) \times \mathbf{\hat{r}=\hat{\theta}}d\theta +\boldsymbol{%
\hat{\varphi}}\sin \theta \text{ }d\varphi , \\
d\mathbf{\hat{\theta}} &=&\left( \boldsymbol{\hat{\varphi}}d\theta +\mathbf{%
\hat{z}}d\varphi \right) \times \mathbf{\hat{\theta}=-\hat{r}}d\theta +%
\boldsymbol{\hat{\varphi}}\cos \theta \text{ }d\varphi , \\
d\boldsymbol{\hat{\varphi}} &=&\left( \boldsymbol{\hat{\varphi}}d\theta +%
\mathbf{\hat{z}}d\varphi \right) \times \boldsymbol{\hat{\varphi}}\mathbf{=-}%
\left( \mathbf{\hat{r}}\sin \theta +\mathbf{\hat{\theta}}\cos \theta \right)
\text{ }d\varphi .
\end{eqnarray*}

The description of the particle velocity $\mathbf{v}$ can use the global
Cartesian coordinates $\left( \mathbf{\hat{x}}\text{, }\mathbf{\hat{y}}\text{%
, }\mathbf{\hat{z}}\right) $. It can also use the local Cartesian coodinates
$\left( \mathbf{\hat{r}}\text{, }\mathbf{\hat{\theta}}\text{, }\boldsymbol{%
\hat{\varphi}}\right) $.
\begin{eqnarray}
\mathbf{v} &\mathbf{=}&v_{x}\mathbf{\hat{x}+}v_{y}\mathbf{\hat{y}+}v_{z}%
\mathbf{\hat{z}}  \notag \\
&=&v_{r}\mathbf{\hat{r}+}v_{\theta }\mathbf{\hat{\theta}+}v_{\varphi }%
\boldsymbol{\hat{\varphi}}  \notag \\
&=&\mathbf{v}\cdot \mathbf{\hat{r}\hat{r}+\mathbf{v}\cdot \hat{\theta}\hat{%
\theta}+\mathbf{v}\cdot }\boldsymbol{\hat{\varphi}\hat{\varphi}}\text{.}
\label{eq4}
\end{eqnarray}%
The distribution function $f$ can also be described in two different forms,
\begin{equation}
f=f\left( r\text{, }\theta \text{, }\varphi \text{, }v_{x}\text{, }v_{y}%
\text{, }v_{z}\text{, }t\right)  \label{eq6}
\end{equation}%
or
\begin{equation}
f=f\left( r\text{, }\theta \text{, }\varphi \text{, }v_{r}\text{, }v_{\theta
}\text{, }v_{\varphi }\text{, }t\right) \text{.}  \label{eq7}
\end{equation}%
In this work we will use the latter form, Eq. (\ref{eq7}). Because it is
more convenient when describing problems with cylindrical symmetry. Under
the definition (\ref{eq7}), it should be stressed that the particle velocity
$\mathbf{v}$ is fixed, while its three components, $v_{r}$, $v_{\theta }$, $%
v_{\varphi }$,varies with the position $\mathbf{R}$, when calculating the
spatial derivative, $\nabla f$. We use the symbol \textquotedblleft $\nabla
|_{\mathbf{v}}$\textquotedblright\ to replace \textquotedblleft $\nabla $%
\textquotedblright\ in Eq. (\ref{e1}) to stress that $\mathbf{v}$ is fixed
when calculating the spatial derivatves. Thus, Eq. (\ref{e1}) is rewritten
as
\begin{equation}
\partial _{t}f+\mathbf{v}\cdot \nabla |_{\mathbf{v}}f=-\frac{1}{\tau }\left(
f-f^{eq}\right) .  \label{eq8}
\end{equation}%
According to the definition (\ref{eq7}),
\begin{equation}
\nabla |_{\mathbf{v}}f=\nabla |_{v_{r},v_{\theta }}\text{ }f+\nabla v_{r}%
\frac{\partial f}{\partial v_{r}}+\nabla v_{\theta }\frac{\partial f}{%
\partial v_{\theta }}+\nabla v_{\varphi }\frac{\partial f}{\partial
v_{\varphi }}\text{,}  \label{eq9}
\end{equation}%
where
\begin{eqnarray}
\nabla v_{r} &=&\nabla \left( \mathbf{\hat{r}\cdot v}\right)  \notag \\
&=&\left( \nabla \mathbf{\hat{r}}\right) \mathbf{\cdot v}  \notag \\
&=&\left( \mathbf{\hat{r}}\frac{\partial \mathbf{\hat{r}}}{\partial r}%
\mathbf{+\hat{\theta}}\frac{1}{r}\frac{\partial \mathbf{\hat{r}}}{\partial
\theta }+\mathbf{\hat{\varphi}}\frac{1}{r\sin \theta }\frac{\partial \mathbf{%
\hat{r}}}{\partial \varphi }\right) \mathbf{\cdot v}  \notag \\
&=&\left( \frac{\mathbf{\hat{\theta}\hat{\theta}}}{r}\mathbf{+}\frac{\mathbf{%
\hat{\varphi}\hat{\varphi}}}{r}\right) \mathbf{\cdot v}  \notag \\
&=&\frac{v_{\theta }}{r}\mathbf{\hat{\theta}}+\frac{v_{\varphi }}{r}\mathbf{%
\hat{\varphi}} \mathtt{,}  \label{eq10}
\end{eqnarray}%
and similarly,
\begin{equation}
\nabla v_{\theta }=-\frac{v_{r}}{r}\mathbf{\hat{\theta}+}\frac{v_{\varphi
}\cos \theta }{r\sin \theta }\mathbf{\hat{\varphi}}\text{,}  \label{eq11}
\end{equation}%
\begin{equation}
\nabla v_{\varphi }=-\left( \frac{v_{r}}{r}\mathbf{+}\frac{v_{\theta }\cos
\theta }{r\sin \theta }\right) \mathbf{\hat{\varphi}}\text{.}  \label{eq12}
\end{equation}%
Substituting Eqs. (\ref{eq9})-(\ref{eq12}) into Eq. (\ref{eq8}) gives the
Boltzmann equation in spherical coordinates,
\begin{eqnarray}
&& \partial _{t}f+\mathbf{v}\cdot \nabla |_{v_{r},v_{\theta },v_{\varphi }}%
\text{ }f+\frac{v_{\theta }^{2}+v_{\varphi }^{2}}{r}\frac{\partial f}{%
\partial v_{r}}+\left( \frac{v_{\varphi }^{2}\cos \theta }{r\sin \theta }-%
\frac{v_{r}v_{\theta }}{r}\right) \frac{\partial f}{\partial v_{\theta }}
-\left( \frac{v_{r}v_{\varphi }}{r}+\frac{v_{\varphi }v_{\theta }\cos \theta
}{r\sin \theta }\right) \frac{\partial f}{\partial v_{\varphi }}  \notag \\
&& =-\frac{1}{\tau }\left( f-f^{eq}\right) .  \label{eq14}
\end{eqnarray}%
For macroscopic system with spherical symmetry, the distribution function $f
$ does not depend explicitly on the angles $\theta $ and $\varphi $, i.e.,
\begin{equation}
f=f\left( r\text{, }v_{r}\text{, }v_{\theta }\text{, }v_{\varphi }\text{, }%
t\right) ;  \label{eq15}
\end{equation}%
and $f$ is invariant under the rotation in the subspace of $\left( v_{\theta
}\text{, }v_{\varphi }\right) $, i.e.,
\begin{equation}
f\left( r\text{, }v_{r}\text{, }v_{\theta }\text{, }v_{\varphi }\text{, }%
t\right) =f\left( r\text{, }v_{r}\text{, }v_{\theta }^{\prime }\text{, }%
v_{\varphi }^{\prime }\text{, }t\right)  \label{eq16}
\end{equation}%
only when
\begin{equation}
v_{\theta }^{2}+v_{\varphi }^{2}=v_{\theta }^{\prime 2}+v_{\varphi }^{\prime
2}\text{.}  \label{eq17}
\end{equation}%
So,
\begin{equation}
\left( v_{\theta }\frac{\partial f}{\partial v_{\varphi }}-v_{\varphi }\frac{%
\partial f}{\partial v_{\theta }}\right) f=0\text{.}  \label{eq18}
\end{equation}%
Using the conditions, (\ref{eq15}) and (\ref{eq18}), in Eq. (\ref{eq14})
gives
\begin{equation}
\partial _{t}f+\mathbf{v}\cdot \nabla |_{v_{r},v_{\theta },v_{\varphi }}%
\text{ }f +\left( \frac{v_{\theta }^{2}+v_{\varphi }^{2}}{r}\frac{\partial f%
}{\partial v_{r}}-\frac{v_{r}v_{\theta }}{r}\frac{\partial f}{\partial
v_{\theta }}-\frac{v_{r}v_{\varphi }}{r}\frac{\partial f}{\partial
v_{\varphi }}\right) = -\frac{1}{\tau }\left( f-f^{eq}\right) .  \label{eq19}
\end{equation}%
It is clear that the term%
\begin{equation*}
\left( \frac{v_{\theta }^{2}+v_{\varphi }^{2}}{r}\frac{\partial f}{\partial
v_{r}}-\frac{v_{r}v_{\theta }}{r}\frac{\partial f}{\partial v_{\theta }}-%
\frac{v_{r}v_{\varphi }}{r}\frac{\partial f}{\partial v_{\varphi }}\right)
\end{equation*}%
plays the role of the force term in the Boltzmann equation in Cartesian
coordinates. It creates the divergence or convergence effects in the flow
system. If the system is not far from its thermodynamic equilibrum, we can
use the approximation, $f=f^{eq}$ when calculating the force term. If
further use the macroscopic condition, $u_{\theta }=u_{\varphi }=0$, the
final Boltzmann model for the flow system with spherical symmetry becomes,%
\begin{equation}
\partial _{t}f+v_{r}\frac{\partial f}{\partial r}+\left[ \frac{%
v_{r}v_{\theta }^{2}}{rT}+\frac{v_{r}v_{\varphi }^{2}}{rT}-\frac{\left(
v_{\theta }^{2}+v_{\varphi }^{2}\right) \left( v_{r}-u_{r}\right) }{rT}%
\right] f^{eq}=-\frac{1}{\tau }\left( f-f^{eq}\right) .  \label{eq22}
\end{equation}

\subsection{Hydrodynamic models and their correlations to kinetic models}

The Navier-Stokes equations in Cartesian coordinates read
\begin{subequations}
\begin{eqnarray}
\frac{\partial \rho }{\partial t}+\frac{\partial (\rho u_{\alpha })}{%
\partial x_{\alpha }} &=&0\text{,}  \label{HydroCart1} \\
\frac{\partial (\rho u_{\alpha })}{\partial t}+\frac{\partial (\rho
u_{\alpha }u_{\beta })}{\partial x_{\beta }}+\frac{\partial P\delta _{\alpha
\beta }}{\partial x_{\beta }} &=&\frac{\partial \sigma _{\alpha \beta }}{%
\partial x_{\beta }}\text{,}  \label{HydroCart2} \\
\frac{\partial }{\partial t}(\rho E+\frac{1}{2}\rho u^{2})+\frac{\partial }{%
\partial x_{\alpha }}[u_{\alpha }(\rho E+\frac{1}{2}\rho u^{2}+P)] &=&\frac{%
\partial }{\partial x_{\alpha }}\left[ -Q_{\alpha }+u_{\beta }\sigma
_{\alpha \beta }\right] \text{.}  \label{HydroCart3}
\end{eqnarray}%
In the left-hand side of Eqs. (\ref{HydroCart1})- (\ref{HydroCart3}), $\rho $%
, $\mathbf{u}$, $P$ $=\rho RT$, $E=\left( D+n\right) RT/2$, $T$, $\gamma
=\left( D+n+2\right) /\left( D+n\right) $ are the hydrodynamic density, flow
velocity, pressure, internal energy, temperature and specific-heat ratio,
respectively. $D$ is the space dimension and $n$ is the number of extra
degrees of freedom whose energy level is $\eta ^{2}/2$. $u^{2}=\mathbf{u}%
\cdot \mathbf{u}$. In the right-hand side of Eqs. (\ref{HydroCart2})- (\ref%
{HydroCart3}),
\end{subequations}
\begin{equation}
\sigma _{\alpha \beta }=\mu \left[ \frac{\partial u_{\beta }}{\partial
r_{\alpha }}+\frac{\partial u_{\alpha }}{\partial r_{\beta }}-\left(
1-\lambda \right) \frac{\partial u_{\gamma }}{\partial r_{\gamma }}\delta
_{\alpha \beta }\right]  \label{stress}
\end{equation}%
is the viscous stress and
\begin{equation}
Q_{\alpha }=-k\frac{\partial E}{\partial r_{\alpha }}  \label{heatflux}
\end{equation}%
is the heat flux. The two parameters,
\begin{eqnarray*}
\mu &=&\frac{2}{D}\rho e\tau \text{,} \\
\lambda &=&\frac{D+n-2}{D+n}\text{,}
\end{eqnarray*}%
are viscosities and the parameter,
\begin{equation*}
k=\frac{2\left( D+n+2\right) }{D\left( D+n\right) }\rho e\tau \text{,}
\end{equation*}%
is the heat conductivity. $e=DRT/2$ is the translational internal energy. It
is clear that $P=2\rho e/D$. When the viscosities and heat conductivity
vanish, the hydrodynamic equations, (\ref{HydroCart1}) - (\ref{HydroCart3}),
become the Euler equations.

The Chapamn-Enskog multiscale analysis shows that, in the process of
recovering the Navier-Stokes equations, (\ref{HydroCart1}) - (\ref%
{HydroCart3}), from the Boltzmann BGK equation, (\ref{e1}), the following
seven moment relations,
\begin{subequations}
\begin{equation}
\int \int f^{eq}d\mathbf{v}d\mathbf{\eta }=\rho \text{,}  \label{Mr1}
\end{equation}%
\begin{equation}
\int \int f^{eq}v_{\alpha }d\mathbf{v}d\mathbf{\eta }=\rho u_{\alpha }\text{,%
}  \label{Mr2}
\end{equation}%
\begin{equation}
\int \int f^{eq}\left( v^{2}+\eta ^{2}\right) d\mathbf{v}d\mathbf{\eta }%
=2\rho \left( \frac{D+n}{D}e+\frac{u^{2}}{2}\right) \text{,}  \label{Mr3}
\end{equation}%
\begin{equation}
\int \int f^{eq}v_{\alpha }v_{\beta }d\mathbf{v}d\mathbf{\eta }=P\delta
_{\alpha \beta }+\rho u_{\alpha }u_{\beta }\text{,}  \label{Mr4}
\end{equation}%
\begin{equation}
\int \int f^{eq}\left( v^{2}+\eta ^{2}\right) v_{\alpha }d\mathbf{v}d\mathbf{%
\eta }=2\rho \left( \frac{D+n+2}{D}e+\frac{u^{2}}{2}\right) u_{\alpha }\text{%
,}  \label{Mr5}
\end{equation}%
\begin{equation}
\int \int f^{eq}v_{\alpha }v_{\beta }v_{\chi }d\mathbf{v}d\mathbf{\eta }%
=\rho RT\left( u_{\alpha }\delta _{\beta \chi }+u_{\beta }\delta _{\alpha
\chi }+u_{\chi }\delta _{\alpha \beta }\right) +\rho u_{\alpha }u_{\beta
}u_{\chi }\text{,}  \label{Mr6}
\end{equation}%
\begin{equation}
\int \int f^{eq}\left( v^{2}+\eta ^{2}\right) v_{\alpha }v_{\beta }d\mathbf{v%
}d\mathbf{\eta }=\frac{4}{D}\rho e\left( \frac{D+n+2}{D}e+\frac{u^{2}}{2}%
\right) \delta _{\alpha \beta }+2\rho u_{\alpha }u_{\beta }\left( \frac{D+n+4%
}{D}e+\frac{u^{2}}{2}\right) \text{,}  \label{Mr7}
\end{equation}%
are used, where $\mathbf{\eta }$ is the velocity in the $n$ extra degrees of
freedom, $\eta ^{2}=\mathbf{\eta }\cdot \mathbf{\eta }$, and
\end{subequations}
\begin{equation}
f^{eq}=g^{eq}\left( \mathbf{v}\right) h^{eq}\left( \mathbf{\eta }\right)
\end{equation}%
with
\begin{subequations}
\begin{eqnarray}
g^{eq}\left( \mathbf{v}\right) &=&\rho \left( \frac{D}{4\pi e}\right)
^{D/2}\exp \left[ -\frac{D}{4e}\left( \mathbf{v}-\mathbf{u}\right) ^{2}%
\right] \text{,} \\
h^{eq}\left( \mathbf{\eta }\right) &=&\left( \frac{D}{4\pi ne}\right)
^{D/2}\exp \left( -\frac{D}{4ne}\mathbf{\eta }^{2}\right) \text{.}
\end{eqnarray}%
We require $\int d\mathbf{\eta }$ $h^{eq}\left( \mathbf{\eta }\right) =1$,
and $n\rightarrow 0$ when $\mathbf{\eta \rightarrow 0}$.

Converting the form of equations, (\ref{HydroCart1}) - (\ref{HydroCart3}),
from the Cartesian coordinates to the cylindrical or spherical coordinates
gives the hydrodynamic model which can be recovered from the kinetic model (%
\ref{e12}) or (\ref{eq14}) in the continnum limit. The Navier-Stokes
equations with spherical symmetry are only one dimensional and read
\end{subequations}
\begin{subequations}
\begin{eqnarray}
&&{{\partial }_{t}}\rho +({{\partial }_{r}}+\frac{2}{r})\left( \rho u\right)
=0\text{,}
\end{eqnarray}
\begin{eqnarray}
&&{{\partial }_{t}}u+u{{\partial }_{r}}u+\frac{1}{\rho }{{\partial }_{r}}p=%
\frac{4}{\rho r}\mu ({{\partial }_{r}}u_{r}-\frac{u_{r}}{r}) -\frac{1}{\rho }%
{{\partial }_{r}}\left[ \mu (1-\lambda )({{\partial }_{r}}+\frac{2}{r}%
)u_{r}-2\mu {{\partial }_{r}}u_{r}\right] \text{,}
\end{eqnarray}
\begin{eqnarray}
&&{{\partial }_{t}}e+u{{\partial }_{r}}e+\frac{p}{\rho }\left[ ({{\partial }%
_{r}}+\frac{2}{r})u\right]  \notag \\
&=&\frac{1}{\rho }({{\partial }_{r}}+\frac{2}{r})(k{{\partial }_{r}}T)+\frac{%
2\mu }{\rho }\left[ {{({{\partial }_{r}}u)}^{2}}+2{{(\frac{u}{r})}^{2}}%
\right] -\frac{1}{\rho }\mu (1-\lambda )\left[ ({{\partial }_{r}}+\frac{2}{r}%
)u\right] ^{2}\text{,}
\end{eqnarray}%
which can be recovered from the kinetic model (\ref{eq22}). The
Navier-Stokes equations with local cylindrical symmetry are two dimensional
and read
\end{subequations}
\begin{subequations}
\begin{equation}
{{\partial }_{t}}\rho +({{\partial }_{r}}+1/r)\left( \rho {{u}_{r}}\right) +{%
{\partial }_{z}}\left( \rho {{u}_{z}}\right) =0
\end{equation}%
\begin{eqnarray}
&&{{\partial }_{t}}{{u}_{r}}+{{u}_{r}}{{\partial }_{r}}{{u}_{r}}+{{u}_{z}}{{%
\partial }_{z}}{{u}_{r}+}\frac{1}{\rho }{\partial }_{r}p  \notag \\
&=&\frac{1}{\rho }{{\partial }_{r}}\left[ 2\mu {{\partial }_{r}}{{u}_{r}}%
-\mu (1-\lambda )({{\partial }_{r}}{{u}_{r}}+\frac{{{u}_{r}}}{r}+{{\partial }%
_{z}}{{u}_{z}})\right]  \notag \\
&&+\frac{\mu }{\rho }\left[ \frac{2{{\partial }_{r}}{{u}_{r}}-2{{u}_{r}}/r}{r%
}+{{\partial }_{z}}({{\partial }_{z}}{{u}_{r}}+{{\partial }_{r}}{{u}_{z}})%
\right]
\end{eqnarray}%
\begin{eqnarray}
&&{{\partial }_{t}}{{u}_{z}}+{{u}_{r}}{{\partial }_{r}}{{u}_{z}}+{{u}_{z}}{{%
\partial }_{z}}{{u}_{z}+\frac{1}{\rho }{\partial }_{z}p}  \notag \\
&=&\frac{\mu }{\rho }{{\partial }_{r}}({{\partial }_{z}}{{u}_{r}}+{{\partial
}_{r}}{{u}_{z}})-\frac{\mu }{\rho }\frac{({{\partial }_{z}}{{u}_{r}}+{{%
\partial }_{r}}{{u}_{z}})}{r}  \notag \\
&&+\frac{1}{\rho }{{\partial }_{z}}\left[ 2\mu {{\partial }_{z}}{{u}_{z}}%
-\mu (1-\lambda )({{\partial }_{r}}{{u}_{r}}+\frac{{{u}_{r}}}{r}+{{\partial }%
_{z}}{{u}_{z}})\right]
\end{eqnarray}%
\begin{eqnarray}
&&{{\partial }_{t}}e+{{u}_{r}}{{\partial }_{r}}e+{{u}_{z}}{{\partial }_{z}}e+%
\frac{1}{\rho }p({{\partial }_{r}}{{u}_{r}}+\frac{{{u}_{r}}}{r}+{{\partial }%
_{z}}{{u}_{z}})  \notag \\
&=&-\frac{1}{\rho }\mu (1-\lambda )({{\partial }_{r}}{{u}_{r}}+\frac{{{u}_{r}%
}}{r}+{{\partial }_{z}}{{u}_{z}})^{2}+\frac{2}{\rho }\mu {{\partial }_{r}}{{u%
}_{r}}{{\partial }_{r}}{u}_{r}  \notag \\
&&+\frac{1}{\rho }\frac{2\mu {{u}_{r}}}{r}\frac{{{u}_{r}}}{r}+\frac{1}{\rho }%
\mu {{({{\partial }_{z}}{{u}_{r}}+{{\partial }_{r}}{{u}_{z}})}^{2}}  \notag
\\
&&+\frac{2}{\rho }\mu {{\partial }_{z}}{{u}_{z}}{{\partial }_{z}}{{u}_{z}+}%
\frac{1}{\rho }\left[ ({{\partial }_{r}}+\frac{1}{r})k{{\partial }_{r}}T+{{%
\partial }_{z}}k{{\partial }_{z}}T\right]
\end{eqnarray}%
which can be recovered from the kinetic model (\ref{e18}).

\subsection{Measurements of nonequilibrium effects}

The Chapman-Enskog multiscale analysis tells that, as the simplest
hydrodynamic model of fluid system, the Euler equations ignore completely
the Thermodynamic NonEquilibrium (TNE) behavior. The Navier-Stokes equations
describe the TNE behavior via the terms in viscosity and heat conductivity.
The Euler model works successfully when we consider the fluid system in a
time scale $t_{0}$ which is large enough compared with thermodynamic
relaxation time $\tau $. Besides the normal high speed compressible flows,
the Euler model works also for solid materials under strong shock. From the
mechanical side, compared with the shocking strength, the material strength,
for example, the yield, and viscous stress are negligible. Consequently, the
Euler model works better with increasing the shock strength. From the side
of time scales, when study the shocking procedure, the used time scale $%
t_{0} $ is generally small enough compared with the time scale $t_{h}$ for
heat conduction and large enough compared with the thermodynamic relaxation
time $\tau $. In other words, during the time interval under investigation,
the heat conduction does not have time to occur significantly and
consequently its effects are negligible. For the objective system where the
thermodynamic relaxation time $\tau $ is fixed, if we decreases the
observing time scale $t_{0}$, we find more TNE effects. The Boltzmann
kinetic model can be used to investigate both the hydrodynamic and
thermodynamic behaviors.

Following the seven moment relations, (\ref{Mr1})-(\ref{Mr7}), used in
recovering the Navier-Stokes equations, we define the following moments,
\end{subequations}
\begin{subequations}
\begin{equation}
\mathbf{M}_{0}^{\ast }(f\text{,}\mathbf{v})=\int \int f\text{ }d\mathbf{v}d%
\mathbf{\eta }\text{,}  \label{Meq1}
\end{equation}%
\begin{equation}
\mathbf{M}_{1}^{\ast }(f\text{,}\mathbf{v})=\int \int f\text{ }\mathbf{v}d%
\mathbf{v}d\mathbf{\eta }\text{,}  \label{Meq2}
\end{equation}%
\begin{equation}
\mathbf{M}_{2,0}^{\ast }(f\text{,}\mathbf{v})=\int \int f\text{ }\left(
\mathbf{v}\cdot \mathbf{v}+\eta ^{2}\right) d\mathbf{v}d\mathbf{\eta }\text{,%
}  \label{Meq3}
\end{equation}%
\begin{equation}
\mathbf{M}_{2}^{\ast }(f\text{,}\mathbf{v})=\int \int f\text{ }\mathbf{vv}d%
\mathbf{v}d\mathbf{\eta }\text{,}  \label{Meq4}
\end{equation}%
\begin{equation}
\mathbf{M}_{3,1}^{\ast }(f\text{,}\mathbf{v})=\int \int f\text{ }\left(
\mathbf{v}\cdot \mathbf{v}+\eta ^{2}\right) \mathbf{v}d\mathbf{v}d\mathbf{%
\eta }\text{,}  \label{Meq5}
\end{equation}%
\begin{equation}
\mathbf{M}_{3}^{\ast }(f\text{,}\mathbf{v})=\int \int f\text{ }\mathbf{vvv}d%
\mathbf{v}d\mathbf{\eta }\text{,}  \label{Meq6}
\end{equation}%
\begin{equation}
\mathbf{M}_{4,2}^{\ast }(f\text{,}\mathbf{v})=\int \int f\text{ }\left(
\mathbf{v}\cdot \mathbf{v}+\eta ^{2}\right) \mathbf{vv}d\mathbf{v}d\mathbf{%
\eta }\text{,}  \label{Meq7}
\end{equation}
where $\mathbf{M}_{n}^{\ast }$ means a $n$-th order tensor and $\mathbf{M}%
_{m,n}^{\ast }$ means a $n$-th-order tensor contracted from a $m$-th order
tensor. For the case of central moments, the variable $\mathbf{v}$ is
replaced with $\mathbf{v}^{\ast }=\left( \mathbf{v}-\mathbf{u}\right) $. It
is clear $\mathbf{M}_{0}^{\ast }$ and $\mathbf{M}_{2,0}^{\ast }$ are
scalars. Each of them has only $1$ component. $\mathbf{M}_{1}^{\ast }$ and $%
\mathbf{M}_{3,1}^{\ast }$ are vectors. Each of them has $2$ independent
components in $2$-dimensional case or $3$ independent components in $3$%
-dimensional case. $\mathbf{M}_{2}^{\ast }$ and $\mathbf{M}_{4,2}^{\ast }$
are 2nd order tensors. Each of them has $3$ independent components in $2$%
-dimensional case or $6$ independent components in $3$-dimensional case. $%
\mathbf{M}_{3}^{\ast }$ is 3rd tensor and has $4$ independent components in $%
2$-dimensional case or $10$ independent components in $3$-dimensional case.
Therefore, the constraints, (\ref{Mr1}) - (\ref{Mr7}), are in fact $16$
linear equations in $f^{eq}$ in $2$-dimensional case and $30$ linear
equations in $f^{eq}$ in $3$-dimensional case. We further define
\end{subequations}
\begin{equation}
\boldsymbol{\Delta }_{m,n}^{\ast }\left( \mathbf{v}\right) =\mathbf{M}%
_{m,n}^{\ast }(f\text{,}\mathbf{v})-\mathbf{M}_{m,n}^{\ast }(f^{eq}\text{,}%
\mathbf{v})\text{.}  \label{Deq1}
\end{equation}%
It is clear that $\boldsymbol{\Delta }_{0}^{\ast }\left( \mathbf{v}\right) =%
\mathbf{0}$, $\boldsymbol{\Delta }_{1}^{\ast }\left( \mathbf{v}\right) =%
\mathbf{0}$ and $\boldsymbol{\Delta }_{2,0}^{\ast }\left( \mathbf{v}\right) =%
\mathbf{0}$, which is due to the mass, momentum and energy conservations.
Except for the three, the quantity $\boldsymbol{\Delta }_{m,n}^{\ast }\left(
\mathbf{v}\right) $ works as a measure for the deviation of the system from
its thermodynamic equilibrium. The information of flow velocity $\mathbf{u}$
is taken into account in the definition (\ref{Deq1}). Similarly,
\begin{equation}
\boldsymbol{\Delta }_{m,n}^{\ast }\left( \mathbf{v}^{\ast }\right) =\mathbf{M%
}_{m,n}^{\ast }(f\text{,}\mathbf{v}^{\ast })-\mathbf{M}_{m,n}^{\ast }(f^{eq}%
\text{,}\mathbf{v}^{\ast })\text{.}  \label{Deq2}
\end{equation}%
Except for $\boldsymbol{\Delta }_{0}^{\ast }\left( \mathbf{v}^{\ast }\right)
$, $\boldsymbol{\Delta }_{1}^{\ast }\left( \mathbf{v}^{\ast }\right) $ and $%
\boldsymbol{\Delta }_{2,0}^{\ast }\left( \mathbf{v}^{\ast }\right) $, the
quantity $\boldsymbol{\Delta }_{m,n}^{\ast }\left( \mathbf{v}^{\ast }\right)
$ works as a measure for the deviation of the system from its thermodynamic
equilibrium, where only the thermal fluctuations of the molecules are
considered.

\section{Discrete Boltzmann models}

There are two key techniques in constructing DBM with force terms. The first
is to approximate $f$ by $f^{eq}$ in the force term so that the velocity
derivative of $f$, $\partial f/\partial \mathbf{v}$, can be analytically
calculated before introducing the Discrete Velocity Model(DVM). The second
is that the DVM can be fixed before the main interaction in the code. In
other words, the discrete velocities $\mathbf{v}_{i}$ do not need to be
recalculated in each iteration step.

For constructing the DBM for systems with cylindrical symmetry, we use Eq. (%
\ref{e18}). We have
\begin{equation}
\partial _{t}f_{i}+\left( v_{ir}\frac{\partial f_{i}}{\partial r}+v_{iz}%
\frac{\partial f_{i}}{\partial z}\right) +\left[ \frac{v_{ir}v_{i\theta }^{2}%
}{rT}-\frac{v_{i\theta }^{2}\left( v_{ir}-u_{r}\right) }{rT}\right]
f_{i}^{eq}=-\frac{1}{\tau }\left( f_{i}-f_{i}^{eq}\right) .  \label{DBC1}
\end{equation}%
where $f_{i}$ ($f_{i}^{eq}$) is the discrete (equilibrium) distribution
function; $v_{i}$ is the $i$-th discrete velocity, $i=1$, $...$, $N$; $N$ is
the total number of the discrete velocity. For constructing the DBM for
systems with spherical symmetry, we use Eq. (\ref{eq22}). We have
\begin{equation}
\partial _{t}f_{i}+v_{ir}\frac{\partial f_{i}}{\partial r}+\left[ \frac{%
v_{ir}v_{i\theta }^{2}}{rT}+\frac{v_{ir}v_{i\varphi }^{2}}{rT}-\frac{\left(
v_{i\theta }^{2}+v_{i\varphi }^{2}\right) \left( v_{ir}-u_{ir}\right) }{rT}%
\right] f_{i}^{eq}=-\frac{1}{\tau }\left( f_{i}-f_{i}^{eq}\right) .
\label{DBS1}
\end{equation}

\emph{The fundamental requirement for a DBM is that it should recover the
same set of hydrodynamic equations as those given by the original continuous
Boltzmann equation}. The Chapman-Enskog multiscale analysis shows that, to
formulate a DBM, we need only require that \emph{the used moment relations
in continuous form can be rewritten in discrete form}. To formulate a DBM
which recovers the Euler equations, the following five constraints are
needed,
\begin{subequations}
\begin{equation}
\rho =\sum_{i=1}^{N}f_{i}^{eq}=\sum_{i=1}^{N}f_{i}\text{,}  \label{m1}
\end{equation}%
\begin{equation}
\rho u_{\alpha }=\sum_{i=1}^{N}f_{i}^{eq}v_{i\alpha
}=\sum_{i=1}^{N}f_{i}v_{i\alpha }\text{,}  \label{m2}
\end{equation}%
\begin{equation}
2\rho \left( \frac{D+n}{D}e+\frac{u^{2}}{2}\right)
=\sum_{i=1}^{N}f_{i}^{eq}(v_{i}^{2}+\eta
_{i}^{2})=\sum_{i=1}^{N}f_{i}(v_{i}^{2}+\eta _{i}^{2})\text{,}  \label{m3}
\end{equation}%
\begin{equation}
P\delta _{\alpha \beta }+\rho u_{\alpha }u_{\beta
}=\sum_{i=1}^{N}f_{i}^{eq}v_{i\alpha }v_{i\beta }\text{,}  \label{m4}
\end{equation}%
\begin{equation}
2\rho \left( \frac{D+n+2}{D}e+\frac{u^{2}}{2}\right) u_{\alpha
}=\sum_{i=1}^{N}f_{i}^{eq}(v_{i}^{2}+\eta _{i}^{2})v_{i\alpha }\text{,}
\label{m5}
\end{equation}%
To formulate a DBM which recovers the Navier-Stokes equations, two more
constraints,
\begin{equation}
\rho RT\left( u_{\alpha }\delta _{\beta \chi }+u_{\beta }\delta _{\alpha
\chi }+u_{\chi }\delta _{\alpha \beta }\right) +\rho u_{\alpha }u_{\beta
}u_{\chi }=\sum f_{i}^{eq}v_{i\alpha }v_{i\beta }v_{i\chi },  \label{m6}
\end{equation}%
\begin{equation}
\frac{4}{D}\rho e\left( \frac{D+n+2}{D}e+\frac{u^{2}}{2}\right) \delta
_{\alpha \beta }+2\rho u_{\alpha }u_{\beta }\left( \frac{D+n+4}{D}e+\frac{%
u^{2}}{2}\right) =\sum f_{i}^{eq}\left( v_{i}^{2}+\eta _{i}^{2}\right)
v_{i\alpha }v_{i\beta }\text{,}  \label{meq7}
\end{equation}%
are needed. Following the same idea as in the definitions, (\ref{Meq1}) - (%
\ref{Meq7}), we define the following moments of the discrete distribution
function $f_{i}$,
\end{subequations}
\begin{subequations}
\begin{equation}
\mathbf{M}_{0}(f_{i}\text{,}\mathbf{v}_{i})=\sum_{i=1}^{N}f_{i}\text{,}
\label{Me1}
\end{equation}%
\begin{equation}
\mathbf{M}_{1}(f_{i}\text{,}\mathbf{v}_{i})=\sum_{i=1}^{N}f_{i}\mathbf{v}_{i}%
\text{,}  \label{Me2}
\end{equation}%
\begin{equation}
\mathbf{M}_{2,0}(f_{i}\text{,}\mathbf{v}_{i})=\sum_{i=1}^{N}f_{i}(\mathbf{v}%
_{i}\cdot \mathbf{v}_{i}+\eta _{i}^{2})\text{,}  \label{Me3}
\end{equation}%
\begin{equation}
\mathbf{M}_{2}(f_{i}\text{,}\mathbf{v}_{i})=\sum_{i=1}^{N}f_{i}\mathbf{v}_{i}%
\mathbf{v}_{i}\text{,}  \label{Me4}
\end{equation}%
\begin{equation}
\mathbf{M}_{3,1}(f_{i}\text{,}\mathbf{v}_{i})=\sum_{i=1}^{N}f_{i}(\mathbf{v}%
_{i}\cdot \mathbf{v}_{i}+\eta _{i}^{2})\mathbf{v}_{i}\text{,}  \label{Me5}
\end{equation}%
\begin{equation}
\mathbf{M}_{3}(f_{i}\text{,}\mathbf{v}_{i})=\sum f_{i}\mathbf{v}_{i}\mathbf{v%
}_{i}\mathbf{v}_{i},  \label{Me6}
\end{equation}%
\begin{equation}
\mathbf{M}_{4,2}(f_{i}\text{,}\mathbf{v}_{i})=\sum f_{i}\left( \mathbf{v}%
_{i}\cdot \mathbf{v}_{i}+\eta _{i}^{2}\right) \mathbf{v}_{i}\mathbf{v}_{i}%
\text{,}  \label{Me7}
\end{equation}%
where $\mathbf{M}_{n}$ means a $n$-th order tensor and $\mathbf{M}_{m,n}$
means a $n$-th-order tensor contracted from a $m$-th order tensor. For the
case of central moments, the variable $\mathbf{v}$ is replaced with $\mathbf{%
v}^{\ast }=\left( \mathbf{v}-\mathbf{u}\right) $. The constraints, (\ref{m1}%
) - (\ref{meq7}), are in fact $16$ linear equations in $f_{i}^{eq}$ in $2$%
-dimensional case and $30$ linear equations in $f_{i}^{eq}$ in $3$%
-dimensional case. Following the same idea as in the definitions, (\ref{Deq1}%
) - (\ref{Deq2}), we further define
\end{subequations}
\begin{equation}
\boldsymbol{\Delta }_{m,n}\left( \mathbf{v}_{i}\right) =\mathbf{M}%
_{m,n}(f_{i}\text{,}\mathbf{v}_{i})-\mathbf{M}_{m,n}(f_{i}^{eq}\text{,}%
\mathbf{v}_{i})\text{.}  \label{De1}
\end{equation}
\begin{equation}
\boldsymbol{\Delta }_{m,n}\left( \mathbf{v}_{i}^{\ast }\right) =\mathbf{M}%
_{m,n}(f_{i}\text{,}\mathbf{v}_{i}^{\ast })-\mathbf{M}_{m,n}(f_{i}^{eq}\text{%
,}\mathbf{v}_{i}^{\ast })\text{.}  \label{De2}
\end{equation}%
Except for $\boldsymbol{\Delta }_{0}$, $\boldsymbol{\Delta }_{1}$and $%
\boldsymbol{\Delta }_{2,0}$, the quantity $\boldsymbol{\Delta }_{m,n}$%
\textbf{\ }works as a measure for the deviation of the system from its
thermodynamic equilibrium.

A key step in formulating a DBM is to find a solution for the discrete
equlibrium distribution function $f_{i}^{eq}$. The constraints, (\ref{m1}) -
(\ref{meq7}) can also be rewritten as
\begin{equation}
\mathbf{\hat{f}}^{eq}=\mathbf{Cf}^{eq}  \label{meq1}
\end{equation}%
where $\mathbf{\hat{f}}^{eq}=\left[ \hat{f}_{k}^{eq}\right] ^{T}$ and $%
\mathbf{f}^{eq}=\left[ f_{k}^{eq}\right] ^{T}$ are column vectors with $k=1$,%
$2$,$\cdots $,$N$, $\mathbf{C}$ is $N\times N$ matrix whose components are
determined by $\mathbf{v}_{i}$ if the parameter $\eta _{i}$ is fixed. It is
clear that
\begin{equation}
\mathbf{f}^{eq}=\mathbf{C}^{-1}\mathbf{\hat{f}}^{eq}\text{.}  \label{meq2}
\end{equation}%
Obviously, the choosing of the DVM must ensure the exsitence of $\mathbf{C}%
^{-1}$. We work in the frame where the particle mass $m=1$ and the constant $%
R=1$.

\subsection{\protect\bigskip DBM for systems with spherical symmetry}

If we require the DBM to recover the Navier-Stokes equations in the
continuum limit, the DBM needs a DVM with 3 dimensions.

\subsubsection{Case with $\protect\gamma =5/3$}

We first consider the simple case where ratio of specific rates is fixed, $%
\gamma =5/3$. We set $\eta _{i}=0$ and $n=0$ in constraint (\ref{meq7}).
Among the seven moment constraints, (\ref{m1}) - (\ref{meq7}), only five are
independent. We do not use the constraints (\ref{m3}) and (\ref{m5}). The
five independent constraints can be rewritten as $26$ independent linear
equations in $f_{i}^{eq}$. Now, we fix the components $\hat{f}_{k}^{eq}$of $%
\mathbf{\hat{f}}^{eq}$. Here $N=26$.

From the constraint (\ref{m1}), we have $\hat{f}_{1}^{eq}=\rho $. From the
constraint (\ref{m2}), we have $\hat{f}_{2}^{eq}=\rho u_{r}$, $\hat{f}%
_{3}^{eq}=\rho u_{\theta }$, $\hat{f}_{4}^{eq}=\rho u_{\varphi }$. From the
constraint (\ref{m4}), we have $\hat{f}_{5}^{eq}=P+\rho u_{r}^{2}$, $\hat{f}%
_{6}^{eq}=\rho u_{r}u_{\theta }$, $\hat{f}_{7}^{eq}=\rho u_{r}u_{\varphi }$,
$\hat{f}_{8}^{eq}=P+\rho u_{\theta }^{2}$, $\hat{f}_{9}^{eq}=\rho u_{\theta
}u_{\varphi }$, $\hat{f}_{10}^{eq}=P+\rho u_{\varphi }^{2}$. From the
constraint (\ref{m6}), we have $\hat{f}_{11}^{eq}=\rho \left[ T\left(
3u_{r}\right) +u_{r}^{3}\right] $, $\hat{f}_{12}^{eq}=\rho \left( Tu_{\theta
}+u_{r}^{2}u_{\theta }\right) $, $\hat{f}_{13}^{eq}=\rho \left( Tu_{\varphi
}+u_{r}^{2}u_{\varphi }\right) $, $\hat{f}_{14}^{eq}=\rho \left(
Tu_{r}+u_{r}u_{\theta }^{2}\right) $, $\hat{f}_{15}^{eq}=\rho \left(
u_{r}u_{\theta }u_{\varphi }\right) $, $\hat{f}_{16}^{eq}=\rho \left(
Tu_{r}+u_{r}u_{\varphi }^{2}\right) $, $\hat{f}_{17}^{eq}=\rho \left[
T\left( 3u_{\theta }\right) +u_{\theta }^{3}\right] $, $\hat{f}%
_{18}^{eq}=\rho \left[ Tu_{\varphi }+u_{\theta }^{2}u_{\varphi }\right] $, $%
\hat{f}_{19}^{eq}=\rho \left[ Tu_{\theta }+u_{\varphi }^{2}u_{\theta }\right]
$, $\hat{f}_{20}^{eq}=\rho \left[ T\left( 3u_{\varphi }\right) +u_{\varphi
}^{3}\right] $. From the constraint (\ref{meq7}), we have $\hat{f}%
_{21}^{eq}=\rho T\left( 5T+u^{2}\right) +\rho u_{r}^{2}\left(
7T+u^{2}\right) $, $\hat{f}_{22}^{eq}=\rho u_{r}u_{\theta }\left(
7T+u^{2}\right) $, $\hat{f}_{23}^{eq}=\rho u_{r}u_{\varphi }\left(
7T+u^{2}\right) $, $\hat{f}_{24}^{eq}=\rho T\left( 5T+u^{2}\right) +\rho
u_{\theta }^{2}\left( 7T+u^{2}\right) $, $\hat{f}_{25}^{eq}=\rho u_{\theta
}u_{\varphi }\left( 7T+u^{2}\right) $, $\hat{f}_{26}^{eq}=\rho T\left(
5T+u^{2}\right) +2\rho u_{\varphi }^{2}\left( 7T+u^{2}\right) $.

Since the system is spherically symmetric in macroscopic scale, $u_{\theta
}=u_{\varphi }=0$ and $u^{2}=u_{r}^{2}$ in the above expressions for $%
\mathbf{\hat{f}}^{eq}=\left[ \hat{f}_{1}^{eq}\text{, }\hat{f}_{2}^{eq}\text{,%
}\cdots \text{,}\hat{f}_{N}^{eq}\right] ^{T}$.

The components of the matrix $\mathbf{C=}\left[ \mathbf{C}_{k}\right]
\mathbf{=}\left[ C_{ki}\right] $ should be fixed in the same sequence, where
$k=1$,$2$,$\cdots $,$26$ and $i=1$,$2$,$\cdots $,$26$. From the constraint (%
\ref{m1}),we have $C_{1i}=1$. From the constraint (\ref{m2}), we have $%
C_{2i}=v_{ir}$, $C_{3i}=v_{i\theta }$, $C_{4i}=v_{i\varphi }$. From the
constraint (\ref{m4}), we have $C_{5i}=v_{ir}^{2}$, $C_{6i}=v_{ir}v_{i\theta
}$, $C_{7i}=v_{ir}v_{i\varphi }$, $C_{8i}=v_{i\theta }^{2}$, $%
C_{9i}=v_{i\theta }v_{i\varphi }$, $C_{10i}=v_{i\varphi }^{2}$. From the
constraint (\ref{m6}), we have $C_{11i}=v_{ir}^{3}$, $C_{12i}=v_{ir}^{2}v_{i%
\theta }$, $C_{13i}=v_{ir}^{2}v_{i\varphi }$, $C_{14i}=v_{ir}v_{i\theta
}^{2} $, $C_{15i}=v_{ir}v_{i\theta }v_{i\varphi }$, $C_{16i}=v_{ir}v_{i%
\varphi }^{2}$, $C_{17i}=v_{i\theta }^{3}$, $C_{18i}=v_{i\theta
}^{2}v_{i\varphi }$, $C_{19i}=v_{i\theta }v_{i\varphi }^{2}$, $%
C_{20i}=v_{i\varphi }^{3}$. From the constraint (\ref{meq7}), we have $%
C_{21i}=\left( v_{ir}^{2}+v_{i\theta }^{2}+v_{i\varphi }^{2}\right)
v_{ir}^{2}$, $C_{22i}=\left( v_{ir}^{2}+v_{i\theta }^{2}+v_{i\varphi
}^{2}\right) v_{ir}v_{i\theta }$, $C_{23i}=\left( v_{ir}^{2}+v_{i\theta
}^{2}+v_{i\varphi }^{2}\right) v_{ir}v_{i\varphi }$, $C_{24i}=\left(
v_{ir}^{2}+v_{i\theta }^{2}+v_{i\varphi }^{2}\right) v_{i\theta }^{2}$, $%
C_{25i}=\left( v_{ir}^{2}+v_{i\theta }^{2}+v_{i\varphi }^{2}\right)
v_{i\theta }v_{i\varphi }$, $C_{26i}=\left( v_{ir}^{2}+v_{i\theta
}^{2}+v_{i\varphi }^{2}\right) v_{i\varphi }^{2}$.

An example for the 3-Dimensional 26-Velocity(D3V26) DVM is as below,
\begin{subequations}
\begin{eqnarray}
\mathbf{v}_{i} &=&\left\{
\begin{array}{cc}
\left( 0,\pm 1,\pm 1\right) c_{1} & i=1,\cdots ,4 \\
\left( \pm 1,0,\pm 1\right) c_{1} & i=5,\cdots ,8 \\
\left( \pm 1,\pm 1,0\right) c_{1} & i=9,\cdots ,12%
\end{array}%
\right. \\
\mathbf{v}_{i} &=&\left\{
\begin{array}{cc}
\left( \pm 1,\pm 1,\pm 1\right) c_{2} & i=13,\cdots ,20%
\end{array}%
\right. \\
\mathbf{v}_{i} &=&\left\{
\begin{array}{cc}
\left( \pm 1,0,0\right) c_{3} & i=21,22 \\
\left( 0,\pm 1,0\right) c_{3} & i=23,24 \\
\left( 0,0,\pm 1\right) c_{3} & i=25,26%
\end{array}%
\right. .
\end{eqnarray}
A specific example of the DVM and the expressions for the inverse of the
matrix $\mathbf{C}$, $\mathbf{C}^{-1}=\left[ \mathbf{C}_{k}^{-1}\right] $,
are shown in the appendix A. Up to this step, a the special discretization
of the velocity space has been performed. Consequently, the DBM for system
with spherical symmetry and $\gamma =5/3$ has been constructed. The spatial
and temporal derivatives of the distribution function in the kinetic model
can be calculated in the normal way. If we are not interested in the extra
degrees of freedom other than the translational, the formulated DBM can be
used to study the hydrodynamic and the thermodynamic behaviors of the
compressible flow system. In this case, $E=e$ in the hydrodynamic energy
equation.

\subsubsection{Case with flexible $\protect\gamma $}

For the case with flexible $\gamma $, if we are interested only in the
hydrodynamic behaviors, we can use the simulation results of the DBM
formulated in last subsection. Just get the number $n$ of the extra degree
of freedom using its relation to $\gamma $, then obtain the total internal
energy $E$ using its definition. If we are interested also in the
thermodynamic nonequilibrium behavior, we need continue the formulation of
the DBM.

To model the case with flexible ratio of specific heats, we resort to the
parameter $\eta _{i}$ to describe the contribution of extra degrees of
freedom. From the constraints (\ref{m4}) and (\ref{m3}) we have
\end{subequations}
\begin{subequations}
\begin{equation}
n\rho RT=\sum_{i=1}^{N}f_{i}^{eq}\eta _{i}^{2}\text{,}  \label{m8}
\end{equation}%
From the constraints (\ref{m5}) and (\ref{m6}), we have
\begin{equation}
n\rho RTu_{\alpha }=\sum_{i=1}^{N}f_{i}^{eq}v_{i\alpha }\eta _{i}^{2}\text{,}
\label{m9}
\end{equation}%
From the constraint (\ref{meq7}) we have
\begin{equation}
n\rho RT\left( RT\delta _{\alpha \beta }+u_{\alpha }u_{\beta }\right) =\sum
f_{i}^{eq}v_{i\alpha }v_{i\beta }\eta _{i}^{2}\text{.}  \label{m10}
\end{equation}%
The constraints (\ref{m8}) - (\ref{m10}) compose $10$ linear equations for
the variable $\eta _{i}^{2}$. It is clear that we can set $\eta _{i}=0$ for $%
i=11$, $\cdots $,$26$.Therefore, once the DBM for the system with $\gamma
=5/3$ are fixed, we can further fixed $\eta _{i}^{2}$. Via replacing the
translational kinetic energy $e$ with the total internal kinetic energy $E$,
the D2V26-DBM can be used to model system with flexible ratio of specific
heats.

The constraints (\ref{m8}) - (\ref{m10}) can be rewritten as
\end{subequations}
\begin{equation}
\mathbf{\hat{g}}=\mathbf{Dg}  \label{m11}
\end{equation}%
where
\begin{eqnarray}
\mathbf{\hat{g}} &=&\left[ \hat{g}_{k}\right] ^{T}  \notag \\
&=&n\rho RT\left[ 1\text{,}u_{r}\text{,}u_{\theta }\text{,}u_{\varphi }\text{%
,}RT+u_{r}^{2}\text{,}u_{r}u_{\theta }\text{,}u_{r}u_{\varphi }\text{,}%
RT+u_{\theta }^{2}\text{,}u_{\theta }u_{\varphi }\text{,}RT+u_{\varphi }^{2}%
\right]
\end{eqnarray}%
and
\begin{eqnarray}
\mathbf{g} &=&\left[ g_{k}\right] ^{T}  \notag \\
&=&\left[ \eta _{i}^{2}\text{, }i=1\text{, }\cdots \text{,}10\right]
\end{eqnarray}%
are column vectors with $k=1$,$2$,$\cdots $,$10$, $\mathbf{D=}\left[ \mathbf{%
D}_{k}\right] =\left[ D_{ki}\right] $ is $10\times 10$ matrix whose
components are determined by $f_{i}^{eq}$ and $v_{i\alpha }$. Specifically, $%
\mathbf{D}_{1}=\left[ f_{i}^{eq}\right] $, $\mathbf{D}_{2}=\left[
f_{i}^{eq}v_{ir}\right] $, $\mathbf{D}_{3}=\left[ f_{i}^{eq}v_{i\theta }%
\right] $, $\mathbf{D}_{4}=\left[ f_{i}^{eq}v_{i\varphi }\right] $, $\mathbf{%
D}_{5}=\left[ f_{i}^{eq}v_{ir}^{2}\right] $, $\mathbf{D}_{6}=\left[
f_{i}^{eq}v_{ir}v_{i\theta }\right] $, $\mathbf{D}_{7}=\left[
f_{i}^{eq}v_{ir}v_{i\varphi }\right] $, $\mathbf{D}_{8}=\left[
f_{i}^{eq}v_{i\theta }^{2}\right] $, $\mathbf{D}_{9}=\left[
f_{i}^{eq}v_{i\theta }v_{i\varphi }\right] $, $\mathbf{D}_{10}=\left[
f_{i}^{eq}v_{i\varphi }^{2}\right] $

Since $f_{i}^{eq}$ has been determined via%
\begin{equation}
\mathbf{f}^{eq}=\mathbf{C}^{-1}\mathbf{\hat{f}}^{eq}\text{,}
\end{equation}%
finally,
\begin{equation}
\mathbf{g=D}^{-1}\mathbf{\hat{g}}\text{.}  \label{g3d}
\end{equation}

Via substituting the results $\eta _{i}^{2}$ to Eqs. (\ref{Me4}) - (\ref{Me7}%
), then using the Eqs. (\ref{De1}) and (\ref{De2}), the thermodynamic
nonequilibrium behavior can be investigated. Up to this step, a DBM with
D2V26 for system with flexible specific heat ratio $\gamma $ has been
formulated.

\subsection{DBM for systems with cylindrical symmetry}

When the system depends also on the value of $z$, the DBM needs a $3$%
-dimensional DVM with $N=26$. The DVM can be similar to the case with
spherical symmetry. The only difference is that the subscript ``$\varphi $''
should be replaced with ``$z$''.

If the system is also independent of the height $z$, The discrete Boltzmann
equation (\ref{DBC1}) is simplified as
\begin{equation}
\partial _{t}f_{i}+v_{ir}\frac{\partial f_{i}}{\partial r}+\left( \frac{%
v_{ir}v_{i\theta }^{2}}{rT}-\frac{v_{i\theta }^{2}\left( v_{ir}-u_{r}\right)
}{rT}\right) f_{i}^{eq}=-\frac{1}{\tau }\left( f_{i}-f_{i}^{eq}\right) \text{%
.}  \label{DBC2}
\end{equation}%
A complete DBM needs also a DVM with 3 dimensions. But when the degree of
freedom in $z$ can be ignored, to recover the Navier-Stokes equations the
continuum limit, the DVM needs only 2 Dimensions and 13 Velocities(D2V13).
In this formulation scheme, we need the minimum number of discrete
velocities, but have to add extra calculations, as shown by Eq.(\ref{g3d}),
to fix $\eta _{i}^{2}$.

In this section we consider a slightly different scheme. We use a DVM with 2
Dimensions and 16 Velocities (D2V16) to formulate the DBM which recovers the
Navier-Stokes equations in the continuum limit. From the constraints (\ref%
{m1}) - (\ref{meq7}), we choose
\begin{equation}
\mathbf{\hat{f}}^{eq}=\left[ \hat{f}_{1}^{eq}\text{, }\hat{f}_{2}^{eq}\text{%
, }\cdots \text{, }\hat{f}_{16}^{eq}\right] ^{T}\mathtt{,}
\end{equation}%
where  $\hat{f}_{1}^{eq}=\rho $, $\hat{f}_{2}^{eq}=\rho u_{r}$,$\hat{f}%
_{3}^{eq}=\rho u_{\theta }$, $\hat{f}_{4}^{eq}=\rho \left[ \left( n+2\right)
RT+\left( u_{r}^{2}+u_{\theta }^{2}\right) \right] $, $\hat{f}%
_{5}^{eq}=P+\rho u_{r}^{2}$, $\hat{f}_{6}^{eq}=\rho u_{r}u_{\theta }$, $\hat{%
f}_{7}^{eq}=P+\rho u_{\theta }^{2}$, $\hat{f}_{8}^{eq}=\rho \lbrack
(n+4)RT+\left( u_{r}^{2}+u_{\theta }^{2}\right) ]u_{r}$, $\hat{f}%
_{9}^{eq}=\rho \lbrack (n+4)RT+\left( u_{r}^{2}+u_{\theta }^{2}\right)
]u_{\theta }$, $\hat{f}_{10}^{eq}=\rho \left[ RT\left( 3u_{r}\right)
+u_{r}^{3}\right] $, \thinspace\ \thinspace\ \thinspace\ \thinspace\
\thinspace\ \thinspace\ \thinspace\ \thinspace\ $\hat{f}_{11}^{eq}=\rho
\left( RTu_{\theta }+u_{r}^{2}u_{\theta }\right) $, \thinspace\ \thinspace\
\thinspace\ \thinspace\ \thinspace\ \thinspace\ \thinspace\ \thinspace\ $%
\hat{f}_{12}^{eq}=\rho \left( RTu_{r}+u_{\theta }^{2}u_{r}\right) $, \newline
$\hat{f}_{13}^{eq}=\rho \left[ RT\left( 3u_{\theta }\right) +u_{\theta }^{3}%
\right] $, $\hat{f}_{14}^{eq}=\rho RT\left[ \left( n+4\right) RT+u^{2}\right]
+\rho u_{r}^{2}\left[ \left( n+6\right) RT+u^{2}\right] $, $\hat{f}%
_{15}^{eq}=\rho u_{r}u_{\theta }\left[ \left( n+6\right) RT+u^{2}\right] $,
$\hat{f}_{16}^{eq}=\rho RT\left[ \left( n+4\right) RT+u^{2}\right] +\rho
u_{\theta }^{2}\left[ \left( n+6\right) RT+u^{2}\right] $. Since the system
is cylindrically symmetric in macroscopic scale and independent of the value
of $z$, $u_{\theta }=u_{z}=0$ in the above expressions for $\hat{f}_{i}^{eq}$%
.

In the same sequence, we fix the matrix $\mathbf{C=}\left[ \mathbf{C}_{k}%
\right] \mathbf{=}\left[ C_{ki}\right] $ below. $C_{1i}=1$, $C_{2i}=v_{ir}$,
$C_{3i}=v_{i\theta }$, $C_{4i}=\left( v_{ir}^{2}+v_{i\theta }^{2}+\eta
_{i}^{2}\right) $, $C_{5i}=v_{ir}^{2}$, $C_{6i}=v_{ir}v_{i\theta }$, $%
C_{7i}=v_{i\theta }^{2}$, $C_{8i}=\left( v_{ir}^{2}+v_{i\theta }^{2}+\eta
_{i}^{2}\right) v_{ir}$, $C_{9i}=\left( v_{ir}^{2}+v_{i\theta }^{2}+\eta
_{i}^{2}\right) v_{i\theta }$, $C_{10i}=v_{ir}^{3}$, $C_{11i}=v_{ir}^{2}v_{i%
\theta }$, $C_{12i}=v_{ir}v_{i\theta }^{2}$, $C_{13i}=v_{i\theta }^{3}$, $%
C_{14i}=\left( v_{ir}^{2}+v_{i\theta }^{2}+\eta _{i}^{2}\right) v_{ir}^{2}$,
$C_{15i}=\left( v_{ir}^{2}+v_{i\theta }^{2}+\eta _{i}^{2}\right)
v_{ir}v_{i\theta }$, $C_{16i}=\left( v_{ir}^{2}+v_{i\theta }^{2}+\eta
_{i}^{2}\right) v_{i\theta }^{2}$.

It should be pointed out that, in this formulation scheme, the values of $%
\eta _{i}$ have to be chosen in such a way that the inverse of the matrix $%
\mathbf{C}$ exists.

An example of the D2V16 is as below,
\begin{subequations}
\begin{eqnarray}
\mathbf{v}_{i} &=&\left\{
\begin{array}{cc}
\left( \pm 1,0\right) c_{1} &  \\
\left( 0,\pm 1\right) c_{1} & , \\
\left( \pm 1,\pm 1\right) c_{2} &
\end{array}%
i=1,\cdots ,8\right. \\
\mathbf{v}_{i} &=&\left\{
\begin{array}{cc}
\left( \pm 1,0\right) c_{3} &  \\
\left( 0,\pm 1\right) c_{3} & , \\
\left( \pm 1,\pm 1\right) c_{4} &
\end{array}%
i=9,\cdots ,16\text{.}\right.
\end{eqnarray}
A specific example of the DVM with specific $\eta _{i}$ and the
corresponding components of $\mathbf{C}^{-1}=\left[ \mathbf{C}_{k}^{-1}%
\right] $ are shown in appendix B.

\section{Conclusion and discussions}

We present a theoretical framework for constructing discrete Boltzmann model
in spherical or cylindrical coordinates for the compressible flow systems
with spherical or cylindrical symmetry. A common property of the two kinds
symmetric systems is that the behavior of the system is independent of the
azimuthal coordinate. A key technique here is to use \emph{local} Cartesian
coordinates to describe the particle velocity. Thus, in the Boltzmann
equation of such a system the geometric effects, like the divergence and
convergence, are described as a \textquotedblleft force
term\textquotedblright .

For a system with spherical or \emph{global} cylindrical symmetry, the
hydrodynamic equations depend only on the radial coordinate. The
hydrodynamic equations for system with local cylindrical symmetry may depend
also on the height $z$. For such a system, even though the hydrodynamic
models are one- or two-dimensional, the discrete Boltzmann model needs a DVM
with 3 dimensions. We use a DVM with 26 velocities to formulate the DBM
which recovers the Navier-Stokes equations with spherical or cylindrical
symmetry in the hydrodynamic limit.

For the system with \emph{global} cylindrical symmetry, we formulate also a
DBM based on a DVM with 2 dimensions and 16 velocities. But it should be
pointed out that, the system described by the DBM based on D2V16 is not the
same as that described by the DBM based on D3V26. For example, the equation
of state, $P=2\rho e/D$, is different for the two kinds of systems.

Besides recovering the hydrodynamic equations in the continuum limit, the
more important point for a DBM is that, in terms of the nonconserved
moments, we can define two sets of measures for the deviation of the system
from its thermodynamic equilibrium state. The measurements are clarified.
With the DBM we can study \emph{simultaneously} both the hydrodynamic and
thermodynamic behaviors. Since the inverse of the transformation matrix $%
\mathbf{C}$ connecting the discrete equilibrium distribution function $%
\mathbf{f}^{eq}$ and corresponding moments $\mathbf{\hat{f}}^{eq}$ has been
fixed, the extension to multiple-relaxation-time DBM\cite%
{MRT2010EPL,XuChen2014FoP} is straightforward.
It should also be pointed out
that, fixing the transformation matrix $\mathbf{C}$ and its inverse is only
one of the possible schemes to get a solution for the discrete equilibrium
distribution function $\mathbf{f}^{eq}$. A second way to find a solution for
the discrete equilibrium distribution function $\mathbf{f}^{eq}$ is to follow the ideas used in
 Refs. \cite{Watari2003,Watari2004,Watari2007}. A difference is that the scheme introduced in this work needs
  the minimum number of discrete velocities.
  
\section*{Acknowledgements}

The authors acknowledge support of the Science Foundations of CAEP [under
Grant No. 2012B0101014] and the Foundation of State Key Laboratory of
Explosion Science and Technology [under Grant No. KFJJ14-1M],National
Natural Science Foundation of China [under Grant Nos. 11074300 and
91130020], National Basic Research Program of China (Grant No. 2013CBA01504)
and the Open Project Program of State Key Laboratory of Theoretical Physics,
Institute of Theoretical Physics, Chinese Academy of Sciences, China (under
Grant No.Y4KF151CJ1).


\appendix

\section{$\mathbf{C}^{-1}$ for the D3V26}


Specifically, we can choose
\end{subequations}
\begin{subequations}
\begin{equation}
\left.
\begin{array}{c}
v_{ir}=c_{1}[0,0,0,0,1,1,-1,-1,1,1,-1,-1] \\
v_{i\theta }=c_{1}[1,1,-1,-1,0,0,0,0,1,-1,-1,1] \\
v_{i\varphi }=c_{1}[1,-1,-1,1,1,-1,-1,1,0,0,0,0]%
\end{array}%
\right\} \text{ for }i=1\text{,}\cdots \text{,}12\text{,}
\end{equation}
\begin{equation}
\left.
\begin{array}{c}
v_{ir}=c_{2}[1,1,1,1,-1,-1,-1,-1] \\
v_{i\theta }=c_{2}[1,1,-1,-1,1,1,-1,-1] \\
v_{i\varphi }=c_{2}[1,-1,-1,1,1,-1,-1,1]%
\end{array}%
\right\} \text{ for }i=13\text{,}\cdots \text{,}20\text{,}
\end{equation}
\begin{equation}
\left.
\begin{array}{c}
v_{ir}=c_{3}[1,-1,0,0,0,0] \\
v_{i\theta }=c_{3}[0,0,1,-1,0,0] \\
v_{i\varphi }=c_{3}[0,0,0,0,1,-1]%
\end{array}%
\right\} \text{ for }i=21\text{,}\cdots \text{,}26\text{,}
\end{equation}
The components of the inverse of $\mathbf{C}$, $\mathbf{C}^{-1}=\left[
\mathbf{C}_{k}^{-1}\right] $, are shown below.
\end{subequations}
\begin{subequations}
\begin{eqnarray}
\mathbf{C}_{1}^{-1} &=&[1/4\,{\frac{{c_{3}}^{2}{c_{2}}^{2}}{\left( 2\,{c_{1}}%
^{2}-{c_{3}}^{2}\right) \left( 2\,{c_{1}}^{2}-3\,{c_{2}}^{2}\right) }}\text{,%
}0\text{,}-1/4\,{\frac{{c_{3}}^{2}{c_{2}}^{2}}{c_{1}\,\left( {c_{1}}^{2}{%
c_{2}}^{2}+{c_{1}}^{2}{c_{3}}^{2}-2\,{c_{3}}^{2}{c_{2}}^{2}\right) }}\text{,}
\notag \\
&&-1/4\,{\frac{{c_{3}}^{2}{c_{2}}^{2}}{c_{1}\,\left( {c_{1}}^{2}{c_{2}}^{2}+{%
c_{1}}^{2}{c_{3}}^{2}-2\,{c_{3}}^{2}{c_{2}}^{2}\right) }}\text{,}-1/4\,{%
\frac{{c_{1}}^{2}{c_{2}}^{2}-{c_{1}}^{2}{c_{3}}^{2}+2\,{c_{3}}^{2}{c_{2}}^{2}%
}{{c_{1}}^{2}\left( 2\,{c_{1}}^{2}-{c_{3}}^{2}\right) \left( 2\,{c_{1}}%
^{2}-3\,{c_{2}}^{2}\right) }}\text{,}  \notag \\
&&0\text{,}0\text{,}-1/4\,{\frac{{c_{1}}^{2}{c_{2}}^{2}+{c_{1}}^{2}{c_{3}}%
^{2}-{c_{3}}^{2}{c_{2}}^{2}}{{c_{1}}^{2}\left( 2\,{c_{1}}^{2}-{c_{3}}%
^{2}\right) \left( 2\,{c_{1}}^{2}-3\,{c_{2}}^{2}\right) }}\text{,}-3/4\,{%
\frac{{c_{2}}^{2}}{{c_{1}}^{2}\left( 2\,{c_{1}}^{2}-3\,{c_{2}}^{2}\right) }}%
\text{,}  \notag \\
&&-1/4\,{\frac{{c_{1}}^{2}{c_{2}}^{2}+{c_{1}}^{2}{c_{3}}^{2}-{c_{3}}^{2}{%
c_{2}}^{2}}{{c_{1}}^{2}\left( 2\,{c_{1}}^{2}-{c_{3}}^{2}\right) \left( 2\,{%
c_{1}}^{2}-3\,{c_{2}}^{2}\right) }}\text{,}0\text{,}-1/4\,{\frac{\left( {%
c_{1}}^{2}-{c_{3}}^{2}\right) {c_{2}}^{2}}{{c_{1}}^{3}\left( {c_{1}}^{2}{%
c_{2}}^{2}+{c_{1}}^{2}{c_{3}}^{2}-2\,{c_{3}}^{2}{c_{2}}^{2}\right) }}\text{,}
\notag \\
&&-1/4\,{\frac{\left( {c_{1}}^{2}-{c_{3}}^{2}\right) {c_{2}}^{2}}{{c_{1}}%
^{3}\left( {c_{1}}^{2}{c_{2}}^{2}+{c_{1}}^{2}{c_{3}}^{2}-2\,{c_{3}}^{2}{c_{2}%
}^{2}\right) }}\text{,}0\text{,}0\text{,}0\text{,}1/4\,{\frac{{c_{2}}^{2}}{%
c_{1}\,\left( {c_{1}}^{2}{c_{2}}^{2}+{c_{1}}^{2}{c_{3}}^{2}-2\,{c_{3}}^{2}{%
c_{2}}^{2}\right) }}\text{,}  \notag \\
&&1/4\,{\frac{\left( {c_{1}}^{2}-{c_{2}}^{2}\right) {c_{3}}^{2}}{{c_{1}}%
^{3}\left( {c_{1}}^{2}{c_{2}}^{2}+{c_{1}}^{2}{c_{3}}^{2}-2\,{c_{3}}^{2}{c_{2}%
}^{2}\right) }}\text{,}1/4\,{\frac{\left( {c_{1}}^{2}-{c_{2}}^{2}\right) {%
c_{3}}^{2}}{{c_{1}}^{3}\left( {c_{1}}^{2}{c_{2}}^{2}+{c_{1}}^{2}{c_{3}}%
^{2}-2\,{c_{3}}^{2}{c_{2}}^{2}\right) }}\text{,}  \notag \\
&&1/4\,{\frac{{c_{2}}^{2}}{c_{1}\,\left( {c_{1}}^{2}{c_{2}}^{2}+{c_{1}}^{2}{%
c_{3}}^{2}-2\,{c_{3}}^{2}{c_{2}}^{2}\right) }}\text{,}-1/4\,{\frac{{c_{1}}%
^{2}-2\,{c_{2}}^{2}}{{c_{1}}^{2}\left( 2\,{c_{1}}^{2}-{c_{3}}^{2}\right)
\left( 2\,{c_{1}}^{2}-3\,{c_{2}}^{2}\right) }}\text{,}  \notag \\
&&0\text{,}0\text{,}1/4\,{\frac{{c_{1}}^{2}-{c_{2}}^{2}}{{c_{1}}^{2}\left(
2\,{c_{1}}^{2}-{c_{3}}^{2}\right) \left( 2\,{c_{1}}^{2}-3\,{c_{2}}%
^{2}\right) }}\text{,}1/4\,{\frac{1}{{c_{1}}^{2}\left( 2\,{c_{1}}^{2}-3\,{%
c_{2}}^{2}\right) }}\text{,}  \notag \\
&&1/4\,{\frac{{c_{1}}^{2}-{c_{2}}^{2}}{{c_{1}}^{2}\left( 2\,{c_{1}}^{2}-{%
c_{3}}^{2}\right) \left( 2\,{c_{1}}^{2}-3\,{c_{2}}^{2}\right) }}]
\end{eqnarray}

\begin{eqnarray}
\mathbf{C}_{2}^{-1} &=&[1/4\,{\frac{{c_{3}}^{2}{c_{2}}^{2}}{\left( 2\,{c_{1}}%
^{2}-{c_{3}}^{2}\right) \left( 2\,{c_{1}}^{2}-3\,{c_{2}}^{2}\right) }}\text{,%
}0\text{,}-1/4\,{\frac{{c_{3}}^{2}{c_{2}}^{2}}{c_{1}\,\left( {c_{1}}^{2}{%
c_{2}}^{2}+{c_{1}}^{2}{c_{3}}^{2}-2\,{c_{3}}^{2}{c_{2}}^{2}\right) }}\text{,}
\notag \\
&&1/4\,{\frac{{c_{3}}^{2}{c_{2}}^{2}}{c_{1}\,\left( {c_{1}}^{2}{c_{2}}^{2}+{%
c_{1}}^{2}{c_{3}}^{2}-2\,{c_{3}}^{2}{c_{2}}^{2}\right) }}\text{,}-1/4\,{%
\frac{{c_{1}}^{2}{c_{2}}^{2}-{c_{1}}^{2}{c_{3}}^{2}+2\,{c_{3}}^{2}{c_{2}}^{2}%
}{{c_{1}}^{2}\left( 2\,{c_{1}}^{2}-{c_{3}}^{2}\right) \left( 2\,{c_{1}}%
^{2}-3\,{c_{2}}^{2}\right) }}\text{,}  \notag \\
&&0\text{,}0\text{,}-1/4\,{\frac{{c_{1}}^{2}{c_{2}}^{2}+{c_{1}}^{2}{c_{3}}%
^{2}-{c_{3}}^{2}{c_{2}}^{2}}{{c_{1}}^{2}\left( 2\,{c_{1}}^{2}-{c_{3}}%
^{2}\right) \left( 2\,{c_{1}}^{2}-3\,{c_{2}}^{2}\right) }}\text{,}3/4\,{%
\frac{{c_{2}}^{2}}{{c_{1}}^{2}\left( 2\,{c_{1}}^{2}-3\,{c_{2}}^{2}\right) }}%
\text{,}  \notag \\
&&-1/4\,{\frac{{c_{1}}^{2}{c_{2}}^{2}+{c_{1}}^{2}{c_{3}}^{2}-{c_{3}}^{2}{%
c_{2}}^{2}}{{c_{1}}^{2}\left( 2\,{c_{1}}^{2}-{c_{3}}^{2}\right) \left( 2\,{%
c_{1}}^{2}-3\,{c_{2}}^{2}\right) }}\text{,}0\text{,}-1/4\,{\frac{\left( {%
c_{1}}^{2}-{c_{3}}^{2}\right) {c_{2}}^{2}}{{c_{1}}^{3}\left( {c_{1}}^{2}{%
c_{2}}^{2}+{c_{1}}^{2}{c_{3}}^{2}-2\,{c_{3}}^{2}{c_{2}}^{2}\right) }}\text{,}
\notag \\
&&1/4\,{\frac{\left( {c_{1}}^{2}-{c_{3}}^{2}\right) {c_{2}}^{2}}{{c_{1}}%
^{3}\left( {c_{1}}^{2}{c_{2}}^{2}+{c_{1}}^{2}{c_{3}}^{2}-2\,{c_{3}}^{2}{c_{2}%
}^{2}\right) }}\text{,}0\text{,}0\text{,}0\text{,}1/4\,{\frac{{c_{2}}^{2}}{%
c_{1}\,\left( {c_{1}}^{2}{c_{2}}^{2}+{c_{1}}^{2}{c_{3}}^{2}-2\,{c_{3}}^{2}{%
c_{2}}^{2}\right) }}\text{,}  \notag \\
&&-1/4\,{\frac{\left( {c_{1}}^{2}-{c_{2}}^{2}\right) {c_{3}}^{2}}{{c_{1}}%
^{3}\left( {c_{1}}^{2}{c_{2}}^{2}+{c_{1}}^{2}{c_{3}}^{2}-2\,{c_{3}}^{2}{c_{2}%
}^{2}\right) }}\text{,}1/4\,{\frac{\left( {c_{1}}^{2}-{c_{2}}^{2}\right) {%
c_{3}}^{2}}{{c_{1}}^{3}\left( {c_{1}}^{2}{c_{2}}^{2}+{c_{1}}^{2}{c_{3}}%
^{2}-2\,{c_{3}}^{2}{c_{2}}^{2}\right) }}\text{,}  \notag \\
&&-1/4\,{\frac{{c_{2}}^{2}}{c_{1}\,\left( {c_{1}}^{2}{c_{2}}^{2}+{c_{1}}^{2}{%
c_{3}}^{2}-2\,{c_{3}}^{2}{c_{2}}^{2}\right) }}\text{,}-1/4\,{\frac{{c_{1}}%
^{2}-2\,{c_{2}}^{2}}{{c_{1}}^{2}\left( 2\,{c_{1}}^{2}-{c_{3}}^{2}\right)
\left( 2\,{c_{1}}^{2}-3\,{c_{2}}^{2}\right) }}\text{,}  \notag \\
&&0\text{,}0\text{,}1/4\,{\frac{{c_{1}}^{2}-{c_{2}}^{2}}{{c_{1}}^{2}\left(
2\,{c_{1}}^{2}-{c_{3}}^{2}\right) \left( 2\,{c_{1}}^{2}-3\,{c_{2}}%
^{2}\right) }}\text{,}-1/4\,{\frac{1}{{c_{1}}^{2}\left( 2\,{c_{1}}^{2}-3\,{%
c_{2}}^{2}\right) }}\text{,}  \notag \\
&&1/4\,{\frac{{c_{1}}^{2}-{c_{2}}^{2}}{{c_{1}}^{2}\left( 2\,{c_{1}}^{2}-{%
c_{3}}^{2}\right) \left( 2\,{c_{1}}^{2}-3\,{c_{2}}^{2}\right) }}]
\end{eqnarray}

\begin{eqnarray}
\mathbf{C}_{3}^{-1} &=&[1/4\,{\frac{{c_{3}}^{2}{c_{2}}^{2}}{\left( 2\,{c_{1}}%
^{2}-{c_{3}}^{2}\right) \left( 2\,{c_{1}}^{2}-3\,{c_{2}}^{2}\right) }}\text{,%
}0\text{,}1/4\,{\frac{{c_{3}}^{2}{c_{2}}^{2}}{c_{1}\,\left( {c_{1}}^{2}{c_{2}%
}^{2}+{c_{1}}^{2}{c_{3}}^{2}-2\,{c_{3}}^{2}{c_{2}}^{2}\right) }}\text{,}
\notag \\
&&1/4\,{\frac{{c_{3}}^{2}{c_{2}}^{2}}{c_{1}\,\left( {c_{1}}^{2}{c_{2}}^{2}+{%
c_{1}}^{2}{c_{3}}^{2}-2\,{c_{3}}^{2}{c_{2}}^{2}\right) }}\text{,}-1/4\,{%
\frac{{c_{1}}^{2}{c_{2}}^{2}-{c_{1}}^{2}{c_{3}}^{2}+2\,{c_{3}}^{2}{c_{2}}^{2}%
}{{c_{1}}^{2}\left( 2\,{c_{1}}^{2}-{c_{3}}^{2}\right) \left( 2\,{c_{1}}%
^{2}-3\,{c_{2}}^{2}\right) }}\text{,}  \notag \\
&&0\text{,}0\text{,}-1/4\,{\frac{{c_{1}}^{2}{c_{2}}^{2}+{c_{1}}^{2}{c_{3}}%
^{2}-{c_{3}}^{2}{c_{2}}^{2}}{{c_{1}}^{2}\left( 2\,{c_{1}}^{2}-{c_{3}}%
^{2}\right) \left( 2\,{c_{1}}^{2}-3\,{c_{2}}^{2}\right) }}\text{,}-3/4\,{%
\frac{{c_{2}}^{2}}{{c_{1}}^{2}\left( 2\,{c_{1}}^{2}-3\,{c_{2}}^{2}\right) }}%
\text{,}  \notag \\
&&-1/4\,{\frac{{c_{1}}^{2}{c_{2}}^{2}+{c_{1}}^{2}{c_{3}}^{2}-{c_{3}}^{2}{%
c_{2}}^{2}}{{c_{1}}^{2}\left( 2\,{c_{1}}^{2}-{c_{3}}^{2}\right) \left( 2\,{%
c_{1}}^{2}-3\,{c_{2}}^{2}\right) }}\text{,}0\text{,}1/4\,{\frac{\left( {c_{1}%
}^{2}-{c_{3}}^{2}\right) {c_{2}}^{2}}{{c_{1}}^{3}\left( {c_{1}}^{2}{c_{2}}%
^{2}+{c_{1}}^{2}{c_{3}}^{2}-2\,{c_{3}}^{2}{c_{2}}^{2}\right) }}\text{,}
\notag \\
&&1/4\,{\frac{\left( {c_{1}}^{2}-{c_{3}}^{2}\right) {c_{2}}^{2}}{{c_{1}}%
^{3}\left( {c_{1}}^{2}{c_{2}}^{2}+{c_{1}}^{2}{c_{3}}^{2}-2\,{c_{3}}^{2}{c_{2}%
}^{2}\right) }}\text{,}0\text{,}0\text{,}0\text{,}-1/4\,{\frac{{c_{2}}^{2}}{%
c_{1}\,\left( {c_{1}}^{2}{c_{2}}^{2}+{c_{1}}^{2}{c_{3}}^{2}-2\,{c_{3}}^{2}{%
c_{2}}^{2}\right) }}\text{,}  \notag \\
&&-1/4\,{\frac{\left( {c_{1}}^{2}-{c_{2}}^{2}\right) {c_{3}}^{2}}{{c_{1}}%
^{3}\left( {c_{1}}^{2}{c_{2}}^{2}+{c_{1}}^{2}{c_{3}}^{2}-2\,{c_{3}}^{2}{c_{2}%
}^{2}\right) }}\text{,}-1/4\,{\frac{\left( {c_{1}}^{2}-{c_{2}}^{2}\right) {%
c_{3}}^{2}}{{c_{1}}^{3}\left( {c_{1}}^{2}{c_{2}}^{2}+{c_{1}}^{2}{c_{3}}%
^{2}-2\,{c_{3}}^{2}{c_{2}}^{2}\right) }}\text{,}  \notag \\
&&-1/4\,{\frac{{c_{2}}^{2}}{c_{1}\,\left( {c_{1}}^{2}{c_{2}}^{2}+{c_{1}}^{2}{%
c_{3}}^{2}-2\,{c_{3}}^{2}{c_{2}}^{2}\right) }}\text{,}-1/4\,{\frac{{c_{1}}%
^{2}-2\,{c_{2}}^{2}}{{c_{1}}^{2}\left( 2\,{c_{1}}^{2}-{c_{3}}^{2}\right)
\left( 2\,{c_{1}}^{2}-3\,{c_{2}}^{2}\right) }}\text{,}  \notag \\
&&0\text{,}0\text{,}1/4\,{\frac{{c_{1}}^{2}-{c_{2}}^{2}}{{c_{1}}^{2}\left(
2\,{c_{1}}^{2}-{c_{3}}^{2}\right) \left( 2\,{c_{1}}^{2}-3\,{c_{2}}%
^{2}\right) }}\text{,}1/4\,{\frac{1}{{c_{1}}^{2}\left( 2\,{c_{1}}^{2}-3\,{%
c_{2}}^{2}\right) }}\text{,}  \notag \\
&&1/4\,{\frac{{c_{1}}^{2}-{c_{2}}^{2}}{{c_{1}}^{2}\left( 2\,{c_{1}}^{2}-{%
c_{3}}^{2}\right) \left( 2\,{c_{1}}^{2}-3\,{c_{2}}^{2}\right) }}]
\end{eqnarray}

\begin{eqnarray}
\mathbf{C}_{4}^{-1} &=&[1/4\,{\frac{{c_{3}}^{2}{c_{2}}^{2}}{\left( 2\,{c_{1}}%
^{2}-{c_{3}}^{2}\right) \left( 2\,{c_{1}}^{2}-3\,{c_{2}}^{2}\right) }}\text{,%
}0\text{,}1/4\,{\frac{{c_{3}}^{2}{c_{2}}^{2}}{c_{1}\,\left( {c_{1}}^{2}{c_{2}%
}^{2}+{c_{1}}^{2}{c_{3}}^{2}-2\,{c_{3}}^{2}{c_{2}}^{2}\right) }}\text{,}
\notag \\
&&-1/4\,{\frac{{c_{3}}^{2}{c_{2}}^{2}}{c_{1}\,\left( {c_{1}}^{2}{c_{2}}^{2}+{%
c_{1}}^{2}{c_{3}}^{2}-2\,{c_{3}}^{2}{c_{2}}^{2}\right) }}\text{,}-1/4\,{%
\frac{{c_{1}}^{2}{c_{2}}^{2}-{c_{1}}^{2}{c_{3}}^{2}+2\,{c_{3}}^{2}{c_{2}}^{2}%
}{{c_{1}}^{2}\left( 2\,{c_{1}}^{2}-{c_{3}}^{2}\right) \left( 2\,{c_{1}}%
^{2}-3\,{c_{2}}^{2}\right) }}\text{,}  \notag \\
&&0\text{,}0\text{,}-1/4\,{\frac{{c_{1}}^{2}{c_{2}}^{2}+{c_{1}}^{2}{c_{3}}%
^{2}-{c_{3}}^{2}{c_{2}}^{2}}{{c_{1}}^{2}\left( 2\,{c_{1}}^{2}-{c_{3}}%
^{2}\right) \left( 2\,{c_{1}}^{2}-3\,{c_{2}}^{2}\right) }}\text{,}3/4\,{%
\frac{{c_{2}}^{2}}{{c_{1}}^{2}\left( 2\,{c_{1}}^{2}-3\,{c_{2}}^{2}\right) }}%
\text{,}  \notag \\
&&-1/4\,{\frac{{c_{1}}^{2}{c_{2}}^{2}+{c_{1}}^{2}{c_{3}}^{2}-{c_{3}}^{2}{%
c_{2}}^{2}}{{c_{1}}^{2}\left( 2\,{c_{1}}^{2}-{c_{3}}^{2}\right) \left( 2\,{%
c_{1}}^{2}-3\,{c_{2}}^{2}\right) }}\text{,}0\text{,}1/4\,{\frac{\left( {c_{1}%
}^{2}-{c_{3}}^{2}\right) {c_{2}}^{2}}{{c_{1}}^{3}\left( {c_{1}}^{2}{c_{2}}%
^{2}+{c_{1}}^{2}{c_{3}}^{2}-2\,{c_{3}}^{2}{c_{2}}^{2}\right) }}\text{,}
\notag \\
&&-1/4\,{\frac{\left( {c_{1}}^{2}-{c_{3}}^{2}\right) {c_{2}}^{2}}{{c_{1}}%
^{3}\left( {c_{1}}^{2}{c_{2}}^{2}+{c_{1}}^{2}{c_{3}}^{2}-2\,{c_{3}}^{2}{c_{2}%
}^{2}\right) }}\text{,}0\text{,}0\text{,}0\text{,}-1/4\,{\frac{{c_{2}}^{2}}{%
c_{1}\,\left( {c_{1}}^{2}{c_{2}}^{2}+{c_{1}}^{2}{c_{3}}^{2}-2\,{c_{3}}^{2}{%
c_{2}}^{2}\right) }}\text{,}  \notag \\
&&1/4\,{\frac{\left( {c_{1}}^{2}-{c_{2}}^{2}\right) {c_{3}}^{2}}{{c_{1}}%
^{3}\left( {c_{1}}^{2}{c_{2}}^{2}+{c_{1}}^{2}{c_{3}}^{2}-2\,{c_{3}}^{2}{c_{2}%
}^{2}\right) }}\text{,}-1/4\,{\frac{\left( {c_{1}}^{2}-{c_{2}}^{2}\right) {%
c_{3}}^{2}}{{c_{1}}^{3}\left( {c_{1}}^{2}{c_{2}}^{2}+{c_{1}}^{2}{c_{3}}%
^{2}-2\,{c_{3}}^{2}{c_{2}}^{2}\right) }}\text{,}  \notag \\
&&1/4\,{\frac{{c_{2}}^{2}}{c_{1}\,\left( {c_{1}}^{2}{c_{2}}^{2}+{c_{1}}^{2}{%
c_{3}}^{2}-2\,{c_{3}}^{2}{c_{2}}^{2}\right) }}\text{,}-1/4\,{\frac{{c_{1}}%
^{2}-2\,{c_{2}}^{2}}{{c_{1}}^{2}\left( 2\,{c_{1}}^{2}-{c_{3}}^{2}\right)
\left( 2\,{c_{1}}^{2}-3\,{c_{2}}^{2}\right) }}\text{,}  \notag \\
&&0\text{,}0\text{,}1/4\,{\frac{{c_{1}}^{2}-{c_{2}}^{2}}{{c_{1}}^{2}\left(
2\,{c_{1}}^{2}-{c_{3}}^{2}\right) \left( 2\,{c_{1}}^{2}-3\,{c_{2}}%
^{2}\right) }}\text{,}  \notag \\
&&-1/4\,{\frac{1}{{c_{1}}^{2}\left( 2\,{c_{1}}^{2}-3\,{c_{2}}^{2}\right) }}%
\text{,}1/4\,{\frac{{c_{1}}^{2}-{c_{2}}^{2}}{{c_{1}}^{2}\left( 2\,{c_{1}}%
^{2}-{c_{3}}^{2}\right) \left( 2\,{c_{1}}^{2}-3\,{c_{2}}^{2}\right) }}]
\end{eqnarray}

\begin{eqnarray}
\mathbf{C}_{5}^{-1} &=&[1/4\,{\frac{{c_{3}}^{2}{c_{2}}^{2}}{\left( 2\,{c_{1}}%
^{2}-{c_{3}}^{2}\right) \left( 2\,{c_{1}}^{2}-3\,{c_{2}}^{2}\right) }}\text{,%
}-1/4\,{\frac{{c_{3}}^{2}{c_{2}}^{2}}{c_{1}\,\left( {c_{1}}^{2}{c_{2}}^{2}+{%
c_{1}}^{2}{c_{3}}^{2}-2\,{c_{3}}^{2}{c_{2}}^{2}\right) }}\text{,}0\text{,}
\notag \\
&&-1/4\,{\frac{{c_{3}}^{2}{c_{2}}^{2}}{c_{1}\,\left( {c_{1}}^{2}{c_{2}}^{2}+{%
c_{1}}^{2}{c_{3}}^{2}-2\,{c_{3}}^{2}{c_{2}}^{2}\right) }}\text{,}-1/4\,{%
\frac{{c_{1}}^{2}{c_{2}}^{2}+{c_{1}}^{2}{c_{3}}^{2}-{c_{3}}^{2}{c_{2}}^{2}}{{%
c_{1}}^{2}\left( 2\,{c_{1}}^{2}-{c_{3}}^{2}\right) \left( 2\,{c_{1}}^{2}-3\,{%
c_{2}}^{2}\right) }}\text{,}  \notag \\
&&0\text{,}-3/4\,{\frac{{c_{2}}^{2}}{{c_{1}}^{2}\left( 2\,{c_{1}}^{2}-3\,{%
c_{2}}^{2}\right) }}\text{,}-1/4\,{\frac{{c_{1}}^{2}{c_{2}}^{2}-{c_{1}}^{2}{%
c_{3}}^{2}+2\,{c_{3}}^{2}{c_{2}}^{2}}{{c_{1}}^{2}\left( 2\,{c_{1}}^{2}-{c_{3}%
}^{2}\right) \left( 2\,{c_{1}}^{2}-3\,{c_{2}}^{2}\right) }}\text{,}0\text{,}
\notag \\
&&-1/4\,{\frac{{c_{1}}^{2}{c_{2}}^{2}+{c_{1}}^{2}{c_{3}}^{2}-{c_{3}}^{2}{%
c_{2}}^{2}}{{c_{1}}^{2}\left( 2\,{c_{1}}^{2}-{c_{3}}^{2}\right) \left( 2\,{%
c_{1}}^{2}-3\,{c_{2}}^{2}\right) }}\text{,}1/4\,{\frac{{c_{2}}^{2}}{%
c_{1}\,\left( {c_{1}}^{2}{c_{2}}^{2}+{c_{1}}^{2}{c_{3}}^{2}-2\,{c_{3}}^{2}{%
c_{2}}^{2}\right) }}\text{,}  \notag \\
&&0\text{,}1/4\,{\frac{\left( {c_{1}}^{2}-{c_{2}}^{2}\right) {c_{3}}^{2}}{{%
c_{1}}^{3}\left( {c_{1}}^{2}{c_{2}}^{2}+{c_{1}}^{2}{c_{3}}^{2}-2\,{c_{3}}^{2}%
{c_{2}}^{2}\right) }}\text{,}-1/4\,{\frac{\left( {c_{1}}^{2}-{c_{3}}%
^{2}\right) {c_{2}}^{2}}{{c_{1}}^{3}\left( {c_{1}}^{2}{c_{2}}^{2}+{c_{1}}^{2}%
{c_{3}}^{2}-2\,{c_{3}}^{2}{c_{2}}^{2}\right) }}\text{,}  \notag \\
&&0\text{,}1/4\,{\frac{\left( {c_{1}}^{2}-{c_{2}}^{2}\right) {c_{3}}^{2}}{{%
c_{1}}^{3}\left( {c_{1}}^{2}{c_{2}}^{2}+{c_{1}}^{2}{c_{3}}^{2}-2\,{c_{3}}^{2}%
{c_{2}}^{2}\right) }}\text{,}0\text{,}-1/4\,{\frac{\left( {c_{1}}^{2}-{c_{3}}%
^{2}\right) {c_{2}}^{2}}{{c_{1}}^{3}\left( {c_{1}}^{2}{c_{2}}^{2}+{c_{1}}^{2}%
{c_{3}}^{2}-2\,{c_{3}}^{2}{c_{2}}^{2}\right) }}\text{,}  \notag \\
&&0\text{,}1/4\,{\frac{{c_{2}}^{2}}{c_{1}\,\left( {c_{1}}^{2}{c_{2}}^{2}+{%
c_{1}}^{2}{c_{3}}^{2}-2\,{c_{3}}^{2}{c_{2}}^{2}\right) }}\text{,}1/4\,{\frac{%
{c_{1}}^{2}-{c_{2}}^{2}}{{c_{1}}^{2}\left( 2\,{c_{1}}^{2}-{c_{3}}^{2}\right)
\left( 2\,{c_{1}}^{2}-3\,{c_{2}}^{2}\right) }}\text{,}  \notag \\
&&0\text{,}1/4\,{\frac{1}{{c_{1}}^{2}\left( 2\,{c_{1}}^{2}-3\,{c_{2}}%
^{2}\right) }}\text{,}-1/4\,{\frac{{c_{1}}^{2}-2\,{c_{2}}^{2}}{{c_{1}}%
^{2}\left( 2\,{c_{1}}^{2}-{c_{3}}^{2}\right) \left( 2\,{c_{1}}^{2}-3\,{c_{2}}%
^{2}\right) }}\text{,}0\text{,}  \notag \\
&&1/4\,{\frac{{c_{1}}^{2}-{c_{2}}^{2}}{{c_{1}}^{2}\left( 2\,{c_{1}}^{2}-{%
c_{3}}^{2}\right) \left( 2\,{c_{1}}^{2}-3\,{c_{2}}^{2}\right) }}]
\end{eqnarray}

\begin{eqnarray}
\mathbf{C}_{6}^{-1} &=&[1/4\,{\frac{{c_{3}}^{2}{c_{2}}^{2}}{\left( 2\,{c_{1}}%
^{2}-{c_{3}}^{2}\right) \left( 2\,{c_{1}}^{2}-3\,{c_{2}}^{2}\right) }}\text{,%
}-1/4\,{\frac{{c_{3}}^{2}{c_{2}}^{2}}{c_{1}\,\left( {c_{1}}^{2}{c_{2}}^{2}+{%
c_{1}}^{2}{c_{3}}^{2}-2\,{c_{3}}^{2}{c_{2}}^{2}\right) }}\text{,}0\text{,}
\notag \\
&&1/4\,{\frac{{c_{3}}^{2}{c_{2}}^{2}}{c_{1}\,\left( {c_{1}}^{2}{c_{2}}^{2}+{%
c_{1}}^{2}{c_{3}}^{2}-2\,{c_{3}}^{2}{c_{2}}^{2}\right) }}\text{,}-1/4\,{%
\frac{{c_{1}}^{2}{c_{2}}^{2}+{c_{1}}^{2}{c_{3}}^{2}-{c_{3}}^{2}{c_{2}}^{2}}{{%
c_{1}}^{2}\left( 2\,{c_{1}}^{2}-{c_{3}}^{2}\right) \left( 2\,{c_{1}}^{2}-3\,{%
c_{2}}^{2}\right) }}\text{,}  \notag \\
&&0\text{,}3/4\,{\frac{{c_{2}}^{2}}{{c_{1}}^{2}\left( 2\,{c_{1}}^{2}-3\,{%
c_{2}}^{2}\right) }}\text{,}-1/4\,{\frac{{c_{1}}^{2}{c_{2}}^{2}-{c_{1}}^{2}{%
c_{3}}^{2}+2\,{c_{3}}^{2}{c_{2}}^{2}}{{c_{1}}^{2}\left( 2\,{c_{1}}^{2}-{c_{3}%
}^{2}\right) \left( 2\,{c_{1}}^{2}-3\,{c_{2}}^{2}\right) }}\text{,}0\text{,}
\notag \\
&&-1/4\,{\frac{{c_{1}}^{2}{c_{2}}^{2}+{c_{1}}^{2}{c_{3}}^{2}-{c_{3}}^{2}{%
c_{2}}^{2}}{{c_{1}}^{2}\left( 2\,{c_{1}}^{2}-{c_{3}}^{2}\right) \left( 2\,{%
c_{1}}^{2}-3\,{c_{2}}^{2}\right) }}\text{,}1/4\,{\frac{{c_{2}}^{2}}{%
c_{1}\,\left( {c_{1}}^{2}{c_{2}}^{2}+{c_{1}}^{2}{c_{3}}^{2}-2\,{c_{3}}^{2}{%
c_{2}}^{2}\right) }}\text{,}  \notag \\
&&0\text{,}-1/4\,{\frac{\left( {c_{1}}^{2}-{c_{2}}^{2}\right) {c_{3}}^{2}}{{%
c_{1}}^{3}\left( {c_{1}}^{2}{c_{2}}^{2}+{c_{1}}^{2}{c_{3}}^{2}-2\,{c_{3}}^{2}%
{c_{2}}^{2}\right) }}\text{,}-1/4\,{\frac{\left( {c_{1}}^{2}-{c_{3}}%
^{2}\right) {c_{2}}^{2}}{{c_{1}}^{3}\left( {c_{1}}^{2}{c_{2}}^{2}+{c_{1}}^{2}%
{c_{3}}^{2}-2\,{c_{3}}^{2}{c_{2}}^{2}\right) }}\text{,}  \notag \\
&&0\text{,}1/4\,{\frac{\left( {c_{1}}^{2}-{c_{2}}^{2}\right) {c_{3}}^{2}}{{%
c_{1}}^{3}\left( {c_{1}}^{2}{c_{2}}^{2}+{c_{1}}^{2}{c_{3}}^{2}-2\,{c_{3}}^{2}%
{c_{2}}^{2}\right) }}\text{,}0\text{,}1/4\,{\frac{\left( {c_{1}}^{2}-{c_{3}}%
^{2}\right) {c_{2}}^{2}}{{c_{1}}^{3}\left( {c_{1}}^{2}{c_{2}}^{2}+{c_{1}}^{2}%
{c_{3}}^{2}-2\,{c_{3}}^{2}{c_{2}}^{2}\right) }}\text{,}  \notag \\
&&0\text{,}-1/4\,{\frac{{c_{2}}^{2}}{c_{1}\,\left( {c_{1}}^{2}{c_{2}}^{2}+{%
c_{1}}^{2}{c_{3}}^{2}-2\,{c_{3}}^{2}{c_{2}}^{2}\right) }}\text{,}1/4\,{\frac{%
{c_{1}}^{2}-{c_{2}}^{2}}{{c_{1}}^{2}\left( 2\,{c_{1}}^{2}-{c_{3}}^{2}\right)
\left( 2\,{c_{1}}^{2}-3\,{c_{2}}^{2}\right) }}\text{,}  \notag \\
&&0\text{,}-1/4\,{\frac{1}{{c_{1}}^{2}\left( 2\,{c_{1}}^{2}-3\,{c_{2}}%
^{2}\right) }}\text{,}-1/4\,{\frac{{c_{1}}^{2}-2\,{c_{2}}^{2}}{{c_{1}}%
^{2}\left( 2\,{c_{1}}^{2}-{c_{3}}^{2}\right) \left( 2\,{c_{1}}^{2}-3\,{c_{2}}%
^{2}\right) }}\text{,}0\text{,}  \notag \\
&&1/4\,{\frac{{c_{1}}^{2}-{c_{2}}^{2}}{{c_{1}}^{2}\left( 2\,{c_{1}}^{2}-{%
c_{3}}^{2}\right) \left( 2\,{c_{1}}^{2}-3\,{c_{2}}^{2}\right) }}]
\end{eqnarray}

\begin{eqnarray}
\mathbf{C}_{7}^{-1} &=&[1/4\,{\frac{{c_{3}}^{2}{c_{2}}^{2}}{\left( 2\,{c_{1}}%
^{2}-{c_{3}}^{2}\right) \left( 2\,{c_{1}}^{2}-3\,{c_{2}}^{2}\right) }}\text{,%
}1/4\,{\frac{{c_{3}}^{2}{c_{2}}^{2}}{c_{1}\,\left( {c_{1}}^{2}{c_{2}}^{2}+{%
c_{1}}^{2}{c_{3}}^{2}-2\,{c_{3}}^{2}{c_{2}}^{2}\right) }}\text{,}0\text{,}
\notag \\
&&1/4\,{\frac{{c_{3}}^{2}{c_{2}}^{2}}{c_{1}\,\left( {c_{1}}^{2}{c_{2}}^{2}+{%
c_{1}}^{2}{c_{3}}^{2}-2\,{c_{3}}^{2}{c_{2}}^{2}\right) }}\text{,}-1/4\,{%
\frac{{c_{1}}^{2}{c_{2}}^{2}+{c_{1}}^{2}{c_{3}}^{2}-{c_{3}}^{2}{c_{2}}^{2}}{{%
c_{1}}^{2}\left( 2\,{c_{1}}^{2}-{c_{3}}^{2}\right) \left( 2\,{c_{1}}^{2}-3\,{%
c_{2}}^{2}\right) }}\text{,}  \notag \\
&&0\text{,}-3/4\,{\frac{{c_{2}}^{2}}{{c_{1}}^{2}\left( 2\,{c_{1}}^{2}-3\,{%
c_{2}}^{2}\right) }}\text{,}-1/4\,{\frac{{c_{1}}^{2}{c_{2}}^{2}-{c_{1}}^{2}{%
c_{3}}^{2}+2\,{c_{3}}^{2}{c_{2}}^{2}}{{c_{1}}^{2}\left( 2\,{c_{1}}^{2}-{c_{3}%
}^{2}\right) \left( 2\,{c_{1}}^{2}-3\,{c_{2}}^{2}\right) }}\text{,}0\text{,}
\notag \\
&&-1/4\,{\frac{{c_{1}}^{2}{c_{2}}^{2}+{c_{1}}^{2}{c_{3}}^{2}-{c_{3}}^{2}{%
c_{2}}^{2}}{{c_{1}}^{2}\left( 2\,{c_{1}}^{2}-{c_{3}}^{2}\right) \left( 2\,{%
c_{1}}^{2}-3\,{c_{2}}^{2}\right) }}\text{,}-1/4\,{\frac{{c_{2}}^{2}}{%
c_{1}\,\left( {c_{1}}^{2}{c_{2}}^{2}+{c_{1}}^{2}{c_{3}}^{2}-2\,{c_{3}}^{2}{%
c_{2}}^{2}\right) }}\text{,}  \notag \\
&&0\text{,}-1/4\,{\frac{\left( {c_{1}}^{2}-{c_{2}}^{2}\right) {c_{3}}^{2}}{{%
c_{1}}^{3}\left( {c_{1}}^{2}{c_{2}}^{2}+{c_{1}}^{2}{c_{3}}^{2}-2\,{c_{3}}^{2}%
{c_{2}}^{2}\right) }}\text{,}1/4\,{\frac{\left( {c_{1}}^{2}-{c_{3}}%
^{2}\right) {c_{2}}^{2}}{{c_{1}}^{3}\left( {c_{1}}^{2}{c_{2}}^{2}+{c_{1}}^{2}%
{c_{3}}^{2}-2\,{c_{3}}^{2}{c_{2}}^{2}\right) }}\text{,}  \notag \\
&&0\text{,}-1/4\,{\frac{\left( {c_{1}}^{2}-{c_{2}}^{2}\right) {c_{3}}^{2}}{{%
c_{1}}^{3}\left( {c_{1}}^{2}{c_{2}}^{2}+{c_{1}}^{2}{c_{3}}^{2}-2\,{c_{3}}^{2}%
{c_{2}}^{2}\right) }}\text{,}0\text{,}1/4\,{\frac{\left( {c_{1}}^{2}-{c_{3}}%
^{2}\right) {c_{2}}^{2}}{{c_{1}}^{3}\left( {c_{1}}^{2}{c_{2}}^{2}+{c_{1}}^{2}%
{c_{3}}^{2}-2\,{c_{3}}^{2}{c_{2}}^{2}\right) }}\text{,}  \notag \\
&&0\text{,}-1/4\,{\frac{{c_{2}}^{2}}{c_{1}\,\left( {c_{1}}^{2}{c_{2}}^{2}+{%
c_{1}}^{2}{c_{3}}^{2}-2\,{c_{3}}^{2}{c_{2}}^{2}\right) }}\text{,}1/4\,{\frac{%
{c_{1}}^{2}-{c_{2}}^{2}}{{c_{1}}^{2}\left( 2\,{c_{1}}^{2}-{c_{3}}^{2}\right)
\left( 2\,{c_{1}}^{2}-3\,{c_{2}}^{2}\right) }}\text{,}0\text{,}  \notag \\
&&1/4\,{\frac{1}{{c_{1}}^{2}\left( 2\,{c_{1}}^{2}-3\,{c_{2}}^{2}\right) }}%
\text{,}-1/4\,{\frac{{c_{1}}^{2}-2\,{c_{2}}^{2}}{{c_{1}}^{2}\left( 2\,{c_{1}}%
^{2}-{c_{3}}^{2}\right) \left( 2\,{c_{1}}^{2}-3\,{c_{2}}^{2}\right) }}\text{,%
}0\text{,}  \notag \\
&&1/4\,{\frac{{c_{1}}^{2}-{c_{2}}^{2}}{{c_{1}}^{2}\left( 2\,{c_{1}}^{2}-{%
c_{3}}^{2}\right) \left( 2\,{c_{1}}^{2}-3\,{c_{2}}^{2}\right) }}]
\end{eqnarray}

\begin{eqnarray}
\mathbf{C}_{8}^{-1} &=&[1/4\,{\frac{{c_{3}}^{2}{c_{2}}^{2}}{\left( 2\,{c_{1}}%
^{2}-{c_{3}}^{2}\right) \left( 2\,{c_{1}}^{2}-3\,{c_{2}}^{2}\right) }}\text{,%
}1/4\,{\frac{{c_{3}}^{2}{c_{2}}^{2}}{c_{1}\,\left( {c_{1}}^{2}{c_{2}}^{2}+{%
c_{1}}^{2}{c_{3}}^{2}-2\,{c_{3}}^{2}{c_{2}}^{2}\right) }}\text{,}0\text{,}
\notag \\
&&-1/4\,{\frac{{c_{3}}^{2}{c_{2}}^{2}}{c_{1}\,\left( {c_{1}}^{2}{c_{2}}^{2}+{%
c_{1}}^{2}{c_{3}}^{2}-2\,{c_{3}}^{2}{c_{2}}^{2}\right) }}\text{,}-1/4\,{%
\frac{{c_{1}}^{2}{c_{2}}^{2}+{c_{1}}^{2}{c_{3}}^{2}-{c_{3}}^{2}{c_{2}}^{2}}{{%
c_{1}}^{2}\left( 2\,{c_{1}}^{2}-{c_{3}}^{2}\right) \left( 2\,{c_{1}}^{2}-3\,{%
c_{2}}^{2}\right) }}\text{,}  \notag \\
&&0\text{,}3/4\,{\frac{{c_{2}}^{2}}{{c_{1}}^{2}\left( 2\,{c_{1}}^{2}-3\,{%
c_{2}}^{2}\right) }}\text{,}-1/4\,{\frac{{c_{1}}^{2}{c_{2}}^{2}-{c_{1}}^{2}{%
c_{3}}^{2}+2\,{c_{3}}^{2}{c_{2}}^{2}}{{c_{1}}^{2}\left( 2\,{c_{1}}^{2}-{c_{3}%
}^{2}\right) \left( 2\,{c_{1}}^{2}-3\,{c_{2}}^{2}\right) }}\text{,}0\text{,}
\notag \\
&&-1/4\,{\frac{{c_{1}}^{2}{c_{2}}^{2}+{c_{1}}^{2}{c_{3}}^{2}-{c_{3}}^{2}{%
c_{2}}^{2}}{{c_{1}}^{2}\left( 2\,{c_{1}}^{2}-{c_{3}}^{2}\right) \left( 2\,{%
c_{1}}^{2}-3\,{c_{2}}^{2}\right) }}\text{,}-1/4\,{\frac{{c_{2}}^{2}}{%
c_{1}\,\left( {c_{1}}^{2}{c_{2}}^{2}+{c_{1}}^{2}{c_{3}}^{2}-2\,{c_{3}}^{2}{%
c_{2}}^{2}\right) }}\text{,}  \notag \\
&&0\text{,}1/4\,{\frac{\left( {c_{1}}^{2}-{c_{2}}^{2}\right) {c_{3}}^{2}}{{%
c_{1}}^{3}\left( {c_{1}}^{2}{c_{2}}^{2}+{c_{1}}^{2}{c_{3}}^{2}-2\,{c_{3}}^{2}%
{c_{2}}^{2}\right) }}\text{,}1/4\,{\frac{\left( {c_{1}}^{2}-{c_{3}}%
^{2}\right) {c_{2}}^{2}}{{c_{1}}^{3}\left( {c_{1}}^{2}{c_{2}}^{2}+{c_{1}}^{2}%
{c_{3}}^{2}-2\,{c_{3}}^{2}{c_{2}}^{2}\right) }}\text{,}  \notag \\
&&0\text{,}-1/4\,{\frac{\left( {c_{1}}^{2}-{c_{2}}^{2}\right) {c_{3}}^{2}}{{%
c_{1}}^{3}\left( {c_{1}}^{2}{c_{2}}^{2}+{c_{1}}^{2}{c_{3}}^{2}-2\,{c_{3}}^{2}%
{c_{2}}^{2}\right) }}\text{,}0\text{,}-1/4\,{\frac{\left( {c_{1}}^{2}-{c_{3}}%
^{2}\right) {c_{2}}^{2}}{{c_{1}}^{3}\left( {c_{1}}^{2}{c_{2}}^{2}+{c_{1}}^{2}%
{c_{3}}^{2}-2\,{c_{3}}^{2}{c_{2}}^{2}\right) }}\text{,}  \notag \\
&&0\text{,}1/4\,{\frac{{c_{2}}^{2}}{c_{1}\,\left( {c_{1}}^{2}{c_{2}}^{2}+{%
c_{1}}^{2}{c_{3}}^{2}-2\,{c_{3}}^{2}{c_{2}}^{2}\right) }}\text{,}1/4\,{\frac{%
{c_{1}}^{2}-{c_{2}}^{2}}{{c_{1}}^{2}\left( 2\,{c_{1}}^{2}-{c_{3}}^{2}\right)
\left( 2\,{c_{1}}^{2}-3\,{c_{2}}^{2}\right) }}\text{,}0\text{,}  \notag \\
&&-1/4\,{\frac{1}{{c_{1}}^{2}\left( 2\,{c_{1}}^{2}-3\,{c_{2}}^{2}\right) }}%
\text{,}-1/4\,{\frac{{c_{1}}^{2}-2\,{c_{2}}^{2}}{{c_{1}}^{2}\left( 2\,{c_{1}}%
^{2}-{c_{3}}^{2}\right) \left( 2\,{c_{1}}^{2}-3\,{c_{2}}^{2}\right) }}\text{,%
}0\text{,}  \notag \\
&&1/4\,{\frac{{c_{1}}^{2}-{c_{2}}^{2}}{{c_{1}}^{2}\left( 2\,{c_{1}}^{2}-{%
c_{3}}^{2}\right) \left( 2\,{c_{1}}^{2}-3\,{c_{2}}^{2}\right) }}]
\end{eqnarray}

\begin{eqnarray}
\mathbf{C}_{9}^{-1} &=&[1/4\,{\frac{{c_{3}}^{2}{c_{2}}^{2}}{\left( 2\,{c_{1}}%
^{2}-{c_{3}}^{2}\right) \left( 2\,{c_{1}}^{2}-3\,{c_{2}}^{2}\right) }}\text{,%
}-1/4\,{\frac{{c_{3}}^{2}{c_{2}}^{2}}{c_{1}\,\left( {c_{1}}^{2}{c_{2}}^{2}+{%
c_{1}}^{2}{c_{3}}^{2}-2\,{c_{3}}^{2}{c_{2}}^{2}\right) }}\text{,}  \notag \\
&&-1/4\,{\frac{{c_{3}}^{2}{c_{2}}^{2}}{c_{1}\,\left( {c_{1}}^{2}{c_{2}}^{2}+{%
c_{1}}^{2}{c_{3}}^{2}-2\,{c_{3}}^{2}{c_{2}}^{2}\right) }}\text{,}0\text{,}%
-1/4\,{\frac{{c_{1}}^{2}{c_{2}}^{2}+{c_{1}}^{2}{c_{3}}^{2}-{c_{3}}^{2}{c_{2}}%
^{2}}{{c_{1}}^{2}\left( 2\,{c_{1}}^{2}-{c_{3}}^{2}\right) \left( 2\,{c_{1}}%
^{2}-3\,{c_{2}}^{2}\right) }}\text{,}  \notag \\
&&-3/4\,{\frac{{c_{2}}^{2}}{{c_{1}}^{2}\left( 2\,{c_{1}}^{2}-3\,{c_{2}}%
^{2}\right) }}\text{,}0\text{,}-1/4\,{\frac{{c_{1}}^{2}{c_{2}}^{2}+{c_{1}}%
^{2}{c_{3}}^{2}-{c_{3}}^{2}{c_{2}}^{2}}{{c_{1}}^{2}\left( 2\,{c_{1}}^{2}-{%
c_{3}}^{2}\right) \left( 2\,{c_{1}}^{2}-3\,{c_{2}}^{2}\right) }}\text{,}0%
\text{,}  \notag \\
&&-1/4\,{\frac{{c_{1}}^{2}{c_{2}}^{2}-{c_{1}}^{2}{c_{3}}^{2}+2\,{c_{3}}^{2}{%
c_{2}}^{2}}{{c_{1}}^{2}\left( 2\,{c_{1}}^{2}-{c_{3}}^{2}\right) \left( 2\,{%
c_{1}}^{2}-3\,{c_{2}}^{2}\right) }}\text{,}1/4\,{\frac{{c_{2}}^{2}}{%
c_{1}\,\left( {c_{1}}^{2}{c_{2}}^{2}+{c_{1}}^{2}{c_{3}}^{2}-2\,{c_{3}}^{2}{%
c_{2}}^{2}\right) }}\text{,}  \notag \\
&&1/4\,{\frac{\left( {c_{1}}^{2}-{c_{2}}^{2}\right) {c_{3}}^{2}}{{c_{1}}%
^{3}\left( {c_{1}}^{2}{c_{2}}^{2}+{c_{1}}^{2}{c_{3}}^{2}-2\,{c_{3}}^{2}{c_{2}%
}^{2}\right) }}\text{,}0\text{,}1/4\,{\frac{\left( {c_{1}}^{2}-{c_{2}}%
^{2}\right) {c_{3}}^{2}}{{c_{1}}^{3}\left( {c_{1}}^{2}{c_{2}}^{2}+{c_{1}}^{2}%
{c_{3}}^{2}-2\,{c_{3}}^{2}{c_{2}}^{2}\right) }}\text{,}  \notag \\
&&0\text{,}-1/4\,{\frac{\left( {c_{1}}^{2}-{c_{3}}^{2}\right) {c_{2}}^{2}}{{%
c_{1}}^{3}\left( {c_{1}}^{2}{c_{2}}^{2}+{c_{1}}^{2}{c_{3}}^{2}-2\,{c_{3}}^{2}%
{c_{2}}^{2}\right) }}\text{,}1/4\,{\frac{{c_{2}}^{2}}{c_{1}\,\left( {c_{1}}%
^{2}{c_{2}}^{2}+{c_{1}}^{2}{c_{3}}^{2}-2\,{c_{3}}^{2}{c_{2}}^{2}\right) }}%
\text{,}  \notag \\
&&0\text{,}-1/4\,{\frac{\left( {c_{1}}^{2}-{c_{3}}^{2}\right) {c_{2}}^{2}}{{%
c_{1}}^{3}\left( {c_{1}}^{2}{c_{2}}^{2}+{c_{1}}^{2}{c_{3}}^{2}-2\,{c_{3}}^{2}%
{c_{2}}^{2}\right) }}\text{,}0\text{,}1/4\,{\frac{{c_{1}}^{2}-{c_{2}}^{2}}{{%
c_{1}}^{2}\left( 2\,{c_{1}}^{2}-{c_{3}}^{2}\right) \left( 2\,{c_{1}}^{2}-3\,{%
c_{2}}^{2}\right) }}\text{,}  \notag \\
&&1/4\,{\frac{1}{{c_{1}}^{2}\left( 2\,{c_{1}}^{2}-3\,{c_{2}}^{2}\right) }}%
\text{,}0\text{,}1/4\,{\frac{{c_{1}}^{2}-{c_{2}}^{2}}{{c_{1}}^{2}\left( 2\,{%
c_{1}}^{2}-{c_{3}}^{2}\right) \left( 2\,{c_{1}}^{2}-3\,{c_{2}}^{2}\right) }}%
\text{,}0\text{,}  \notag \\
&&-1/4\,{\frac{{c_{1}}^{2}-2\,{c_{2}}^{2}}{{c_{1}}^{2}\left( 2\,{c_{1}}^{2}-{%
c_{3}}^{2}\right) \left( 2\,{c_{1}}^{2}-3\,{c_{2}}^{2}\right) }}]
\end{eqnarray}

\begin{eqnarray}
\mathbf{C}_{10}^{-1} &=&[1/4\,{\frac{{c_{3}}^{2}{c_{2}}^{2}}{\left( 2\,{c_{1}%
}^{2}-{c_{3}}^{2}\right) \left( 2\,{c_{1}}^{2}-3\,{c_{2}}^{2}\right) }}\text{%
,}-1/4\,{\frac{{c_{3}}^{2}{c_{2}}^{2}}{c_{1}\,\left( {c_{1}}^{2}{c_{2}}^{2}+{%
c_{1}}^{2}{c_{3}}^{2}-2\,{c_{3}}^{2}{c_{2}}^{2}\right) }}\text{,}  \notag \\
&&1/4\,{\frac{{c_{3}}^{2}{c_{2}}^{2}}{c_{1}\,\left( {c_{1}}^{2}{c_{2}}^{2}+{%
c_{1}}^{2}{c_{3}}^{2}-2\,{c_{3}}^{2}{c_{2}}^{2}\right) }}\text{,}0\text{,}%
-1/4\,{\frac{{c_{1}}^{2}{c_{2}}^{2}+{c_{1}}^{2}{c_{3}}^{2}-{c_{3}}^{2}{c_{2}}%
^{2}}{{c_{1}}^{2}\left( 2\,{c_{1}}^{2}-{c_{3}}^{2}\right) \left( 2\,{c_{1}}%
^{2}-3\,{c_{2}}^{2}\right) }}\text{,}  \notag \\
&&3/4\,{\frac{{c_{2}}^{2}}{{c_{1}}^{2}\left( 2\,{c_{1}}^{2}-3\,{c_{2}}%
^{2}\right) }}\text{,}0\text{,}-1/4\,{\frac{{c_{1}}^{2}{c_{2}}^{2}+{c_{1}}%
^{2}{c_{3}}^{2}-{c_{3}}^{2}{c_{2}}^{2}}{{c_{1}}^{2}\left( 2\,{c_{1}}^{2}-{%
c_{3}}^{2}\right) \left( 2\,{c_{1}}^{2}-3\,{c_{2}}^{2}\right) }}\text{,}0%
\text{,}  \notag \\
&&-1/4\,{\frac{{c_{1}}^{2}{c_{2}}^{2}-{c_{1}}^{2}{c_{3}}^{2}+2\,{c_{3}}^{2}{%
c_{2}}^{2}}{{c_{1}}^{2}\left( 2\,{c_{1}}^{2}-{c_{3}}^{2}\right) \left( 2\,{%
c_{1}}^{2}-3\,{c_{2}}^{2}\right) }}\text{,}1/4\,{\frac{{c_{2}}^{2}}{%
c_{1}\,\left( {c_{1}}^{2}{c_{2}}^{2}+{c_{1}}^{2}{c_{3}}^{2}-2\,{c_{3}}^{2}{%
c_{2}}^{2}\right) }}\text{,}  \notag \\
&&-1/4\,{\frac{\left( {c_{1}}^{2}-{c_{2}}^{2}\right) {c_{3}}^{2}}{{c_{1}}%
^{3}\left( {c_{1}}^{2}{c_{2}}^{2}+{c_{1}}^{2}{c_{3}}^{2}-2\,{c_{3}}^{2}{c_{2}%
}^{2}\right) }}\text{,}0\text{,}1/4\,{\frac{\left( {c_{1}}^{2}-{c_{2}}%
^{2}\right) {c_{3}}^{2}}{{c_{1}}^{3}\left( {c_{1}}^{2}{c_{2}}^{2}+{c_{1}}^{2}%
{c_{3}}^{2}-2\,{c_{3}}^{2}{c_{2}}^{2}\right) }}\text{,}  \notag \\
&&0\text{,}-1/4\,{\frac{\left( {c_{1}}^{2}-{c_{3}}^{2}\right) {c_{2}}^{2}}{{%
c_{1}}^{3}\left( {c_{1}}^{2}{c_{2}}^{2}+{c_{1}}^{2}{c_{3}}^{2}-2\,{c_{3}}^{2}%
{c_{2}}^{2}\right) }}\text{,}-1/4\,{\frac{{c_{2}}^{2}}{c_{1}\,\left( {c_{1}}%
^{2}{c_{2}}^{2}+{c_{1}}^{2}{c_{3}}^{2}-2\,{c_{3}}^{2}{c_{2}}^{2}\right) }}%
\text{,}  \notag \\
&&0\text{,}1/4\,{\frac{\left( {c_{1}}^{2}-{c_{3}}^{2}\right) {c_{2}}^{2}}{{%
c_{1}}^{3}\left( {c_{1}}^{2}{c_{2}}^{2}+{c_{1}}^{2}{c_{3}}^{2}-2\,{c_{3}}^{2}%
{c_{2}}^{2}\right) }}\text{,}0\text{,}1/4\,{\frac{{c_{1}}^{2}-{c_{2}}^{2}}{{%
c_{1}}^{2}\left( 2\,{c_{1}}^{2}-{c_{3}}^{2}\right) \left( 2\,{c_{1}}^{2}-3\,{%
c_{2}}^{2}\right) }}\text{,}  \notag \\
&&-1/4\,{\frac{1}{{c_{1}}^{2}\left( 2\,{c_{1}}^{2}-3\,{c_{2}}^{2}\right) }}%
\text{,}0\text{,}1/4\,{\frac{{c_{1}}^{2}-{c_{2}}^{2}}{{c_{1}}^{2}\left( 2\,{%
c_{1}}^{2}-{c_{3}}^{2}\right) \left( 2\,{c_{1}}^{2}-3\,{c_{2}}^{2}\right) }}%
\text{,}0\text{,}  \notag \\
&&-1/4\,{\frac{{c_{1}}^{2}-2\,{c_{2}}^{2}}{{c_{1}}^{2}\left( 2\,{c_{1}}^{2}-{%
c_{3}}^{2}\right) \left( 2\,{c_{1}}^{2}-3\,{c_{2}}^{2}\right) }}]
\end{eqnarray}

\begin{eqnarray}
\mathbf{C}_{11}^{-1} &=&[1/4\,{\frac{{c_{3}}^{2}{c_{2}}^{2}}{\left( 2\,{c_{1}%
}^{2}-{c_{3}}^{2}\right) \left( 2\,{c_{1}}^{2}-3\,{c_{2}}^{2}\right) }}\text{%
,}1/4\,{\frac{{c_{3}}^{2}{c_{2}}^{2}}{c_{1}\,\left( {c_{1}}^{2}{c_{2}}^{2}+{%
c_{1}}^{2}{c_{3}}^{2}-2\,{c_{3}}^{2}{c_{2}}^{2}\right) }}\text{,}  \notag \\
&&1/4\,{\frac{{c_{3}}^{2}{c_{2}}^{2}}{c_{1}\,\left( {c_{1}}^{2}{c_{2}}^{2}+{%
c_{1}}^{2}{c_{3}}^{2}-2\,{c_{3}}^{2}{c_{2}}^{2}\right) }}\text{,}0\text{,}%
-1/4\,{\frac{{c_{1}}^{2}{c_{2}}^{2}+{c_{1}}^{2}{c_{3}}^{2}-{c_{3}}^{2}{c_{2}}%
^{2}}{{c_{1}}^{2}\left( 2\,{c_{1}}^{2}-{c_{3}}^{2}\right) \left( 2\,{c_{1}}%
^{2}-3\,{c_{2}}^{2}\right) }}\text{,}  \notag \\
&&-3/4\,{\frac{{c_{2}}^{2}}{{c_{1}}^{2}\left( 2\,{c_{1}}^{2}-3\,{c_{2}}%
^{2}\right) }}\text{,}0\text{,}-1/4\,{\frac{{c_{1}}^{2}{c_{2}}^{2}+{c_{1}}%
^{2}{c_{3}}^{2}-{c_{3}}^{2}{c_{2}}^{2}}{{c_{1}}^{2}\left( 2\,{c_{1}}^{2}-{%
c_{3}}^{2}\right) \left( 2\,{c_{1}}^{2}-3\,{c_{2}}^{2}\right) }}\text{,}0%
\text{,}  \notag \\
&&-1/4\,{\frac{{c_{1}}^{2}{c_{2}}^{2}-{c_{1}}^{2}{c_{3}}^{2}+2\,{c_{3}}^{2}{%
c_{2}}^{2}}{{c_{1}}^{2}\left( 2\,{c_{1}}^{2}-{c_{3}}^{2}\right) \left( 2\,{%
c_{1}}^{2}-3\,{c_{2}}^{2}\right) }}\text{,}-1/4\,{\frac{{c_{2}}^{2}}{%
c_{1}\,\left( {c_{1}}^{2}{c_{2}}^{2}+{c_{1}}^{2}{c_{3}}^{2}-2\,{c_{3}}^{2}{%
c_{2}}^{2}\right) }}\text{,}  \notag \\
&&-1/4\,{\frac{\left( {c_{1}}^{2}-{c_{2}}^{2}\right) {c_{3}}^{2}}{{c_{1}}%
^{3}\left( {c_{1}}^{2}{c_{2}}^{2}+{c_{1}}^{2}{c_{3}}^{2}-2\,{c_{3}}^{2}{c_{2}%
}^{2}\right) }}\text{,}0\text{,}-1/4\,{\frac{\left( {c_{1}}^{2}-{c_{2}}%
^{2}\right) {c_{3}}^{2}}{{c_{1}}^{3}\left( {c_{1}}^{2}{c_{2}}^{2}+{c_{1}}^{2}%
{c_{3}}^{2}-2\,{c_{3}}^{2}{c_{2}}^{2}\right) }}\text{,}  \notag \\
&&0\text{,}1/4\,{\frac{\left( {c_{1}}^{2}-{c_{3}}^{2}\right) {c_{2}}^{2}}{{%
c_{1}}^{3}\left( {c_{1}}^{2}{c_{2}}^{2}+{c_{1}}^{2}{c_{3}}^{2}-2\,{c_{3}}^{2}%
{c_{2}}^{2}\right) }}\text{,}-1/4\,{\frac{{c_{2}}^{2}}{c_{1}\,\left( {c_{1}}%
^{2}{c_{2}}^{2}+{c_{1}}^{2}{c_{3}}^{2}-2\,{c_{3}}^{2}{c_{2}}^{2}\right) }}%
\text{,}0\text{,}  \notag \\
&&1/4\,{\frac{\left( {c_{1}}^{2}-{c_{3}}^{2}\right) {c_{2}}^{2}}{{c_{1}}%
^{3}\left( {c_{1}}^{2}{c_{2}}^{2}+{c_{1}}^{2}{c_{3}}^{2}-2\,{c_{3}}^{2}{c_{2}%
}^{2}\right) }}\text{,}0\text{,}1/4\,{\frac{{c_{1}}^{2}-{c_{2}}^{2}}{{c_{1}}%
^{2}\left( 2\,{c_{1}}^{2}-{c_{3}}^{2}\right) \left( 2\,{c_{1}}^{2}-3\,{c_{2}}%
^{2}\right) }}\text{,}  \notag \\
&&1/4\,{\frac{1}{{c_{1}}^{2}\left( 2\,{c_{1}}^{2}-3\,{c_{2}}^{2}\right) }}%
\text{,}0\text{,}1/4\,{\frac{{c_{1}}^{2}-{c_{2}}^{2}}{{c_{1}}^{2}\left( 2\,{%
c_{1}}^{2}-{c_{3}}^{2}\right) \left( 2\,{c_{1}}^{2}-3\,{c_{2}}^{2}\right) }}%
\text{,}0\text{,}  \notag \\
&&-1/4\,{\frac{{c_{1}}^{2}-2\,{c_{2}}^{2}}{{c_{1}}^{2}\left( 2\,{c_{1}}^{2}-{%
c_{3}}^{2}\right) \left( 2\,{c_{1}}^{2}-3\,{c_{2}}^{2}\right) }}]
\end{eqnarray}

\begin{eqnarray}
\mathbf{C}_{12}^{-1} &=&[1/4\,{\frac{{c_{3}}^{2}{c_{2}}^{2}}{\left( 2\,{c_{1}%
}^{2}-{c_{3}}^{2}\right) \left( 2\,{c_{1}}^{2}-3\,{c_{2}}^{2}\right) }}\text{%
,}1/4\,{\frac{{c_{3}}^{2}{c_{2}}^{2}}{c_{1}\,\left( {c_{1}}^{2}{c_{2}}^{2}+{%
c_{1}}^{2}{c_{3}}^{2}-2\,{c_{3}}^{2}{c_{2}}^{2}\right) }}\text{,}  \notag \\
&&-1/4\,{\frac{{c_{3}}^{2}{c_{2}}^{2}}{c_{1}\,\left( {c_{1}}^{2}{c_{2}}^{2}+{%
c_{1}}^{2}{c_{3}}^{2}-2\,{c_{3}}^{2}{c_{2}}^{2}\right) }}\text{,}0\text{,}%
-1/4\,{\frac{{c_{1}}^{2}{c_{2}}^{2}+{c_{1}}^{2}{c_{3}}^{2}-{c_{3}}^{2}{c_{2}}%
^{2}}{{c_{1}}^{2}\left( 2\,{c_{1}}^{2}-{c_{3}}^{2}\right) \left( 2\,{c_{1}}%
^{2}-3\,{c_{2}}^{2}\right) }}\text{,}  \notag \\
&&3/4\,{\frac{{c_{2}}^{2}}{{c_{1}}^{2}\left( 2\,{c_{1}}^{2}-3\,{c_{2}}%
^{2}\right) }}\text{,}0\text{,}-1/4\,{\frac{{c_{1}}^{2}{c_{2}}^{2}+{c_{1}}%
^{2}{c_{3}}^{2}-{c_{3}}^{2}{c_{2}}^{2}}{{c_{1}}^{2}\left( 2\,{c_{1}}^{2}-{%
c_{3}}^{2}\right) \left( 2\,{c_{1}}^{2}-3\,{c_{2}}^{2}\right) }}\text{,}0%
\text{,}  \notag \\
&&-1/4\,{\frac{{c_{1}}^{2}{c_{2}}^{2}-{c_{1}}^{2}{c_{3}}^{2}+2\,{c_{3}}^{2}{%
c_{2}}^{2}}{{c_{1}}^{2}\left( 2\,{c_{1}}^{2}-{c_{3}}^{2}\right) \left( 2\,{%
c_{1}}^{2}-3\,{c_{2}}^{2}\right) }}\text{,}-1/4\,{\frac{{c_{2}}^{2}}{%
c_{1}\,\left( {c_{1}}^{2}{c_{2}}^{2}+{c_{1}}^{2}{c_{3}}^{2}-2\,{c_{3}}^{2}{%
c_{2}}^{2}\right) }}\text{,}  \notag \\
&&1/4\,{\frac{\left( {c_{1}}^{2}-{c_{2}}^{2}\right) {c_{3}}^{2}}{{c_{1}}%
^{3}\left( {c_{1}}^{2}{c_{2}}^{2}+{c_{1}}^{2}{c_{3}}^{2}-2\,{c_{3}}^{2}{c_{2}%
}^{2}\right) }}\text{,}0\text{,}-1/4\,{\frac{\left( {c_{1}}^{2}-{c_{2}}%
^{2}\right) {c_{3}}^{2}}{{c_{1}}^{3}\left( {c_{1}}^{2}{c_{2}}^{2}+{c_{1}}^{2}%
{c_{3}}^{2}-2\,{c_{3}}^{2}{c_{2}}^{2}\right) }}\text{,}  \notag \\
&&0\text{,}1/4\,{\frac{\left( {c_{1}}^{2}-{c_{3}}^{2}\right) {c_{2}}^{2}}{{%
c_{1}}^{3}\left( {c_{1}}^{2}{c_{2}}^{2}+{c_{1}}^{2}{c_{3}}^{2}-2\,{c_{3}}^{2}%
{c_{2}}^{2}\right) }}\text{,}1/4\,{\frac{{c_{2}}^{2}}{c_{1}\,\left( {c_{1}}%
^{2}{c_{2}}^{2}+{c_{1}}^{2}{c_{3}}^{2}-2\,{c_{3}}^{2}{c_{2}}^{2}\right) }}%
\text{,}0\text{,}  \notag \\
&&-1/4\,{\frac{\left( {c_{1}}^{2}-{c_{3}}^{2}\right) {c_{2}}^{2}}{{c_{1}}%
^{3}\left( {c_{1}}^{2}{c_{2}}^{2}+{c_{1}}^{2}{c_{3}}^{2}-2\,{c_{3}}^{2}{c_{2}%
}^{2}\right) }}\text{,}0\text{,}1/4\,{\frac{{c_{1}}^{2}-{c_{2}}^{2}}{{c_{1}}%
^{2}\left( 2\,{c_{1}}^{2}-{c_{3}}^{2}\right) \left( 2\,{c_{1}}^{2}-3\,{c_{2}}%
^{2}\right) }}\text{,}  \notag \\
&&-1/4\,{\frac{1}{{c_{1}}^{2}\left( 2\,{c_{1}}^{2}-3\,{c_{2}}^{2}\right) }}%
\text{,}0\text{,}1/4\,{\frac{{c_{1}}^{2}-{c_{2}}^{2}}{{c_{1}}^{2}\left( 2\,{%
c_{1}}^{2}-{c_{3}}^{2}\right) \left( 2\,{c_{1}}^{2}-3\,{c_{2}}^{2}\right) }}%
\text{,}0\text{,}  \notag \\
&&-1/4\,{\frac{{c_{1}}^{2}-2\,{c_{2}}^{2}}{{c_{1}}^{2}\left( 2\,{c_{1}}^{2}-{%
c_{3}}^{2}\right) \left( 2\,{c_{1}}^{2}-3\,{c_{2}}^{2}\right) }}]
\end{eqnarray}

\begin{eqnarray}
\mathbf{C}_{13}^{-1} &=&[-1/4\,{\frac{{c_{1}}^{2}{c_{3}}^{2}}{\left( 3\,{%
c_{2}}^{2}-{c_{3}}^{2}\right) \left( 2\,{c_{1}}^{2}-3\,{c_{2}}^{2}\right) }}%
\text{,}1/8\,{\frac{{c_{1}}^{2}{c_{3}}^{2}}{c_{2}\,\left( {c_{1}}^{2}{c_{2}}%
^{2}+{c_{1}}^{2}{c_{3}}^{2}-2\,{c_{3}}^{2}{c_{2}}^{2}\right) }}\text{,}
\notag \\
&&1/8\,{\frac{{c_{1}}^{2}{c_{3}}^{2}}{c_{2}\,\left( {c_{1}}^{2}{c_{2}}^{2}+{%
c_{1}}^{2}{c_{3}}^{2}-2\,{c_{3}}^{2}{c_{2}}^{2}\right) }}\text{,}1/8\,{\frac{%
{c_{1}}^{2}{c_{3}}^{2}}{c_{2}\,\left( {c_{1}}^{2}{c_{2}}^{2}+{c_{1}}^{2}{%
c_{3}}^{2}-2\,{c_{3}}^{2}{c_{2}}^{2}\right) }}\text{,}  \notag \\
&&1/8\,{\frac{2\,{c_{1}}^{2}+{c_{3}}^{2}}{\left( 3\,{c_{2}}^{2}-{c_{3}}%
^{2}\right) \left( 2\,{c_{1}}^{2}-3\,{c_{2}}^{2}\right) }}\text{,}1/4\,{%
\frac{{c_{1}}^{2}}{{c_{2}}^{2}\left( 2\,{c_{1}}^{2}-3\,{c_{2}}^{2}\right) }}%
\text{,}1/4\,{\frac{{c_{1}}^{2}}{{c_{2}}^{2}\left( 2\,{c_{1}}^{2}-3\,{c_{2}}%
^{2}\right) }}\text{,}  \notag \\
&&1/8\,{\frac{2\,{c_{1}}^{2}+{c_{3}}^{2}}{\left( 3\,{c_{2}}^{2}-{c_{3}}%
^{2}\right) \left( 2\,{c_{1}}^{2}-3\,{c_{2}}^{2}\right) }}\text{,}1/4\,{%
\frac{{c_{1}}^{2}}{{c_{2}}^{2}\left( 2\,{c_{1}}^{2}-3\,{c_{2}}^{2}\right) }}%
\text{,}1/8\,{\frac{2\,{c_{1}}^{2}+{c_{3}}^{2}}{\left( 3\,{c_{2}}^{2}-{c_{3}}%
^{2}\right) \left( 2\,{c_{1}}^{2}-3\,{c_{2}}^{2}\right) }}\text{,}  \notag \\
&&-1/8\,{\frac{{c_{1}}^{2}}{c_{2}\,\left( {c_{1}}^{2}{c_{2}}^{2}+{c_{1}}^{2}{%
c_{3}}^{2}-2\,{c_{3}}^{2}{c_{2}}^{2}\right) }}\text{,}1/8\,{\frac{{c_{1}}%
^{2}-{c_{3}}^{2}}{c_{2}\,\left( {c_{1}}^{2}{c_{2}}^{2}+{c_{1}}^{2}{c_{3}}%
^{2}-2\,{c_{3}}^{2}{c_{2}}^{2}\right) }}\text{,}  \notag \\
&&1/8\,{\frac{{c_{1}}^{2}-{c_{3}}^{2}}{c_{2}\,\left( {c_{1}}^{2}{c_{2}}^{2}+{%
c_{1}}^{2}{c_{3}}^{2}-2\,{c_{3}}^{2}{c_{2}}^{2}\right) }}\text{,}1/8\,{\frac{%
{c_{1}}^{2}-{c_{3}}^{2}}{c_{2}\,\left( {c_{1}}^{2}{c_{2}}^{2}+{c_{1}}^{2}{%
c_{3}}^{2}-2\,{c_{3}}^{2}{c_{2}}^{2}\right) }}\text{,}1/8\,{c_{2}}^{-3}\text{%
,}  \notag \\
&&1/8\,{\frac{{c_{1}}^{2}-{c_{3}}^{2}}{c_{2}\,\left( {c_{1}}^{2}{c_{2}}^{2}+{%
c_{1}}^{2}{c_{3}}^{2}-2\,{c_{3}}^{2}{c_{2}}^{2}\right) }}\text{,}-1/8\,{%
\frac{{c_{1}}^{2}}{c_{2}\,\left( {c_{1}}^{2}{c_{2}}^{2}+{c_{1}}^{2}{c_{3}}%
^{2}-2\,{c_{3}}^{2}{c_{2}}^{2}\right) }}\text{,}  \notag \\
&&1/8\,{\frac{{c_{1}}^{2}-{c_{3}}^{2}}{c_{2}\,\left( {c_{1}}^{2}{c_{2}}^{2}+{%
c_{1}}^{2}{c_{3}}^{2}-2\,{c_{3}}^{2}{c_{2}}^{2}\right) }}\text{,}1/8\,{\frac{%
{c_{1}}^{2}-{c_{3}}^{2}}{c_{2}\,\left( {c_{1}}^{2}{c_{2}}^{2}+{c_{1}}^{2}{%
c_{3}}^{2}-2\,{c_{3}}^{2}{c_{2}}^{2}\right) }}\text{,}  \notag \\
&&-1/8\,{\frac{{c_{1}}^{2}}{c_{2}\,\left( {c_{1}}^{2}{c_{2}}^{2}+{c_{1}}^{2}{%
c_{3}}^{2}-2\,{c_{3}}^{2}{c_{2}}^{2}\right) }}\text{,}-1/8\,{\frac{1}{\left(
3\,{c_{2}}^{2}-{c_{3}}^{2}\right) \left( 2\,{c_{1}}^{2}-3\,{c_{2}}%
^{2}\right) }}\text{,}  \notag \\
&&-1/8\,{\frac{1}{{c_{2}}^{2}\left( 2\,{c_{1}}^{2}-3\,{c_{2}}^{2}\right) }}%
\text{,}-1/8\,{\frac{1}{{c_{2}}^{2}\left( 2\,{c_{1}}^{2}-3\,{c_{2}}%
^{2}\right) }}\text{,}-1/8\,{\frac{1}{\left( 3\,{c_{2}}^{2}-{c_{3}}%
^{2}\right) \left( 2\,{c_{1}}^{2}-3\,{c_{2}}^{2}\right) }}\text{,}  \notag \\
&&-1/8\,{\frac{1}{{c_{2}}^{2}\left( 2\,{c_{1}}^{2}-3\,{c_{2}}^{2}\right) }}%
\text{,}-1/8\,{\frac{1}{\left( 3\,{c_{2}}^{2}-{c_{3}}^{2}\right) \left( 2\,{%
c_{1}}^{2}-3\,{c_{2}}^{2}\right) }}]
\end{eqnarray}

\begin{eqnarray}
\mathbf{C}_{14}^{-1} &=&[-1/4\,{\frac{{c_{1}}^{2}{c_{3}}^{2}}{\left( 3\,{%
c_{2}}^{2}-{c_{3}}^{2}\right) \left( 2\,{c_{1}}^{2}-3\,{c_{2}}^{2}\right) }}%
\text{,}1/8\,{\frac{{c_{1}}^{2}{c_{3}}^{2}}{c_{2}\,\left( {c_{1}}^{2}{c_{2}}%
^{2}+{c_{1}}^{2}{c_{3}}^{2}-2\,{c_{3}}^{2}{c_{2}}^{2}\right) }}\text{,}
\notag \\
&&1/8\,{\frac{{c_{1}}^{2}{c_{3}}^{2}}{c_{2}\,\left( {c_{1}}^{2}{c_{2}}^{2}+{%
c_{1}}^{2}{c_{3}}^{2}-2\,{c_{3}}^{2}{c_{2}}^{2}\right) }}\text{,}-1/8\,{%
\frac{{c_{1}}^{2}{c_{3}}^{2}}{c_{2}\,\left( {c_{1}}^{2}{c_{2}}^{2}+{c_{1}}%
^{2}{c_{3}}^{2}-2\,{c_{3}}^{2}{c_{2}}^{2}\right) }}\text{,}  \notag \\
&&1/8\,{\frac{2\,{c_{1}}^{2}+{c_{3}}^{2}}{\left( 3\,{c_{2}}^{2}-{c_{3}}%
^{2}\right) \left( 2\,{c_{1}}^{2}-3\,{c_{2}}^{2}\right) }}\text{,}1/4\,{%
\frac{{c_{1}}^{2}}{{c_{2}}^{2}\left( 2\,{c_{1}}^{2}-3\,{c_{2}}^{2}\right) }}%
\text{,}-1/4\,{\frac{{c_{1}}^{2}}{{c_{2}}^{2}\left( 2\,{c_{1}}^{2}-3\,{c_{2}}%
^{2}\right) }}\text{,}  \notag \\
&&1/8\,{\frac{2\,{c_{1}}^{2}+{c_{3}}^{2}}{\left( 3\,{c_{2}}^{2}-{c_{3}}%
^{2}\right) \left( 2\,{c_{1}}^{2}-3\,{c_{2}}^{2}\right) }}\text{,}-1/4\,{%
\frac{{c_{1}}^{2}}{{c_{2}}^{2}\left( 2\,{c_{1}}^{2}-3\,{c_{2}}^{2}\right) }}%
\text{,}1/8\,{\frac{2\,{c_{1}}^{2}+{c_{3}}^{2}}{\left( 3\,{c_{2}}^{2}-{c_{3}}%
^{2}\right) \left( 2\,{c_{1}}^{2}-3\,{c_{2}}^{2}\right) }}\text{,}  \notag \\
&&-1/8\,{\frac{{c_{1}}^{2}}{c_{2}\,\left( {c_{1}}^{2}{c_{2}}^{2}+{c_{1}}^{2}{%
c_{3}}^{2}-2\,{c_{3}}^{2}{c_{2}}^{2}\right) }}\text{,}1/8\,{\frac{{c_{1}}%
^{2}-{c_{3}}^{2}}{c_{2}\,\left( {c_{1}}^{2}{c_{2}}^{2}+{c_{1}}^{2}{c_{3}}%
^{2}-2\,{c_{3}}^{2}{c_{2}}^{2}\right) }}\text{,}  \notag \\
&&-1/8\,{\frac{{c_{1}}^{2}-{c_{3}}^{2}}{c_{2}\,\left( {c_{1}}^{2}{c_{2}}^{2}+%
{c_{1}}^{2}{c_{3}}^{2}-2\,{c_{3}}^{2}{c_{2}}^{2}\right) }}\text{,}1/8\,{%
\frac{{c_{1}}^{2}-{c_{3}}^{2}}{c_{2}\,\left( {c_{1}}^{2}{c_{2}}^{2}+{c_{1}}%
^{2}{c_{3}}^{2}-2\,{c_{3}}^{2}{c_{2}}^{2}\right) }}\text{,}  \notag \\
&&-1/8\,{c_{2}}^{-3}\text{,}1/8\,{\frac{{c_{1}}^{2}-{c_{3}}^{2}}{%
c_{2}\,\left( {c_{1}}^{2}{c_{2}}^{2}+{c_{1}}^{2}{c_{3}}^{2}-2\,{c_{3}}^{2}{%
c_{2}}^{2}\right) }}\text{,}-1/8\,{\frac{{c_{1}}^{2}}{c_{2}\,\left( {c_{1}}%
^{2}{c_{2}}^{2}+{c_{1}}^{2}{c_{3}}^{2}-2\,{c_{3}}^{2}{c_{2}}^{2}\right) }}%
\text{,}  \notag \\
&&-1/8\,{\frac{{c_{1}}^{2}-{c_{3}}^{2}}{c_{2}\,\left( {c_{1}}^{2}{c_{2}}^{2}+%
{c_{1}}^{2}{c_{3}}^{2}-2\,{c_{3}}^{2}{c_{2}}^{2}\right) }}\text{,}1/8\,{%
\frac{{c_{1}}^{2}-{c_{3}}^{2}}{c_{2}\,\left( {c_{1}}^{2}{c_{2}}^{2}+{c_{1}}%
^{2}{c_{3}}^{2}-2\,{c_{3}}^{2}{c_{2}}^{2}\right) }}\text{,}  \notag \\
&&1/8\,{\frac{{c_{1}}^{2}}{c_{2}\,\left( {c_{1}}^{2}{c_{2}}^{2}+{c_{1}}^{2}{%
c_{3}}^{2}-2\,{c_{3}}^{2}{c_{2}}^{2}\right) }}\text{,}-1/8\,{\frac{1}{\left(
3\,{c_{2}}^{2}-{c_{3}}^{2}\right) \left( 2\,{c_{1}}^{2}-3\,{c_{2}}%
^{2}\right) }}\text{,}  \notag \\
&&-1/8\,{\frac{1}{{c_{2}}^{2}\left( 2\,{c_{1}}^{2}-3\,{c_{2}}^{2}\right) }}%
\text{,}1/8\,{\frac{1}{{c_{2}}^{2}\left( 2\,{c_{1}}^{2}-3\,{c_{2}}%
^{2}\right) }}\text{,}-1/8\,{\frac{1}{\left( 3\,{c_{2}}^{2}-{c_{3}}%
^{2}\right) \left( 2\,{c_{1}}^{2}-3\,{c_{2}}^{2}\right) }}\text{,}  \notag \\
&&1/8\,{\frac{1}{{c_{2}}^{2}\left( 2\,{c_{1}}^{2}-3\,{c_{2}}^{2}\right) }}%
\text{,}-1/8\,{\frac{1}{\left( 3\,{c_{2}}^{2}-{c_{3}}^{2}\right) \left( 2\,{%
c_{1}}^{2}-3\,{c_{2}}^{2}\right) }}]
\end{eqnarray}

\begin{eqnarray}
\mathbf{C}_{15}^{-1} &=&[-1/4\,{\frac{{c_{1}}^{2}{c_{3}}^{2}}{\left( 3\,{%
c_{2}}^{2}-{c_{3}}^{2}\right) \left( 2\,{c_{1}}^{2}-3\,{c_{2}}^{2}\right) }}%
\text{,}1/8\,{\frac{{c_{1}}^{2}{c_{3}}^{2}}{c_{2}\,\left( {c_{1}}^{2}{c_{2}}%
^{2}+{c_{1}}^{2}{c_{3}}^{2}-2\,{c_{3}}^{2}{c_{2}}^{2}\right) }}\text{,}
\notag \\
&&-1/8\,{\frac{{c_{1}}^{2}{c_{3}}^{2}}{c_{2}\,\left( {c_{1}}^{2}{c_{2}}^{2}+{%
c_{1}}^{2}{c_{3}}^{2}-2\,{c_{3}}^{2}{c_{2}}^{2}\right) }}\text{,}-1/8\,{%
\frac{{c_{1}}^{2}{c_{3}}^{2}}{c_{2}\,\left( {c_{1}}^{2}{c_{2}}^{2}+{c_{1}}%
^{2}{c_{3}}^{2}-2\,{c_{3}}^{2}{c_{2}}^{2}\right) }}\text{,}  \notag \\
&&1/8\,{\frac{2\,{c_{1}}^{2}+{c_{3}}^{2}}{\left( 3\,{c_{2}}^{2}-{c_{3}}%
^{2}\right) \left( 2\,{c_{1}}^{2}-3\,{c_{2}}^{2}\right) }}\text{,}-1/4\,{%
\frac{{c_{1}}^{2}}{{c_{2}}^{2}\left( 2\,{c_{1}}^{2}-3\,{c_{2}}^{2}\right) }}%
\text{,}-1/4\,{\frac{{c_{1}}^{2}}{{c_{2}}^{2}\left( 2\,{c_{1}}^{2}-3\,{c_{2}}%
^{2}\right) }}\text{,}  \notag \\
&&1/8\,{\frac{2\,{c_{1}}^{2}+{c_{3}}^{2}}{\left( 3\,{c_{2}}^{2}-{c_{3}}%
^{2}\right) \left( 2\,{c_{1}}^{2}-3\,{c_{2}}^{2}\right) }}\text{,}1/4\,{%
\frac{{c_{1}}^{2}}{{c_{2}}^{2}\left( 2\,{c_{1}}^{2}-3\,{c_{2}}^{2}\right) }}%
\text{,}1/8\,{\frac{2\,{c_{1}}^{2}+{c_{3}}^{2}}{\left( 3\,{c_{2}}^{2}-{c_{3}}%
^{2}\right) \left( 2\,{c_{1}}^{2}-3\,{c_{2}}^{2}\right) }}\text{,}  \notag \\
&&-1/8\,{\frac{{c_{1}}^{2}}{c_{2}\,\left( {c_{1}}^{2}{c_{2}}^{2}+{c_{1}}^{2}{%
c_{3}}^{2}-2\,{c_{3}}^{2}{c_{2}}^{2}\right) }}\text{,}-1/8\,{\frac{{c_{1}}%
^{2}-{c_{3}}^{2}}{c_{2}\,\left( {c_{1}}^{2}{c_{2}}^{2}+{c_{1}}^{2}{c_{3}}%
^{2}-2\,{c_{3}}^{2}{c_{2}}^{2}\right) }}\text{,}  \notag \\
&&-1/8\,{\frac{{c_{1}}^{2}-{c_{3}}^{2}}{c_{2}\,\left( {c_{1}}^{2}{c_{2}}^{2}+%
{c_{1}}^{2}{c_{3}}^{2}-2\,{c_{3}}^{2}{c_{2}}^{2}\right) }}\text{,}1/8\,{%
\frac{{c_{1}}^{2}-{c_{3}}^{2}}{c_{2}\,\left( {c_{1}}^{2}{c_{2}}^{2}+{c_{1}}%
^{2}{c_{3}}^{2}-2\,{c_{3}}^{2}{c_{2}}^{2}\right) }}\text{,}1/8\,{c_{2}}^{-3}%
\text{,}  \notag \\
&&1/8\,{\frac{{c_{1}}^{2}-{c_{3}}^{2}}{c_{2}\,\left( {c_{1}}^{2}{c_{2}}^{2}+{%
c_{1}}^{2}{c_{3}}^{2}-2\,{c_{3}}^{2}{c_{2}}^{2}\right) }}\text{,}1/8\,{\frac{%
{c_{1}}^{2}}{c_{2}\,\left( {c_{1}}^{2}{c_{2}}^{2}+{c_{1}}^{2}{c_{3}}^{2}-2\,{%
c_{3}}^{2}{c_{2}}^{2}\right) }}\text{,}  \notag \\
&&-1/8\,{\frac{{c_{1}}^{2}-{c_{3}}^{2}}{c_{2}\,\left( {c_{1}}^{2}{c_{2}}^{2}+%
{c_{1}}^{2}{c_{3}}^{2}-2\,{c_{3}}^{2}{c_{2}}^{2}\right) }}\text{,}-1/8\,{%
\frac{{c_{1}}^{2}-{c_{3}}^{2}}{c_{2}\,\left( {c_{1}}^{2}{c_{2}}^{2}+{c_{1}}%
^{2}{c_{3}}^{2}-2\,{c_{3}}^{2}{c_{2}}^{2}\right) }}\text{,}  \notag \\
&&1/8\,{\frac{{c_{1}}^{2}}{c_{2}\,\left( {c_{1}}^{2}{c_{2}}^{2}+{c_{1}}^{2}{%
c_{3}}^{2}-2\,{c_{3}}^{2}{c_{2}}^{2}\right) }}\text{,}-1/8\,{\frac{1}{\left(
3\,{c_{2}}^{2}-{c_{3}}^{2}\right) \left( 2\,{c_{1}}^{2}-3\,{c_{2}}%
^{2}\right) }}\text{,}  \notag \\
&&1/8\,{\frac{1}{{c_{2}}^{2}\left( 2\,{c_{1}}^{2}-3\,{c_{2}}^{2}\right) }}%
\text{,}1/8\,{\frac{1}{{c_{2}}^{2}\left( 2\,{c_{1}}^{2}-3\,{c_{2}}%
^{2}\right) }}\text{,}-1/8\,{\frac{1}{\left( 3\,{c_{2}}^{2}-{c_{3}}%
^{2}\right) \left( 2\,{c_{1}}^{2}-3\,{c_{2}}^{2}\right) }}\text{,}  \notag \\
&&-1/8\,{\frac{1}{{c_{2}}^{2}\left( 2\,{c_{1}}^{2}-3\,{c_{2}}^{2}\right) }}%
\text{,}-1/8\,{\frac{1}{\left( 3\,{c_{2}}^{2}-{c_{3}}^{2}\right) \left( 2\,{%
c_{1}}^{2}-3\,{c_{2}}^{2}\right) }}]
\end{eqnarray}

\begin{eqnarray}
\mathbf{C}_{16}^{-1} &=&[-1/4\,{\frac{{c_{1}}^{2}{c_{3}}^{2}}{\left( 3\,{%
c_{2}}^{2}-{c_{3}}^{2}\right) \left( 2\,{c_{1}}^{2}-3\,{c_{2}}^{2}\right) }}%
\text{,}1/8\,{\frac{{c_{1}}^{2}{c_{3}}^{2}}{c_{2}\,\left( {c_{1}}^{2}{c_{2}}%
^{2}+{c_{1}}^{2}{c_{3}}^{2}-2\,{c_{3}}^{2}{c_{2}}^{2}\right) }}\text{,}
\notag \\
&&-1/8\,{\frac{{c_{1}}^{2}{c_{3}}^{2}}{c_{2}\,\left( {c_{1}}^{2}{c_{2}}^{2}+{%
c_{1}}^{2}{c_{3}}^{2}-2\,{c_{3}}^{2}{c_{2}}^{2}\right) }}\text{,}1/8\,{\frac{%
{c_{1}}^{2}{c_{3}}^{2}}{c_{2}\,\left( {c_{1}}^{2}{c_{2}}^{2}+{c_{1}}^{2}{%
c_{3}}^{2}-2\,{c_{3}}^{2}{c_{2}}^{2}\right) }}\text{,}  \notag \\
&&1/8\,{\frac{2\,{c_{1}}^{2}+{c_{3}}^{2}}{\left( 3\,{c_{2}}^{2}-{c_{3}}%
^{2}\right) \left( 2\,{c_{1}}^{2}-3\,{c_{2}}^{2}\right) }}\text{,}-1/4\,{%
\frac{{c_{1}}^{2}}{{c_{2}}^{2}\left( 2\,{c_{1}}^{2}-3\,{c_{2}}^{2}\right) }}%
\text{,}1/4\,{\frac{{c_{1}}^{2}}{{c_{2}}^{2}\left( 2\,{c_{1}}^{2}-3\,{c_{2}}%
^{2}\right) }}\text{,}  \notag \\
&&1/8\,{\frac{2\,{c_{1}}^{2}+{c_{3}}^{2}}{\left( 3\,{c_{2}}^{2}-{c_{3}}%
^{2}\right) \left( 2\,{c_{1}}^{2}-3\,{c_{2}}^{2}\right) }}\text{,}-1/4\,{%
\frac{{c_{1}}^{2}}{{c_{2}}^{2}\left( 2\,{c_{1}}^{2}-3\,{c_{2}}^{2}\right) }}%
\text{,}1/8\,{\frac{2\,{c_{1}}^{2}+{c_{3}}^{2}}{\left( 3\,{c_{2}}^{2}-{c_{3}}%
^{2}\right) \left( 2\,{c_{1}}^{2}-3\,{c_{2}}^{2}\right) }}\text{,}  \notag \\
&&-1/8\,{\frac{{c_{1}}^{2}}{c_{2}\,\left( {c_{1}}^{2}{c_{2}}^{2}+{c_{1}}^{2}{%
c_{3}}^{2}-2\,{c_{3}}^{2}{c_{2}}^{2}\right) }}\text{,}-1/8\,{\frac{{c_{1}}%
^{2}-{c_{3}}^{2}}{c_{2}\,\left( {c_{1}}^{2}{c_{2}}^{2}+{c_{1}}^{2}{c_{3}}%
^{2}-2\,{c_{3}}^{2}{c_{2}}^{2}\right) }}\text{,}  \notag \\
&&1/8\,{\frac{{c_{1}}^{2}-{c_{3}}^{2}}{c_{2}\,\left( {c_{1}}^{2}{c_{2}}^{2}+{%
c_{1}}^{2}{c_{3}}^{2}-2\,{c_{3}}^{2}{c_{2}}^{2}\right) }}\text{,}1/8\,{\frac{%
{c_{1}}^{2}-{c_{3}}^{2}}{c_{2}\,\left( {c_{1}}^{2}{c_{2}}^{2}+{c_{1}}^{2}{%
c_{3}}^{2}-2\,{c_{3}}^{2}{c_{2}}^{2}\right) }}\text{,}  \notag \\
&&-1/8\,{c_{2}}^{-3}\text{,}1/8\,{\frac{{c_{1}}^{2}-{c_{3}}^{2}}{%
c_{2}\,\left( {c_{1}}^{2}{c_{2}}^{2}+{c_{1}}^{2}{c_{3}}^{2}-2\,{c_{3}}^{2}{%
c_{2}}^{2}\right) }}\text{,}1/8\,{\frac{{c_{1}}^{2}}{c_{2}\,\left( {c_{1}}%
^{2}{c_{2}}^{2}+{c_{1}}^{2}{c_{3}}^{2}-2\,{c_{3}}^{2}{c_{2}}^{2}\right) }}%
\text{,}  \notag \\
&&1/8\,{\frac{{c_{1}}^{2}-{c_{3}}^{2}}{c_{2}\,\left( {c_{1}}^{2}{c_{2}}^{2}+{%
c_{1}}^{2}{c_{3}}^{2}-2\,{c_{3}}^{2}{c_{2}}^{2}\right) }}\text{,}-1/8\,{%
\frac{{c_{1}}^{2}-{c_{3}}^{2}}{c_{2}\,\left( {c_{1}}^{2}{c_{2}}^{2}+{c_{1}}%
^{2}{c_{3}}^{2}-2\,{c_{3}}^{2}{c_{2}}^{2}\right) }}\text{,}  \notag \\
&&-1/8\,{\frac{{c_{1}}^{2}}{c_{2}\,\left( {c_{1}}^{2}{c_{2}}^{2}+{c_{1}}^{2}{%
c_{3}}^{2}-2\,{c_{3}}^{2}{c_{2}}^{2}\right) }}\text{,}-1/8\,{\frac{1}{\left(
3\,{c_{2}}^{2}-{c_{3}}^{2}\right) \left( 2\,{c_{1}}^{2}-3\,{c_{2}}%
^{2}\right) }}\text{,}  \notag \\
&&1/8\,{\frac{1}{{c_{2}}^{2}\left( 2\,{c_{1}}^{2}-3\,{c_{2}}^{2}\right) }}%
\text{,}-1/8\,{\frac{1}{{c_{2}}^{2}\left( 2\,{c_{1}}^{2}-3\,{c_{2}}%
^{2}\right) }}\text{,}-1/8\,{\frac{1}{\left( 3\,{c_{2}}^{2}-{c_{3}}%
^{2}\right) \left( 2\,{c_{1}}^{2}-3\,{c_{2}}^{2}\right) }}\text{,}  \notag \\
&&1/8\,{\frac{1}{{c_{2}}^{2}\left( 2\,{c_{1}}^{2}-3\,{c_{2}}^{2}\right) }}%
\text{,}-1/8\,{\frac{1}{\left( 3\,{c_{2}}^{2}-{c_{3}}^{2}\right) \left( 2\,{%
c_{1}}^{2}-3\,{c_{2}}^{2}\right) }}]
\end{eqnarray}

\begin{eqnarray}
\mathbf{C}_{17}^{-1} &=&[-1/4\,{\frac{{c_{1}}^{2}{c_{3}}^{2}}{\left( 3\,{%
c_{2}}^{2}-{c_{3}}^{2}\right) \left( 2\,{c_{1}}^{2}-3\,{c_{2}}^{2}\right) }}%
\text{,}-1/8\,{\frac{{c_{1}}^{2}{c_{3}}^{2}}{c_{2}\,\left( {c_{1}}^{2}{c_{2}}%
^{2}+{c_{1}}^{2}{c_{3}}^{2}-2\,{c_{3}}^{2}{c_{2}}^{2}\right) }}\text{,}
\notag \\
&&1/8\,{\frac{{c_{1}}^{2}{c_{3}}^{2}}{c_{2}\,\left( {c_{1}}^{2}{c_{2}}^{2}+{%
c_{1}}^{2}{c_{3}}^{2}-2\,{c_{3}}^{2}{c_{2}}^{2}\right) }}\text{,}1/8\,{\frac{%
{c_{1}}^{2}{c_{3}}^{2}}{c_{2}\,\left( {c_{1}}^{2}{c_{2}}^{2}+{c_{1}}^{2}{%
c_{3}}^{2}-2\,{c_{3}}^{2}{c_{2}}^{2}\right) }}\text{,}  \notag \\
&&1/8\,{\frac{2\,{c_{1}}^{2}+{c_{3}}^{2}}{\left( 3\,{c_{2}}^{2}-{c_{3}}%
^{2}\right) \left( 2\,{c_{1}}^{2}-3\,{c_{2}}^{2}\right) }}\text{,}-1/4\,{%
\frac{{c_{1}}^{2}}{{c_{2}}^{2}\left( 2\,{c_{1}}^{2}-3\,{c_{2}}^{2}\right) }}%
\text{,}-1/4\,{\frac{{c_{1}}^{2}}{{c_{2}}^{2}\left( 2\,{c_{1}}^{2}-3\,{c_{2}}%
^{2}\right) }}\text{,}  \notag \\
&&1/8\,{\frac{2\,{c_{1}}^{2}+{c_{3}}^{2}}{\left( 3\,{c_{2}}^{2}-{c_{3}}%
^{2}\right) \left( 2\,{c_{1}}^{2}-3\,{c_{2}}^{2}\right) }}\text{,}1/4\,{%
\frac{{c_{1}}^{2}}{{c_{2}}^{2}\left( 2\,{c_{1}}^{2}-3\,{c_{2}}^{2}\right) }}%
\text{,}1/8\,{\frac{2\,{c_{1}}^{2}+{c_{3}}^{2}}{\left( 3\,{c_{2}}^{2}-{c_{3}}%
^{2}\right) \left( 2\,{c_{1}}^{2}-3\,{c_{2}}^{2}\right) }}\text{,}  \notag \\
&&1/8\,{\frac{{c_{1}}^{2}}{c_{2}\,\left( {c_{1}}^{2}{c_{2}}^{2}+{c_{1}}^{2}{%
c_{3}}^{2}-2\,{c_{3}}^{2}{c_{2}}^{2}\right) }}\text{,}1/8\,{\frac{{c_{1}}%
^{2}-{c_{3}}^{2}}{c_{2}\,\left( {c_{1}}^{2}{c_{2}}^{2}+{c_{1}}^{2}{c_{3}}%
^{2}-2\,{c_{3}}^{2}{c_{2}}^{2}\right) }}\text{,}  \notag \\
&&1/8\,{\frac{{c_{1}}^{2}-{c_{3}}^{2}}{c_{2}\,\left( {c_{1}}^{2}{c_{2}}^{2}+{%
c_{1}}^{2}{c_{3}}^{2}-2\,{c_{3}}^{2}{c_{2}}^{2}\right) }}\text{,}-1/8\,{%
\frac{{c_{1}}^{2}-{c_{3}}^{2}}{c_{2}\,\left( {c_{1}}^{2}{c_{2}}^{2}+{c_{1}}%
^{2}{c_{3}}^{2}-2\,{c_{3}}^{2}{c_{2}}^{2}\right) }}\text{,}  \notag \\
&&-1/8\,{c_{2}}^{-3}\text{,}-1/8\,{\frac{{c_{1}}^{2}-{c_{3}}^{2}}{%
c_{2}\,\left( {c_{1}}^{2}{c_{2}}^{2}+{c_{1}}^{2}{c_{3}}^{2}-2\,{c_{3}}^{2}{%
c_{2}}^{2}\right) }}\text{,}-1/8\,{\frac{{c_{1}}^{2}}{c_{2}\,\left( {c_{1}}%
^{2}{c_{2}}^{2}+{c_{1}}^{2}{c_{3}}^{2}-2\,{c_{3}}^{2}{c_{2}}^{2}\right) }}%
\text{,}  \notag \\
&&1/8\,{\frac{{c_{1}}^{2}-{c_{3}}^{2}}{c_{2}\,\left( {c_{1}}^{2}{c_{2}}^{2}+{%
c_{1}}^{2}{c_{3}}^{2}-2\,{c_{3}}^{2}{c_{2}}^{2}\right) }}\text{,}1/8\,{\frac{%
{c_{1}}^{2}-{c_{3}}^{2}}{c_{2}\,\left( {c_{1}}^{2}{c_{2}}^{2}+{c_{1}}^{2}{%
c_{3}}^{2}-2\,{c_{3}}^{2}{c_{2}}^{2}\right) }}\text{,}  \notag \\
&&-1/8\,{\frac{{c_{1}}^{2}}{c_{2}\,\left( {c_{1}}^{2}{c_{2}}^{2}+{c_{1}}^{2}{%
c_{3}}^{2}-2\,{c_{3}}^{2}{c_{2}}^{2}\right) }}\text{,}-1/8\,{\frac{1}{\left(
3\,{c_{2}}^{2}-{c_{3}}^{2}\right) \left( 2\,{c_{1}}^{2}-3\,{c_{2}}%
^{2}\right) }}\text{,}  \notag \\
&&1/8\,{\frac{1}{{c_{2}}^{2}\left( 2\,{c_{1}}^{2}-3\,{c_{2}}^{2}\right) }}%
\text{,}1/8\,{\frac{1}{{c_{2}}^{2}\left( 2\,{c_{1}}^{2}-3\,{c_{2}}%
^{2}\right) }}\text{,}-1/8\,{\frac{1}{\left( 3\,{c_{2}}^{2}-{c_{3}}%
^{2}\right) \left( 2\,{c_{1}}^{2}-3\,{c_{2}}^{2}\right) }}\text{,}  \notag \\
&&-1/8\,{\frac{1}{{c_{2}}^{2}\left( 2\,{c_{1}}^{2}-3\,{c_{2}}^{2}\right) }}%
\text{,}-1/8\,{\frac{1}{\left( 3\,{c_{2}}^{2}-{c_{3}}^{2}\right) \left( 2\,{%
c_{1}}^{2}-3\,{c_{2}}^{2}\right) }}]
\end{eqnarray}

\begin{eqnarray}
\mathbf{C}_{18}^{-1} &=&[-1/4\,{\frac{{c_{1}}^{2}{c_{3}}^{2}}{\left( 3\,{%
c_{2}}^{2}-{c_{3}}^{2}\right) \left( 2\,{c_{1}}^{2}-3\,{c_{2}}^{2}\right) }}%
\text{,}-1/8\,{\frac{{c_{1}}^{2}{c_{3}}^{2}}{c_{2}\,\left( {c_{1}}^{2}{c_{2}}%
^{2}+{c_{1}}^{2}{c_{3}}^{2}-2\,{c_{3}}^{2}{c_{2}}^{2}\right) }}\text{,}
\notag \\
&&1/8\,{\frac{{c_{1}}^{2}{c_{3}}^{2}}{c_{2}\,\left( {c_{1}}^{2}{c_{2}}^{2}+{%
c_{1}}^{2}{c_{3}}^{2}-2\,{c_{3}}^{2}{c_{2}}^{2}\right) }}\text{,}-1/8\,{%
\frac{{c_{1}}^{2}{c_{3}}^{2}}{c_{2}\,\left( {c_{1}}^{2}{c_{2}}^{2}+{c_{1}}%
^{2}{c_{3}}^{2}-2\,{c_{3}}^{2}{c_{2}}^{2}\right) }}\text{,}  \notag \\
&&1/8\,{\frac{2\,{c_{1}}^{2}+{c_{3}}^{2}}{\left( 3\,{c_{2}}^{2}-{c_{3}}%
^{2}\right) \left( 2\,{c_{1}}^{2}-3\,{c_{2}}^{2}\right) }}\text{,}-1/4\,{%
\frac{{c_{1}}^{2}}{{c_{2}}^{2}\left( 2\,{c_{1}}^{2}-3\,{c_{2}}^{2}\right) }}%
\text{,}1/4\,{\frac{{c_{1}}^{2}}{{c_{2}}^{2}\left( 2\,{c_{1}}^{2}-3\,{c_{2}}%
^{2}\right) }}\text{,}  \notag \\
&&1/8\,{\frac{2\,{c_{1}}^{2}+{c_{3}}^{2}}{\left( 3\,{c_{2}}^{2}-{c_{3}}%
^{2}\right) \left( 2\,{c_{1}}^{2}-3\,{c_{2}}^{2}\right) }}\text{,}-1/4\,{%
\frac{{c_{1}}^{2}}{{c_{2}}^{2}\left( 2\,{c_{1}}^{2}-3\,{c_{2}}^{2}\right) }}%
\text{,}1/8\,{\frac{2\,{c_{1}}^{2}+{c_{3}}^{2}}{\left( 3\,{c_{2}}^{2}-{c_{3}}%
^{2}\right) \left( 2\,{c_{1}}^{2}-3\,{c_{2}}^{2}\right) }}\text{,}  \notag \\
&&1/8\,{\frac{{c_{1}}^{2}}{c_{2}\,\left( {c_{1}}^{2}{c_{2}}^{2}+{c_{1}}^{2}{%
c_{3}}^{2}-2\,{c_{3}}^{2}{c_{2}}^{2}\right) }}\text{,}1/8\,{\frac{{c_{1}}%
^{2}-{c_{3}}^{2}}{c_{2}\,\left( {c_{1}}^{2}{c_{2}}^{2}+{c_{1}}^{2}{c_{3}}%
^{2}-2\,{c_{3}}^{2}{c_{2}}^{2}\right) }}\text{,}  \notag \\
&&-1/8\,{\frac{{c_{1}}^{2}-{c_{3}}^{2}}{c_{2}\,\left( {c_{1}}^{2}{c_{2}}^{2}+%
{c_{1}}^{2}{c_{3}}^{2}-2\,{c_{3}}^{2}{c_{2}}^{2}\right) }}\text{,}-1/8\,{%
\frac{{c_{1}}^{2}-{c_{3}}^{2}}{c_{2}\,\left( {c_{1}}^{2}{c_{2}}^{2}+{c_{1}}%
^{2}{c_{3}}^{2}-2\,{c_{3}}^{2}{c_{2}}^{2}\right) }}\text{,}  \notag \\
&&1/8\,{c_{2}}^{-3}\text{,}-1/8\,{\frac{{c_{1}}^{2}-{c_{3}}^{2}}{%
c_{2}\,\left( {c_{1}}^{2}{c_{2}}^{2}+{c_{1}}^{2}{c_{3}}^{2}-2\,{c_{3}}^{2}{%
c_{2}}^{2}\right) }}\text{,}-1/8\,{\frac{{c_{1}}^{2}}{c_{2}\,\left( {c_{1}}%
^{2}{c_{2}}^{2}+{c_{1}}^{2}{c_{3}}^{2}-2\,{c_{3}}^{2}{c_{2}}^{2}\right) }}%
\text{,}  \notag \\
&&-1/8\,{\frac{{c_{1}}^{2}-{c_{3}}^{2}}{c_{2}\,\left( {c_{1}}^{2}{c_{2}}^{2}+%
{c_{1}}^{2}{c_{3}}^{2}-2\,{c_{3}}^{2}{c_{2}}^{2}\right) }}\text{,}1/8\,{%
\frac{{c_{1}}^{2}-{c_{3}}^{2}}{c_{2}\,\left( {c_{1}}^{2}{c_{2}}^{2}+{c_{1}}%
^{2}{c_{3}}^{2}-2\,{c_{3}}^{2}{c_{2}}^{2}\right) }}\text{,}  \notag \\
&&1/8\,{\frac{{c_{1}}^{2}}{c_{2}\,\left( {c_{1}}^{2}{c_{2}}^{2}+{c_{1}}^{2}{%
c_{3}}^{2}-2\,{c_{3}}^{2}{c_{2}}^{2}\right) }}\text{,}-1/8\,{\frac{1}{\left(
3\,{c_{2}}^{2}-{c_{3}}^{2}\right) \left( 2\,{c_{1}}^{2}-3\,{c_{2}}%
^{2}\right) }}\text{,}  \notag \\
&&1/8\,{\frac{1}{{c_{2}}^{2}\left( 2\,{c_{1}}^{2}-3\,{c_{2}}^{2}\right) }}%
\text{,}-1/8\,{\frac{1}{{c_{2}}^{2}\left( 2\,{c_{1}}^{2}-3\,{c_{2}}%
^{2}\right) }}\text{,}-1/8\,{\frac{1}{\left( 3\,{c_{2}}^{2}-{c_{3}}%
^{2}\right) \left( 2\,{c_{1}}^{2}-3\,{c_{2}}^{2}\right) }}\text{,}  \notag \\
&&1/8\,{\frac{1}{{c_{2}}^{2}\left( 2\,{c_{1}}^{2}-3\,{c_{2}}^{2}\right) }}%
\text{,}-1/8\,{\frac{1}{\left( 3\,{c_{2}}^{2}-{c_{3}}^{2}\right) \left( 2\,{%
c_{1}}^{2}-3\,{c_{2}}^{2}\right) }}]
\end{eqnarray}

\begin{eqnarray}
\mathbf{C}_{19}^{-1} &=&[-1/4\,{\frac{{c_{1}}^{2}{c_{3}}^{2}}{\left( 3\,{%
c_{2}}^{2}-{c_{3}}^{2}\right) \left( 2\,{c_{1}}^{2}-3\,{c_{2}}^{2}\right) }}%
\text{,}-1/8\,{\frac{{c_{1}}^{2}{c_{3}}^{2}}{c_{2}\,\left( {c_{1}}^{2}{c_{2}}%
^{2}+{c_{1}}^{2}{c_{3}}^{2}-2\,{c_{3}}^{2}{c_{2}}^{2}\right) }}\text{,}
\notag \\
&&-1/8\,{\frac{{c_{1}}^{2}{c_{3}}^{2}}{c_{2}\,\left( {c_{1}}^{2}{c_{2}}^{2}+{%
c_{1}}^{2}{c_{3}}^{2}-2\,{c_{3}}^{2}{c_{2}}^{2}\right) }}\text{,}-1/8\,{%
\frac{{c_{1}}^{2}{c_{3}}^{2}}{c_{2}\,\left( {c_{1}}^{2}{c_{2}}^{2}+{c_{1}}%
^{2}{c_{3}}^{2}-2\,{c_{3}}^{2}{c_{2}}^{2}\right) }}\text{,}  \notag \\
&&1/8\,{\frac{2\,{c_{1}}^{2}+{c_{3}}^{2}}{\left( 3\,{c_{2}}^{2}-{c_{3}}%
^{2}\right) \left( 2\,{c_{1}}^{2}-3\,{c_{2}}^{2}\right) }}\text{,}1/4\,{%
\frac{{c_{1}}^{2}}{{c_{2}}^{2}\left( 2\,{c_{1}}^{2}-3\,{c_{2}}^{2}\right) }}%
\text{,}1/4\,{\frac{{c_{1}}^{2}}{{c_{2}}^{2}\left( 2\,{c_{1}}^{2}-3\,{c_{2}}%
^{2}\right) }}\text{,}  \notag \\
&&1/8\,{\frac{2\,{c_{1}}^{2}+{c_{3}}^{2}}{\left( 3\,{c_{2}}^{2}-{c_{3}}%
^{2}\right) \left( 2\,{c_{1}}^{2}-3\,{c_{2}}^{2}\right) }}\text{,}1/4\,{%
\frac{{c_{1}}^{2}}{{c_{2}}^{2}\left( 2\,{c_{1}}^{2}-3\,{c_{2}}^{2}\right) }}%
\text{,}1/8\,{\frac{2\,{c_{1}}^{2}+{c_{3}}^{2}}{\left( 3\,{c_{2}}^{2}-{c_{3}}%
^{2}\right) \left( 2\,{c_{1}}^{2}-3\,{c_{2}}^{2}\right) }}\text{,}  \notag \\
&&1/8\,{\frac{{c_{1}}^{2}}{c_{2}\,\left( {c_{1}}^{2}{c_{2}}^{2}+{c_{1}}^{2}{%
c_{3}}^{2}-2\,{c_{3}}^{2}{c_{2}}^{2}\right) }}\text{,}-1/8\,{\frac{{c_{1}}%
^{2}-{c_{3}}^{2}}{c_{2}\,\left( {c_{1}}^{2}{c_{2}}^{2}+{c_{1}}^{2}{c_{3}}%
^{2}-2\,{c_{3}}^{2}{c_{2}}^{2}\right) }}\text{,}  \notag \\
&&-1/8\,{\frac{{c_{1}}^{2}-{c_{3}}^{2}}{c_{2}\,\left( {c_{1}}^{2}{c_{2}}^{2}+%
{c_{1}}^{2}{c_{3}}^{2}-2\,{c_{3}}^{2}{c_{2}}^{2}\right) }}\text{,}-1/8\,{%
\frac{{c_{1}}^{2}-{c_{3}}^{2}}{c_{2}\,\left( {c_{1}}^{2}{c_{2}}^{2}+{c_{1}}%
^{2}{c_{3}}^{2}-2\,{c_{3}}^{2}{c_{2}}^{2}\right) }}\text{,}  \notag \\
&&-1/8\,{c_{2}}^{-3}\text{,}-1/8\,{\frac{{c_{1}}^{2}-{c_{3}}^{2}}{%
c_{2}\,\left( {c_{1}}^{2}{c_{2}}^{2}+{c_{1}}^{2}{c_{3}}^{2}-2\,{c_{3}}^{2}{%
c_{2}}^{2}\right) }}\text{,}1/8\,{\frac{{c_{1}}^{2}}{c_{2}\,\left( {c_{1}}%
^{2}{c_{2}}^{2}+{c_{1}}^{2}{c_{3}}^{2}-2\,{c_{3}}^{2}{c_{2}}^{2}\right) }}%
\text{,}  \notag \\
&&-1/8\,{\frac{{c_{1}}^{2}-{c_{3}}^{2}}{c_{2}\,\left( {c_{1}}^{2}{c_{2}}^{2}+%
{c_{1}}^{2}{c_{3}}^{2}-2\,{c_{3}}^{2}{c_{2}}^{2}\right) }}\text{,}-1/8\,{%
\frac{{c_{1}}^{2}-{c_{3}}^{2}}{c_{2}\,\left( {c_{1}}^{2}{c_{2}}^{2}+{c_{1}}%
^{2}{c_{3}}^{2}-2\,{c_{3}}^{2}{c_{2}}^{2}\right) }}\text{,}  \notag \\
&&1/8\,{\frac{{c_{1}}^{2}}{c_{2}\,\left( {c_{1}}^{2}{c_{2}}^{2}+{c_{1}}^{2}{%
c_{3}}^{2}-2\,{c_{3}}^{2}{c_{2}}^{2}\right) }}\text{,}-1/8\,{\frac{1}{\left(
3\,{c_{2}}^{2}-{c_{3}}^{2}\right) \left( 2\,{c_{1}}^{2}-3\,{c_{2}}%
^{2}\right) }}\text{,}  \notag \\
&&-1/8\,{\frac{1}{{c_{2}}^{2}\left( 2\,{c_{1}}^{2}-3\,{c_{2}}^{2}\right) }}%
\text{,}-1/8\,{\frac{1}{{c_{2}}^{2}\left( 2\,{c_{1}}^{2}-3\,{c_{2}}%
^{2}\right) }}\text{,}-1/8\,{\frac{1}{\left( 3\,{c_{2}}^{2}-{c_{3}}%
^{2}\right) \left( 2\,{c_{1}}^{2}-3\,{c_{2}}^{2}\right) }}\text{,}  \notag \\
&&-1/8\,{\frac{1}{{c_{2}}^{2}\left( 2\,{c_{1}}^{2}-3\,{c_{2}}^{2}\right) }}%
\text{,}-1/8\,{\frac{1}{\left( 3\,{c_{2}}^{2}-{c_{3}}^{2}\right) \left( 2\,{%
c_{1}}^{2}-3\,{c_{2}}^{2}\right) }}]
\end{eqnarray}

\begin{eqnarray}
\mathbf{C}_{20}^{-1} &=&[-1/4\,{\frac{{c_{1}}^{2}{c_{3}}^{2}}{\left( 3\,{%
c_{2}}^{2}-{c_{3}}^{2}\right) \left( 2\,{c_{1}}^{2}-3\,{c_{2}}^{2}\right) }}%
\text{,}-1/8\,{\frac{{c_{1}}^{2}{c_{3}}^{2}}{c_{2}\,\left( {c_{1}}^{2}{c_{2}}%
^{2}+{c_{1}}^{2}{c_{3}}^{2}-2\,{c_{3}}^{2}{c_{2}}^{2}\right) }}\text{,}
\notag \\
&&-1/8\,{\frac{{c_{1}}^{2}{c_{3}}^{2}}{c_{2}\,\left( {c_{1}}^{2}{c_{2}}^{2}+{%
c_{1}}^{2}{c_{3}}^{2}-2\,{c_{3}}^{2}{c_{2}}^{2}\right) }}\text{,}1/8\,{\frac{%
{c_{1}}^{2}{c_{3}}^{2}}{c_{2}\,\left( {c_{1}}^{2}{c_{2}}^{2}+{c_{1}}^{2}{%
c_{3}}^{2}-2\,{c_{3}}^{2}{c_{2}}^{2}\right) }}\text{,}  \notag \\
&&1/8\,{\frac{2\,{c_{1}}^{2}+{c_{3}}^{2}}{\left( 3\,{c_{2}}^{2}-{c_{3}}%
^{2}\right) \left( 2\,{c_{1}}^{2}-3\,{c_{2}}^{2}\right) }}\text{,}1/4\,{%
\frac{{c_{1}}^{2}}{{c_{2}}^{2}\left( 2\,{c_{1}}^{2}-3\,{c_{2}}^{2}\right) }}%
\text{,}-1/4\,{\frac{{c_{1}}^{2}}{{c_{2}}^{2}\left( 2\,{c_{1}}^{2}-3\,{c_{2}}%
^{2}\right) }}\text{,}  \notag \\
&&1/8\,{\frac{2\,{c_{1}}^{2}+{c_{3}}^{2}}{\left( 3\,{c_{2}}^{2}-{c_{3}}%
^{2}\right) \left( 2\,{c_{1}}^{2}-3\,{c_{2}}^{2}\right) }}\text{,}-1/4\,{%
\frac{{c_{1}}^{2}}{{c_{2}}^{2}\left( 2\,{c_{1}}^{2}-3\,{c_{2}}^{2}\right) }}%
\text{,}1/8\,{\frac{2\,{c_{1}}^{2}+{c_{3}}^{2}}{\left( 3\,{c_{2}}^{2}-{c_{3}}%
^{2}\right) \left( 2\,{c_{1}}^{2}-3\,{c_{2}}^{2}\right) }}\text{,}  \notag \\
&&1/8\,{\frac{{c_{1}}^{2}}{c_{2}\,\left( {c_{1}}^{2}{c_{2}}^{2}+{c_{1}}^{2}{%
c_{3}}^{2}-2\,{c_{3}}^{2}{c_{2}}^{2}\right) }}\text{,}-1/8\,{\frac{{c_{1}}%
^{2}-{c_{3}}^{2}}{c_{2}\,\left( {c_{1}}^{2}{c_{2}}^{2}+{c_{1}}^{2}{c_{3}}%
^{2}-2\,{c_{3}}^{2}{c_{2}}^{2}\right) }}\text{,}  \notag \\
&&1/8\,{\frac{{c_{1}}^{2}-{c_{3}}^{2}}{c_{2}\,\left( {c_{1}}^{2}{c_{2}}^{2}+{%
c_{1}}^{2}{c_{3}}^{2}-2\,{c_{3}}^{2}{c_{2}}^{2}\right) }}\text{,}-1/8\,{%
\frac{{c_{1}}^{2}-{c_{3}}^{2}}{c_{2}\,\left( {c_{1}}^{2}{c_{2}}^{2}+{c_{1}}%
^{2}{c_{3}}^{2}-2\,{c_{3}}^{2}{c_{2}}^{2}\right) }}\text{,}1/8\,{c_{2}}^{-3}%
\text{,}  \notag \\
&&-1/8\,{\frac{{c_{1}}^{2}-{c_{3}}^{2}}{c_{2}\,\left( {c_{1}}^{2}{c_{2}}^{2}+%
{c_{1}}^{2}{c_{3}}^{2}-2\,{c_{3}}^{2}{c_{2}}^{2}\right) }}\text{,}1/8\,{%
\frac{{c_{1}}^{2}}{c_{2}\,\left( {c_{1}}^{2}{c_{2}}^{2}+{c_{1}}^{2}{c_{3}}%
^{2}-2\,{c_{3}}^{2}{c_{2}}^{2}\right) }}\text{,}  \notag \\
&&1/8\,{\frac{{c_{1}}^{2}-{c_{3}}^{2}}{c_{2}\,\left( {c_{1}}^{2}{c_{2}}^{2}+{%
c_{1}}^{2}{c_{3}}^{2}-2\,{c_{3}}^{2}{c_{2}}^{2}\right) }}\text{,}-1/8\,{%
\frac{{c_{1}}^{2}-{c_{3}}^{2}}{c_{2}\,\left( {c_{1}}^{2}{c_{2}}^{2}+{c_{1}}%
^{2}{c_{3}}^{2}-2\,{c_{3}}^{2}{c_{2}}^{2}\right) }}\text{,}  \notag \\
&&-1/8\,{\frac{{c_{1}}^{2}}{c_{2}\,\left( {c_{1}}^{2}{c_{2}}^{2}+{c_{1}}^{2}{%
c_{3}}^{2}-2\,{c_{3}}^{2}{c_{2}}^{2}\right) }}\text{,}-1/8\,{\frac{1}{\left(
3\,{c_{2}}^{2}-{c_{3}}^{2}\right) \left( 2\,{c_{1}}^{2}-3\,{c_{2}}%
^{2}\right) }}\text{,}  \notag \\
&&-1/8\,{\frac{1}{{c_{2}}^{2}\left( 2\,{c_{1}}^{2}-3\,{c_{2}}^{2}\right) }}%
\text{,}1/8\,{\frac{1}{{c_{2}}^{2}\left( 2\,{c_{1}}^{2}-3\,{c_{2}}%
^{2}\right) }}\text{,}-1/8\,{\frac{1}{\left( 3\,{c_{2}}^{2}-{c_{3}}%
^{2}\right) \left( 2\,{c_{1}}^{2}-3\,{c_{2}}^{2}\right) }}\text{,}  \notag \\
&&1/8\,{\frac{1}{{c_{2}}^{2}\left( 2\,{c_{1}}^{2}-3\,{c_{2}}^{2}\right) }}%
\text{,}-1/8\,{\frac{1}{\left( 3\,{c_{2}}^{2}-{c_{3}}^{2}\right) \left( 2\,{%
c_{1}}^{2}-3\,{c_{2}}^{2}\right) }}]
\end{eqnarray}

\begin{eqnarray}
\mathbf{C}_{21}^{-1} &=&[{\frac{{c_{1}}^{2}{c_{2}}^{2}}{\left( 3\,{c_{2}}%
^{2}-{c_{3}}^{2}\right) \left( 2\,{c_{1}}^{2}-{c_{3}}^{2}\right) }}\text{,}%
1/2\,{\frac{{c_{1}}^{2}{c_{2}}^{2}}{c_{3}\,\left( {c_{1}}^{2}{c_{2}}^{2}+{%
c_{1}}^{2}{c_{3}}^{2}-2\,{c_{3}}^{2}{c_{2}}^{2}\right) }}\text{,}0\text{,}0%
\text{,}  \notag \\
&&1/2\,{\frac{4\,{c_{1}}^{2}{c_{2}}^{2}-2\,{c_{1}}^{2}{c_{3}}^{2}-{c_{3}}^{2}%
{c_{2}}^{2}}{{c_{3}}^{2}\left( 3\,{c_{2}}^{2}-{c_{3}}^{2}\right) \left( 2\,{%
c_{1}}^{2}-{c_{3}}^{2}\right) }}\text{,}0\text{,}0\text{,}-1/2\,{\frac{%
\left( 2\,{c_{1}}^{2}+{c_{3}}^{2}\right) {c_{2}}^{2}}{{c_{3}}^{2}\left( 3\,{%
c_{2}}^{2}-{c_{3}}^{2}\right) \left( 2\,{c_{1}}^{2}-{c_{3}}^{2}\right) }}%
\text{,}  \notag \\
&&0\text{,}-1/2\,{\frac{\left( 2\,{c_{1}}^{2}+{c_{3}}^{2}\right) {c_{2}}^{2}%
}{{c_{3}}^{2}\left( 3\,{c_{2}}^{2}-{c_{3}}^{2}\right) \left( 2\,{c_{1}}^{2}-{%
c_{3}}^{2}\right) }}\text{,}1/2\,{\frac{{c_{1}}^{2}-2\,{c_{2}}^{2}}{%
c_{3}\,\left( {c_{1}}^{2}{c_{2}}^{2}+{c_{1}}^{2}{c_{3}}^{2}-2\,{c_{3}}^{2}{%
c_{2}}^{2}\right) }}\text{,}  \notag \\
&&0\text{,}0\text{,}-1/2\,{\frac{{c_{1}}^{2}-{c_{2}}^{2}}{c_{3}\,\left( {%
c_{1}}^{2}{c_{2}}^{2}+{c_{1}}^{2}{c_{3}}^{2}-2\,{c_{3}}^{2}{c_{2}}%
^{2}\right) }}\text{,}0\text{,}-1/2\,{\frac{{c_{1}}^{2}-{c_{2}}^{2}}{%
c_{3}\,\left( {c_{1}}^{2}{c_{2}}^{2}+{c_{1}}^{2}{c_{3}}^{2}-2\,{c_{3}}^{2}{%
c_{2}}^{2}\right) }}\text{,}  \notag \\
&&0\text{,}0\text{,}0\text{,}0\text{,}-1/2\,{\frac{2\,{c_{2}}^{2}-{c_{3}}^{2}%
}{{c_{3}}^{2}\left( 3\,{c_{2}}^{2}-{c_{3}}^{2}\right) \left( 2\,{c_{1}}^{2}-{%
c_{3}}^{2}\right) }}\text{,}0\text{,}0\text{,}1/2\,{\frac{{c_{2}}^{2}}{{c_{3}%
}^{2}\left( 3\,{c_{2}}^{2}-{c_{3}}^{2}\right) \left( 2\,{c_{1}}^{2}-{c_{3}}%
^{2}\right) }}\text{,}  \notag \\
&&0\text{,}1/2\,{\frac{{c_{2}}^{2}}{{c_{3}}^{2}\left( 3\,{c_{2}}^{2}-{c_{3}}%
^{2}\right) \left( 2\,{c_{1}}^{2}-{c_{3}}^{2}\right) }}]
\end{eqnarray}

\begin{eqnarray}
\mathbf{C}_{22}^{-1} &=&[{\frac{{c_{1}}^{2}{c_{2}}^{2}}{\left( 3\,{c_{2}}%
^{2}-{c_{3}}^{2}\right) \left( 2\,{c_{1}}^{2}-{c_{3}}^{2}\right) }}\text{,}%
-1/2\,{\frac{{c_{1}}^{2}{c_{2}}^{2}}{c_{3}\,\left( {c_{1}}^{2}{c_{2}}^{2}+{%
c_{1}}^{2}{c_{3}}^{2}-2\,{c_{3}}^{2}{c_{2}}^{2}\right) }}\text{,}0\text{,}0%
\text{,}  \notag \\
&&1/2\,{\frac{4\,{c_{1}}^{2}{c_{2}}^{2}-2\,{c_{1}}^{2}{c_{3}}^{2}-{c_{3}}^{2}%
{c_{2}}^{2}}{{c_{3}}^{2}\left( 3\,{c_{2}}^{2}-{c_{3}}^{2}\right) \left( 2\,{%
c_{1}}^{2}-{c_{3}}^{2}\right) }}\text{,}0\text{,}0\text{,}-1/2\,{\frac{%
\left( 2\,{c_{1}}^{2}+{c_{3}}^{2}\right) {c_{2}}^{2}}{{c_{3}}^{2}\left( 3\,{%
c_{2}}^{2}-{c_{3}}^{2}\right) \left( 2\,{c_{1}}^{2}-{c_{3}}^{2}\right) }}%
\text{,}  \notag \\
&&0\text{,}-1/2\,{\frac{\left( 2\,{c_{1}}^{2}+{c_{3}}^{2}\right) {c_{2}}^{2}%
}{{c_{3}}^{2}\left( 3\,{c_{2}}^{2}-{c_{3}}^{2}\right) \left( 2\,{c_{1}}^{2}-{%
c_{3}}^{2}\right) }}\text{,}-1/2\,{\frac{{c_{1}}^{2}-2\,{c_{2}}^{2}}{%
c_{3}\,\left( {c_{1}}^{2}{c_{2}}^{2}+{c_{1}}^{2}{c_{3}}^{2}-2\,{c_{3}}^{2}{%
c_{2}}^{2}\right) }}\text{,}  \notag \\
&&0\text{,}0\text{,}1/2\,{\frac{{c_{1}}^{2}-{c_{2}}^{2}}{c_{3}\,\left( {c_{1}%
}^{2}{c_{2}}^{2}+{c_{1}}^{2}{c_{3}}^{2}-2\,{c_{3}}^{2}{c_{2}}^{2}\right) }}%
\text{,}0\text{,}1/2\,{\frac{{c_{1}}^{2}-{c_{2}}^{2}}{c_{3}\,\left( {c_{1}}%
^{2}{c_{2}}^{2}+{c_{1}}^{2}{c_{3}}^{2}-2\,{c_{3}}^{2}{c_{2}}^{2}\right) }}%
\text{,}  \notag \\
&&0\text{,}0\text{,}0\text{,}0\text{,}-1/2\,{\frac{2\,{c_{2}}^{2}-{c_{3}}^{2}%
}{{c_{3}}^{2}\left( 3\,{c_{2}}^{2}-{c_{3}}^{2}\right) \left( 2\,{c_{1}}^{2}-{%
c_{3}}^{2}\right) }}\text{,}0\text{,}0\text{,}1/2\,{\frac{{c_{2}}^{2}}{{c_{3}%
}^{2}\left( 3\,{c_{2}}^{2}-{c_{3}}^{2}\right) \left( 2\,{c_{1}}^{2}-{c_{3}}%
^{2}\right) }}\text{,}  \notag \\
&&0\text{,}1/2\,{\frac{{c_{2}}^{2}}{{c_{3}}^{2}\left( 3\,{c_{2}}^{2}-{c_{3}}%
^{2}\right) \left( 2\,{c_{1}}^{2}-{c_{3}}^{2}\right) }}]
\end{eqnarray}

\begin{eqnarray}
\mathbf{C}_{23}^{-1} &=&[{\frac{{c_{1}}^{2}{c_{2}}^{2}}{\left( 3\,{c_{2}}%
^{2}-{c_{3}}^{2}\right) \left( 2\,{c_{1}}^{2}-{c_{3}}^{2}\right) }}\text{,}0%
\text{,}1/2\,{\frac{{c_{1}}^{2}{c_{2}}^{2}}{c_{3}\,\left( {c_{1}}^{2}{c_{2}}%
^{2}+{c_{1}}^{2}{c_{3}}^{2}-2\,{c_{3}}^{2}{c_{2}}^{2}\right) }}\text{,}0%
\text{,}  \notag \\
&&-1/2\,{\frac{\left( 2\,{c_{1}}^{2}+{c_{3}}^{2}\right) {c_{2}}^{2}}{{c_{3}}%
^{2}\left( 3\,{c_{2}}^{2}-{c_{3}}^{2}\right) \left( 2\,{c_{1}}^{2}-{c_{3}}%
^{2}\right) }}\text{,}0\text{,}0\text{,}1/2\,{\frac{4\,{c_{1}}^{2}{c_{2}}%
^{2}-2\,{c_{1}}^{2}{c_{3}}^{2}-{c_{3}}^{2}{c_{2}}^{2}}{{c_{3}}^{2}\left( 3\,{%
c_{2}}^{2}-{c_{3}}^{2}\right) \left( 2\,{c_{1}}^{2}-{c_{3}}^{2}\right) }}%
\text{,}  \notag \\
&&0\text{,}-1/2\,{\frac{\left( 2\,{c_{1}}^{2}+{c_{3}}^{2}\right) {c_{2}}^{2}%
}{{c_{3}}^{2}\left( 3\,{c_{2}}^{2}-{c_{3}}^{2}\right) \left( 2\,{c_{1}}^{2}-{%
c_{3}}^{2}\right) }}\text{,}0\text{,}-1/2\,{\frac{{c_{1}}^{2}-{c_{2}}^{2}}{%
c_{3}\,\left( {c_{1}}^{2}{c_{2}}^{2}+{c_{1}}^{2}{c_{3}}^{2}-2\,{c_{3}}^{2}{%
c_{2}}^{2}\right) }}\text{,}  \notag \\
&&0\text{,}0\text{,}0\text{,}0\text{,}1/2\,{\frac{{c_{1}}^{2}-2\,{c_{2}}^{2}%
}{c_{3}\,\left( {c_{1}}^{2}{c_{2}}^{2}+{c_{1}}^{2}{c_{3}}^{2}-2\,{c_{3}}^{2}{%
c_{2}}^{2}\right) }}\text{,}0\text{,}-1/2\,{\frac{{c_{1}}^{2}-{c_{2}}^{2}}{%
c_{3}\,\left( {c_{1}}^{2}{c_{2}}^{2}+{c_{1}}^{2}{c_{3}}^{2}-2\,{c_{3}}^{2}{%
c_{2}}^{2}\right) }}\text{,}  \notag \\
&&0\text{,}1/2\,{\frac{{c_{2}}^{2}}{{c_{3}}^{2}\left( 3\,{c_{2}}^{2}-{c_{3}}%
^{2}\right) \left( 2\,{c_{1}}^{2}-{c_{3}}^{2}\right) }}\text{,}0\text{,}0%
\text{,}-1/2\,{\frac{2\,{c_{2}}^{2}-{c_{3}}^{2}}{{c_{3}}^{2}\left( 3\,{c_{2}}%
^{2}-{c_{3}}^{2}\right) \left( 2\,{c_{1}}^{2}-{c_{3}}^{2}\right) }}\text{,}
\notag \\
&&0\text{,}1/2\,{\frac{{c_{2}}^{2}}{{c_{3}}^{2}\left( 3\,{c_{2}}^{2}-{c_{3}}%
^{2}\right) \left( 2\,{c_{1}}^{2}-{c_{3}}^{2}\right) }}]
\end{eqnarray}

\begin{eqnarray}
\mathbf{C}_{24}^{-1} &=&[{\frac{{c_{1}}^{2}{c_{2}}^{2}}{\left( 3\,{c_{2}}%
^{2}-{c_{3}}^{2}\right) \left( 2\,{c_{1}}^{2}-{c_{3}}^{2}\right) }}\text{,}0%
\text{,}-1/2\,{\frac{{c_{1}}^{2}{c_{2}}^{2}}{c_{3}\,\left( {c_{1}}^{2}{c_{2}}%
^{2}+{c_{1}}^{2}{c_{3}}^{2}-2\,{c_{3}}^{2}{c_{2}}^{2}\right) }}\text{,}
\notag \\
&&0\text{,}-1/2\,{\frac{\left( 2\,{c_{1}}^{2}+{c_{3}}^{2}\right) {c_{2}}^{2}%
}{{c_{3}}^{2}\left( 3\,{c_{2}}^{2}-{c_{3}}^{2}\right) \left( 2\,{c_{1}}^{2}-{%
c_{3}}^{2}\right) }}\text{,}0\text{,}0\text{,}1/2\,{\frac{4\,{c_{1}}^{2}{%
c_{2}}^{2}-2\,{c_{1}}^{2}{c_{3}}^{2}-{c_{3}}^{2}{c_{2}}^{2}}{{c_{3}}%
^{2}\left( 3\,{c_{2}}^{2}-{c_{3}}^{2}\right) \left( 2\,{c_{1}}^{2}-{c_{3}}%
^{2}\right) }}\text{,}  \notag \\
&&0\text{,}-1/2\,{\frac{\left( 2\,{c_{1}}^{2}+{c_{3}}^{2}\right) {c_{2}}^{2}%
}{{c_{3}}^{2}\left( 3\,{c_{2}}^{2}-{c_{3}}^{2}\right) \left( 2\,{c_{1}}^{2}-{%
c_{3}}^{2}\right) }}\text{,}0\text{,}1/2\,{\frac{{c_{1}}^{2}-{c_{2}}^{2}}{%
c_{3}\,\left( {c_{1}}^{2}{c_{2}}^{2}+{c_{1}}^{2}{c_{3}}^{2}-2\,{c_{3}}^{2}{%
c_{2}}^{2}\right) }}\text{,}  \notag \\
&&0\text{,}0\text{,}0\text{,}0\text{,}-1/2\,{\frac{{c_{1}}^{2}-2\,{c_{2}}^{2}%
}{c_{3}\,\left( {c_{1}}^{2}{c_{2}}^{2}+{c_{1}}^{2}{c_{3}}^{2}-2\,{c_{3}}^{2}{%
c_{2}}^{2}\right) }}\text{,}0\text{,}1/2\,{\frac{{c_{1}}^{2}-{c_{2}}^{2}}{%
c_{3}\,\left( {c_{1}}^{2}{c_{2}}^{2}+{c_{1}}^{2}{c_{3}}^{2}-2\,{c_{3}}^{2}{%
c_{2}}^{2}\right) }}\text{,}  \notag \\
&&0\text{,}1/2\,{\frac{{c_{2}}^{2}}{{c_{3}}^{2}\left( 3\,{c_{2}}^{2}-{c_{3}}%
^{2}\right) \left( 2\,{c_{1}}^{2}-{c_{3}}^{2}\right) }}\text{,}0\text{,}0%
\text{,}-1/2\,{\frac{2\,{c_{2}}^{2}-{c_{3}}^{2}}{{c_{3}}^{2}\left( 3\,{c_{2}}%
^{2}-{c_{3}}^{2}\right) \left( 2\,{c_{1}}^{2}-{c_{3}}^{2}\right) }}\text{,}
\notag \\
&&0\text{,}1/2\,{\frac{{c_{2}}^{2}}{{c_{3}}^{2}\left( 3\,{c_{2}}^{2}-{c_{3}}%
^{2}\right) \left( 2\,{c_{1}}^{2}-{c_{3}}^{2}\right) }}]
\end{eqnarray}

\begin{eqnarray}
\mathbf{C}_{25}^{-1} &=&[{\frac{{c_{1}}^{2}{c_{2}}^{2}}{\left( 3\,{c_{2}}%
^{2}-{c_{3}}^{2}\right) \left( 2\,{c_{1}}^{2}-{c_{3}}^{2}\right) }}\text{,}0%
\text{,}0\text{,}1/2\,{\frac{{c_{1}}^{2}{c_{2}}^{2}}{c_{3}\,\left( {c_{1}}%
^{2}{c_{2}}^{2}+{c_{1}}^{2}{c_{3}}^{2}-2\,{c_{3}}^{2}{c_{2}}^{2}\right) }}%
\text{,}  \notag \\
&&-1/2\,{\frac{\left( 2\,{c_{1}}^{2}+{c_{3}}^{2}\right) {c_{2}}^{2}}{{c_{3}}%
^{2}\left( 3\,{c_{2}}^{2}-{c_{3}}^{2}\right) \left( 2\,{c_{1}}^{2}-{c_{3}}%
^{2}\right) }}\text{,}0\text{,}0\text{,}-1/2\,{\frac{\left( 2\,{c_{1}}^{2}+{%
c_{3}}^{2}\right) {c_{2}}^{2}}{{c_{3}}^{2}\left( 3\,{c_{2}}^{2}-{c_{3}}%
^{2}\right) \left( 2\,{c_{1}}^{2}-{c_{3}}^{2}\right) }}\text{,}  \notag \\
&&0\text{,}1/2\,{\frac{4\,{c_{1}}^{2}{c_{2}}^{2}-2\,{c_{1}}^{2}{c_{3}}^{2}-{%
c_{3}}^{2}{c_{2}}^{2}}{{c_{3}}^{2}\left( 3\,{c_{2}}^{2}-{c_{3}}^{2}\right)
\left( 2\,{c_{1}}^{2}-{c_{3}}^{2}\right) }}\text{,}0\text{,}0\text{,}-1/2\,{%
\frac{{c_{1}}^{2}-{c_{2}}^{2}}{c_{3}\,\left( {c_{1}}^{2}{c_{2}}^{2}+{c_{1}}%
^{2}{c_{3}}^{2}-2\,{c_{3}}^{2}{c_{2}}^{2}\right) }}\text{,}  \notag \\
&&0\text{,}0\text{,}0\text{,}0\text{,}-1/2\,{\frac{{c_{1}}^{2}-{c_{2}}^{2}}{%
c_{3}\,\left( {c_{1}}^{2}{c_{2}}^{2}+{c_{1}}^{2}{c_{3}}^{2}-2\,{c_{3}}^{2}{%
c_{2}}^{2}\right) }}\text{,}0\text{,}1/2\,{\frac{{c_{1}}^{2}-2\,{c_{2}}^{2}}{%
c_{3}\,\left( {c_{1}}^{2}{c_{2}}^{2}+{c_{1}}^{2}{c_{3}}^{2}-2\,{c_{3}}^{2}{%
c_{2}}^{2}\right) }}\text{,}  \notag \\
&&1/2\,{\frac{{c_{2}}^{2}}{{c_{3}}^{2}\left( 3\,{c_{2}}^{2}-{c_{3}}%
^{2}\right) \left( 2\,{c_{1}}^{2}-{c_{3}}^{2}\right) }}\text{,}0\text{,}0%
\text{,}1/2\,{\frac{{c_{2}}^{2}}{{c_{3}}^{2}\left( 3\,{c_{2}}^{2}-{c_{3}}%
^{2}\right) \left( 2\,{c_{1}}^{2}-{c_{3}}^{2}\right) }}\text{,}0\text{,}
\notag \\
&&-1/2\,{\frac{2\,{c_{2}}^{2}-{c_{3}}^{2}}{{c_{3}}^{2}\left( 3\,{c_{2}}^{2}-{%
c_{3}}^{2}\right) \left( 2\,{c_{1}}^{2}-{c_{3}}^{2}\right) }}]
\end{eqnarray}

\begin{eqnarray}
\mathbf{C}_{26}^{-1} &=&[{\frac{{c_{1}}^{2}{c_{2}}^{2}}{\left( 3\,{c_{2}}%
^{2}-{c_{3}}^{2}\right) \left( 2\,{c_{1}}^{2}-{c_{3}}^{2}\right) }}\text{,}0%
\text{,}0\text{,}-1/2\,{\frac{{c_{1}}^{2}{c_{2}}^{2}}{c_{3}\,\left( {c_{1}}%
^{2}{c_{2}}^{2}+{c_{1}}^{2}{c_{3}}^{2}-2\,{c_{3}}^{2}{c_{2}}^{2}\right) }}%
\text{,}  \notag \\
&&-1/2\,{\frac{\left( 2\,{c_{1}}^{2}+{c_{3}}^{2}\right) {c_{2}}^{2}}{{c_{3}}%
^{2}\left( 3\,{c_{2}}^{2}-{c_{3}}^{2}\right) \left( 2\,{c_{1}}^{2}-{c_{3}}%
^{2}\right) }}\text{,}0\text{,}0\text{,}-1/2\,{\frac{\left( 2\,{c_{1}}^{2}+{%
c_{3}}^{2}\right) {c_{2}}^{2}}{{c_{3}}^{2}\left( 3\,{c_{2}}^{2}-{c_{3}}%
^{2}\right) \left( 2\,{c_{1}}^{2}-{c_{3}}^{2}\right) }}\text{,}  \notag \\
&&0\text{,}1/2\,{\frac{4\,{c_{1}}^{2}{c_{2}}^{2}-2\,{c_{1}}^{2}{c_{3}}^{2}-{%
c_{3}}^{2}{c_{2}}^{2}}{{c_{3}}^{2}\left( 3\,{c_{2}}^{2}-{c_{3}}^{2}\right)
\left( 2\,{c_{1}}^{2}-{c_{3}}^{2}\right) }}\text{,}0\text{,}0\text{,}1/2\,{%
\frac{{c_{1}}^{2}-{c_{2}}^{2}}{c_{3}\,\left( {c_{1}}^{2}{c_{2}}^{2}+{c_{1}}%
^{2}{c_{3}}^{2}-2\,{c_{3}}^{2}{c_{2}}^{2}\right) }}\text{,}  \notag \\
&&0\text{,}0\text{,}0\text{,}0\text{,}1/2\,{\frac{{c_{1}}^{2}-{c_{2}}^{2}}{%
c_{3}\,\left( {c_{1}}^{2}{c_{2}}^{2}+{c_{1}}^{2}{c_{3}}^{2}-2\,{c_{3}}^{2}{%
c_{2}}^{2}\right) }}\text{,}0\text{,}-1/2\,{\frac{{c_{1}}^{2}-2\,{c_{2}}^{2}%
}{c_{3}\,\left( {c_{1}}^{2}{c_{2}}^{2}+{c_{1}}^{2}{c_{3}}^{2}-2\,{c_{3}}^{2}{%
c_{2}}^{2}\right) }}\text{,}  \notag \\
&&1/2\,{\frac{{c_{2}}^{2}}{{c_{3}}^{2}\left( 3\,{c_{2}}^{2}-{c_{3}}%
^{2}\right) \left( 2\,{c_{1}}^{2}-{c_{3}}^{2}\right) }}\text{,}0\text{,}0%
\text{,}1/2\,{\frac{{c_{2}}^{2}}{{c_{3}}^{2}\left( 3\,{c_{2}}^{2}-{c_{3}}%
^{2}\right) \left( 2\,{c_{1}}^{2}-{c_{3}}^{2}\right) }}\text{,}0\text{,}
\notag \\
&&-1/2\,{\frac{2\,{c_{2}}^{2}-{c_{3}}^{2}}{{c_{3}}^{2}\left( 3\,{c_{2}}^{2}-{%
c_{3}}^{2}\right) \left( 2\,{c_{1}}^{2}-{c_{3}}^{2}\right) }}]
\end{eqnarray}

\section{$\mathbf{C}^{-1}$ for the D2V16}

The DVM with 2 dimensions and 16 velocities can have 4 energy levels.
\end{subequations}
\begin{subequations}
\begin{equation}
\left.
\begin{array}{c}
v_{ir}=[1,0,-1,0]c_{1} \\
v_{i\theta }=[0,1,0,-1]c_{1}%
\end{array}%
\right\} i=1,\cdots ,4\text{,}
\end{equation}%
\begin{equation}
\left.
\begin{array}{c}
v_{ir}=[1,-1,-1,1]c_{2} \\
v_{i\theta }=[1,1,-1,-1]c_{2}%
\end{array}%
\right\} i=5,\cdots ,8\text{,}
\end{equation}%
\begin{equation}
\left.
\begin{array}{c}
v_{ir}=[1,0,-1,0]2c_{3} \\
v_{i\theta }=[0,1,0,-1]2c_{3}%
\end{array}%
\right\} i=9,\cdots ,12\text{,}
\end{equation}%
\begin{equation}
\left.
\begin{array}{c}
v_{ir}=[1,-1,-1,1]2\sqrt{2}c_{4} \\
v_{i\theta }=[1,1,-1,-1]2\sqrt{2}c_{4}%
\end{array}%
\right\} i=13,\cdots ,16\text{.}
\end{equation}

If we choose $\eta _{i}=\eta _{0}$ for $i=5$, $\cdots $,$8$, and $\eta
_{i}=0 $ for others, then there are only five variables in the components of
the matrix $\mathbf{C}$. The components of the inverse of $\mathbf{C}$, $%
\mathbf{C}^{-1}=\left[ \mathbf{C}_{k}^{-1}\right] $, are shown below.

\end{subequations}
\begin{subequations}
\begin{eqnarray}
\mathbf{C}_{1}^{-1} &=&[1/2\,{\frac{{c_{3}}^{2}{c_{4}}^{2}}{\left( {c_{1}}%
^{2}-2\,{c_{4}}^{2}\right) \left( {c_{1}}^{2}-{c_{3}}^{2}\right) }}\text{,}%
-1/2\,{\frac{{c_{3}}^{2}}{c_{1}\,\left( {c_{1}}^{2}-{c_{3}}^{2}\right) }}%
\text{,}0\text{,}  \notag \\
&&-1/2\,{\frac{2\,{c_{2}}^{4}-{c_{2}}^{2}{c_{3}}^{2}+{c_{2}}^{2}{\eta _{0}}%
^{2}-2\,{c_{2}}^{2}{c_{4}}^{2}+{c_{3}}^{2}{c_{4}}^{2}}{{\eta _{0}}^{2}\left(
{c_{1}}^{2}-2\,{c_{4}}^{2}\right) \left( {c_{1}}^{2}-{c_{3}}^{2}\right) }}%
\text{,}  \notag \\
&&C_{1\text{,}5}^{-1}\text{, }0\text{, }C_{1\text{,}7}^{-1}\text{,}-1/2\,{%
\frac{\left( {c_{2}}^{2}-{c_{4}}^{2}\right) {c_{3}}^{2}}{c_{1}\,{c_{4}}^{2}{%
\eta _{0}}^{2}\left( {c_{1}}^{2}-{c_{3}}^{2}\right) }}\text{,}0\text{,}1/2\,{%
\frac{{c_{2}}^{2}{c_{3}}^{2}-{c_{3}}^{2}{c_{4}}^{2}+{c_{4}}^{2}{\eta _{0}}%
^{2}}{c_{1}\,{c_{4}}^{2}{\eta _{0}}^{2}\left( {c_{1}}^{2}-{c_{3}}^{2}\right)
}}\text{,}  \notag \\
&&0\text{,}1/2\,{\frac{{c_{2}}^{2}{c_{3}}^{2}+{\eta _{0}}^{2}{c_{3}}^{2}-{%
c_{3}}^{2}{c_{4}}^{2}-{c_{4}}^{2}{\eta _{0}}^{2}}{c_{1}\,{c_{4}}^{2}{\eta
_{0}}^{2}\left( {c_{1}}^{2}-{c_{3}}^{2}\right) }}\text{,}0\text{,}1/2\,{%
\frac{{c_{1}}^{2}-{c_{4}}^{2}}{{c_{1}}^{2}\left( {c_{1}}^{2}-2\,{c_{4}}%
^{2}\right) \left( {c_{1}}^{2}-{c_{3}}^{2}\right) }}\text{,}  \notag \\
&&0\text{,}1/2\,{\frac{{c_{4}}^{2}}{{c_{1}}^{2}\left( {c_{1}}^{2}-2\,{c_{4}}%
^{2}\right) \left( {c_{1}}^{2}-{c_{3}}^{2}\right) }}]
\end{eqnarray}%
\begin{eqnarray}
\mathbf{C}_{2}^{-1} &=&[1/2\,{\frac{{c_{3}}^{2}{c_{4}}^{2}}{\left( {c_{1}}%
^{2}-2\,{c_{4}}^{2}\right) \left( {c_{1}}^{2}-{c_{3}}^{2}\right) }}\text{,}0%
\text{,}-1/2\,{\frac{{c_{3}}^{2}}{c_{1}\,\left( {c_{1}}^{2}-{c_{3}}%
^{2}\right) }}\text{,}  \notag \\
&&-1/2\,{\frac{2\,{c_{2}}^{4}-{c_{2}}^{2}{c_{3}}^{2}+{c_{2}}^{2}{\eta _{0}}%
^{2}-2\,{c_{2}}^{2}{c_{4}}^{2}+{c_{3}}^{2}{c_{4}}^{2}}{{\eta _{0}}^{2}\left(
{c_{1}}^{2}-2\,{c_{4}}^{2}\right) \left( {c_{1}}^{2}-{c_{3}}^{2}\right) }}%
\text{,}C_{2,5}^{-1}\text{,}0\text{,}C_{2,7}^{-1}\text{,}0\text{,}  \notag \\
&&-1/2\,{\frac{\left( {c_{2}}^{2}-{c_{4}}^{2}\right) {c_{3}}^{2}}{c_{1}\,{%
c_{4}}^{2}{\eta _{0}}^{2}\left( {c_{1}}^{2}-{c_{3}}^{2}\right) }}\text{,}0%
\text{,}1/2\,{\frac{{c_{2}}^{2}{c_{3}}^{2}+{\eta _{0}}^{2}{c_{3}}^{2}-{c_{3}}%
^{2}{c_{4}}^{2}-{c_{4}}^{2}{\eta _{0}}^{2}}{c_{1}\,{c_{4}}^{2}{\eta _{0}}%
^{2}\left( {c_{1}}^{2}-{c_{3}}^{2}\right) }}\text{,}  \notag \\
&&0\text{,}1/2\,{\frac{{c_{2}}^{2}{c_{3}}^{2}-{c_{3}}^{2}{c_{4}}^{2}+{c_{4}}%
^{2}{\eta _{0}}^{2}}{c_{1}\,{c_{4}}^{2}{\eta _{0}}^{2}\left( {c_{1}}^{2}-{%
c_{3}}^{2}\right) }}\text{,}1/2\,{\frac{{c_{4}}^{2}}{{c_{1}}^{2}\left( {c_{1}%
}^{2}-2\,{c_{4}}^{2}\right) \left( {c_{1}}^{2}-{c_{3}}^{2}\right) }}\text{,}
\notag \\
&&0\text{,}1/2\,{\frac{{c_{1}}^{2}-{c_{4}}^{2}}{{c_{1}}^{2}\left( {c_{1}}%
^{2}-2\,{c_{4}}^{2}\right) \left( {c_{1}}^{2}-{c_{3}}^{2}\right) }}]
\end{eqnarray}%
\begin{eqnarray}
\mathbf{C}_{3}^{-1} &=&[1/2\,{\frac{{c_{3}}^{2}{c_{4}}^{2}}{\left( {c_{1}}%
^{2}-2\,{c_{4}}^{2}\right) \left( {c_{1}}^{2}-{c_{3}}^{2}\right) }}\text{,}%
1/2\,{\frac{{c_{3}}^{2}}{c_{1}\,\left( {c_{1}}^{2}-{c_{3}}^{2}\right) }}%
\text{,}0\text{,}  \notag \\
&&-1/2\,{\frac{2\,{c_{2}}^{4}-{c_{2}}^{2}{c_{3}}^{2}+{c_{2}}^{2}{\eta _{0}}%
^{2}-2\,{c_{2}}^{2}{c_{4}}^{2}+{c_{3}}^{2}{c_{4}}^{2}}{{\eta _{0}}^{2}\left(
{c_{1}}^{2}-2\,{c_{4}}^{2}\right) \left( {c_{1}}^{2}-{c_{3}}^{2}\right) }}%
\text{,}  \notag \\
&&C_{3,5}^{-1}\text{,}0\text{,}C_{3,7}^{-1}\text{,}1/2\,{\frac{\left( {c_{2}}%
^{2}-{c_{4}}^{2}\right) {c_{3}}^{2}}{c_{1}\,{c_{4}}^{2}{\eta _{0}}^{2}\left(
{c_{1}}^{2}-{c_{3}}^{2}\right) }}\text{,}0\text{,}-1/2\,{\frac{{c_{2}}^{2}{%
c_{3}}^{2}-{c_{3}}^{2}{c_{4}}^{2}+{c_{4}}^{2}{\eta _{0}}^{2}}{c_{1}\,{c_{4}}%
^{2}{\eta _{0}}^{2}\left( {c_{1}}^{2}-{c_{3}}^{2}\right) }}\text{,}  \notag
\\
&&0\text{,}-1/2\,{\frac{{c_{2}}^{2}{c_{3}}^{2}+{\eta _{0}}^{2}{c_{3}}^{2}-{%
c_{3}}^{2}{c_{4}}^{2}-{c_{4}}^{2}{\eta _{0}}^{2}}{c_{1}\,{c_{4}}^{2}{\eta
_{0}}^{2}\left( {c_{1}}^{2}-{c_{3}}^{2}\right) }}\text{,}  \notag \\
&&0\text{,}1/2\,{\frac{{c_{1}}^{2}-{c_{4}}^{2}}{\left( {c_{1}}^{2}-2\,{c_{4}}%
^{2}\right) {c_{1}}^{2}\left( {c_{1}}^{2}-{c_{3}}^{2}\right) }}\text{,}0%
\text{,}1/2\,{\frac{{c_{4}}^{2}}{\left( {c_{1}}^{2}-2\,{c_{4}}^{2}\right) {%
c_{1}}^{2}\left( {c_{1}}^{2}-{c_{3}}^{2}\right) }}]
\end{eqnarray}%
\begin{eqnarray}
\mathbf{C}_{4}^{-1} &=&[1/2\,{\frac{{c_{3}}^{2}{c_{4}}^{2}}{\left( {c_{1}}%
^{2}-2\,{c_{4}}^{2}\right) \left( {c_{1}}^{2}-{c_{3}}^{2}\right) }}\text{,}0%
\text{,}1/2\,{\frac{{c_{3}}^{2}}{c_{1}\,\left( {c_{1}}^{2}-{c_{3}}%
^{2}\right) }}\text{,}  \notag \\
&&-1/2\,{\frac{2\,{c_{2}}^{4}-{c_{2}}^{2}{c_{3}}^{2}+{c_{2}}^{2}{\eta _{0}}%
^{2}-2\,{c_{2}}^{2}{c_{4}}^{2}+{c_{3}}^{2}{c_{4}}^{2}}{{\eta _{0}}^{2}\left(
{c_{1}}^{2}-2\,{c_{4}}^{2}\right) \left( {c_{1}}^{2}-{c_{3}}^{2}\right) }}%
\text{,}C_{4,5}^{-1}\text{,}0\text{,}C_{4,7}^{-1}\text{,}  \notag \\
&&0\text{,}\,{\frac{\left( {c_{2}}^{2}-{c_{4}}^{2}\right) {c_{3}}^{2}}{%
2c_{1}\,{c_{4}}^{2}{\eta _{0}}^{2}\left( {c_{1}}^{2}-{c_{3}}^{2}\right) }}%
\text{,}0\text{,}-{\frac{{c_{2}}^{2}{c_{3}}^{2}+{\eta _{0}}^{2}{c_{3}}^{2}-{%
c_{3}}^{2}{c_{4}}^{2}-{c_{4}}^{2}{\eta _{0}}^{2}}{2c_{1}\,{c_{4}}^{2}{\eta
_{0}}^{2}\left( {c_{1}}^{2}-{c_{3}}^{2}\right) }}\text{,}  \notag \\
&&0\text{,}-{\frac{{c_{2}}^{2}{c_{3}}^{2}-{c_{3}}^{2}{c_{4}}^{2}+{c_{4}}^{2}{%
\eta _{0}}^{2}}{2c_{1}\,{c_{4}}^{2}{\eta _{0}}^{2}\left( {c_{1}}^{2}-{c_{3}}%
^{2}\right) }}\text{,}\,{\frac{{c_{4}}^{2}}{2\left( {c_{1}}^{2}-2\,{c_{4}}%
^{2}\right) {c_{1}}^{2}\left( {c_{1}}^{2}-{c_{3}}^{2}\right) }}\text{,}
\notag \\
&&0\text{,}1/2\,{\frac{{c_{1}}^{2}-{c_{4}}^{2}}{\left( {c_{1}}^{2}-2\,{c_{4}}%
^{2}\right) {c_{1}}^{2}\left( {c_{1}}^{2}-{c_{3}}^{2}\right) }}]
\end{eqnarray}%
\begin{eqnarray}
\mathbf{C}_{5}^{-1} &=&[0\text{,}0\text{,}0\text{,}1/4\,{\eta _{0}}^{-2}%
\text{,}-1/4\,{\eta _{0}}^{-2}\text{,}-1/2\,{\frac{{c_{4}}^{2}}{{c_{2}}%
^{2}\left( {\eta _{0}}^{2}+2\,{c_{2}}^{2}-2\,{c_{4}}^{2}\right) }}\text{,}
\notag \\
&&-1/4\,{\eta _{0}}^{-2}\text{,}1/4\,{\frac{1}{c_{2}\,{\eta _{0}}^{2}}}\text{%
,}1/4\,{\frac{1}{c_{2}\,{\eta _{0}}^{2}}}\text{,}-1/4\,{\frac{1}{c_{2}\,{%
\eta _{0}}^{2}}}\text{,}-1/4\,{\frac{1}{c_{2}\,{\eta _{0}}^{2}}}\text{,}
\notag \\
&&-1/4\,{\frac{1}{c_{2}\,{\eta _{0}}^{2}}}\text{,}-1/4\,{\frac{1}{c_{2}\,{%
\eta _{0}}^{2}}}\text{,}0\text{,}1/4\,{\frac{1}{{c_{2}}^{2}\left( {\eta _{0}}%
^{2}+2\,{c_{2}}^{2}-2\,{c_{4}}^{2}\right) }}\text{,}0]
\end{eqnarray}%
\begin{eqnarray}
\mathbf{C}_{6}^{-1} &=&[0\text{,}0\text{,}0\text{,}1/4\,{\eta _{0}}^{-2}%
\text{,}-1/4\,{\eta _{0}}^{-2}\text{,}1/2\,{\frac{{c_{4}}^{2}}{{c_{2}}%
^{2}\left( {\eta _{0}}^{2}+2\,{c_{2}}^{2}-2\,{c_{4}}^{2}\right) }}\text{,}%
-1/4\,{\eta _{0}}^{-2}\text{,}  \notag \\
&&-1/4\,{\frac{1}{c_{2}\,{\eta _{0}}^{2}}}\text{,}1/4\,{\frac{1}{c_{2}\,{%
\eta _{0}}^{2}}}\text{,}1/4\,{\frac{1}{c_{2}\,{\eta _{0}}^{2}}}\text{,}-1/4\,%
{\frac{1}{c_{2}\,{\eta _{0}}^{2}}}\text{,}1/4\,{\frac{1}{c_{2}\,{\eta _{0}}%
^{2}}}\text{,}  \notag \\
&&-1/4\,{\frac{1}{c_{2}\,{\eta _{0}}^{2}}}\text{,}0\text{,}-1/4\,{\frac{1}{{%
c_{2}}^{2}\left( {\eta _{0}}^{2}+2\,{c_{2}}^{2}-2\,{c_{4}}^{2}\right) }}%
\text{,}0]
\end{eqnarray}%
\begin{eqnarray}
\mathbf{C}_{7}^{-1} &=&[0\text{,}0\text{,}0\text{,}1/4\,{\eta _{0}}^{-2}%
\text{,}-1/4\,{\eta _{0}}^{-2}\text{,}-1/2\,{\frac{{c_{4}}^{2}}{{c_{2}}%
^{2}\left( {\eta _{0}}^{2}+2\,{c_{2}}^{2}-2\,{c_{4}}^{2}\right) }}\text{,}
\notag \\
&&-1/4\,{\eta _{0}}^{-2}\text{,}-1/4\,{\frac{1}{c_{2}\,{\eta _{0}}^{2}}}%
\text{,}-1/4\,{\frac{1}{c_{2}\,{\eta _{0}}^{2}}}\text{,}1/4\,{\frac{1}{%
c_{2}\,{\eta _{0}}^{2}}}\text{,}1/4\,{\frac{1}{c_{2}\,{\eta _{0}}^{2}}}\text{%
,}  \notag \\
&&1/4\,{\frac{1}{c_{2}\,{\eta _{0}}^{2}}}\text{,}1/4\,{\frac{1}{c_{2}\,{\eta
_{0}}^{2}}}\text{,}0\text{,}1/4\,{\frac{1}{{c_{2}}^{2}\left( {\eta _{0}}%
^{2}+2\,{c_{2}}^{2}-2\,{c_{4}}^{2}\right) }}\text{,}0]
\end{eqnarray}%
\begin{eqnarray}
\mathbf{C}_{8}^{-1} &=&[0\text{,}0\text{,}0\text{,}1/4\,{\eta _{0}}^{-2}%
\text{,}-1/4\,{\eta _{0}}^{-2}\text{,}1/2\,{\frac{{c_{4}}^{2}}{{c_{2}}%
^{2}\left( {\eta _{0}}^{2}+2\,{c_{2}}^{2}-2\,{c_{4}}^{2}\right) }}\text{,}%
-1/4\,{\eta _{0}}^{-2}\text{,}  \notag \\
&&1/4\,{\frac{1}{c_{2}\,{\eta _{0}}^{2}}}\text{,}-1/4\,{\frac{1}{c_{2}\,{%
\eta _{0}}^{2}}}\text{,}-1/4\,{\frac{1}{c_{2}\,{\eta _{0}}^{2}}}\text{,}1/4\,%
{\frac{1}{c_{2}\,{\eta _{0}}^{2}}}\text{,}-1/4\,{\frac{1}{c_{2}\,{\eta _{0}}%
^{2}}}\text{,}  \notag \\
&&1/4\,{\frac{1}{c_{2}\,{\eta _{0}}^{2}}}\text{,}0\text{,}-1/4\,{\frac{1}{{%
c_{2}}^{2}\left( {\eta _{0}}^{2}+2\,{c_{2}}^{2}-2\,{c_{4}}^{2}\right) }}%
\text{,}0]
\end{eqnarray}%
\begin{eqnarray}
\mathbf{C}_{9}^{-1} &=&[-1/2\,{\frac{{c_{4}}^{2}{c_{1}}^{2}}{\left( {c_{3}}%
^{2}-2\,{c_{4}}^{2}\right) \left( {c_{1}}^{2}-{c_{3}}^{2}\right) }}\text{,}%
1/2\,{\frac{{c_{1}}^{2}}{c_{3}\,\left( {c_{1}}^{2}-{c_{3}}^{2}\right) }}%
\text{,}0\text{,}  \notag \\
&&-1/2\,{\frac{{c_{1}}^{2}{c_{2}}^{2}-{c_{4}}^{2}{c_{1}}^{2}-2\,{c_{2}}^{4}-{%
c_{2}}^{2}{\eta _{0}}^{2}+2\,{c_{2}}^{2}{c_{4}}^{2}}{{\eta _{0}}^{2}\left( {%
c_{3}}^{2}-2\,{c_{4}}^{2}\right) \left( {c_{1}}^{2}-{c_{3}}^{2}\right) }}%
\text{,}C_{9,5}^{-1}\text{,}0\text{,}C_{9,7}^{-1}\text{,}  \notag \\
&&1/2\,{\frac{\left( {c_{2}}^{2}-{c_{4}}^{2}\right) {c_{1}}^{2}}{{c_{4}}^{2}{%
\eta _{0}}^{2}c_{3}\,\left( {c_{1}}^{2}-{c_{3}}^{2}\right) }}\text{,}0\text{,%
}-1/2\,{\frac{{c_{1}}^{2}{c_{2}}^{2}-{c_{4}}^{2}{c_{1}}^{2}+{c_{4}}^{2}{\eta
_{0}}^{2}}{{c_{4}}^{2}{\eta _{0}}^{2}c_{3}\,\left( {c_{1}}^{2}-{c_{3}}%
^{2}\right) }}\text{,}0\text{,}  \notag \\
&&-1/2\,{\frac{{c_{1}}^{2}{c_{2}}^{2}-{c_{4}}^{2}{\eta _{0}}^{2}-{c_{4}}^{2}{%
c_{1}}^{2}+{c_{1}}^{2}{\eta _{0}}^{2}}{{c_{4}}^{2}{\eta _{0}}%
^{2}c_{3}\,\left( {c_{1}}^{2}-{c_{3}}^{2}\right) }}\text{,}0\text{,}  \notag
\\
&&-\,{\frac{-{c_{4}}^{2}+{c_{3}}^{2}}{2{c_{3}}^{2}\left( {c_{3}}^{2}-2\,{%
c_{4}}^{2}\right) \left( {c_{1}}^{2}-{c_{3}}^{2}\right) }}\text{,}0\text{,}%
-\,{\frac{{c_{4}}^{2}}{2{c_{3}}^{2}\left( {c_{3}}^{2}-2\,{c_{4}}^{2}\right)
\left( {c_{1}}^{2}-{c_{3}}^{2}\right) }}]
\end{eqnarray}%
\begin{eqnarray}
\mathbf{C}_{10}^{-1} &=&[-1/2\,{\frac{{c_{4}}^{2}{c_{1}}^{2}}{\left( {c_{3}}%
^{2}-2\,{c_{4}}^{2}\right) \left( {c_{1}}^{2}-{c_{3}}^{2}\right) }}\text{,}0%
\text{,}1/2\,{\frac{{c_{1}}^{2}}{c_{3}\,\left( {c_{1}}^{2}-{c_{3}}%
^{2}\right) }}\text{,}  \notag \\
&&-1/2\,{\frac{{c_{1}}^{2}{c_{2}}^{2}-{c_{4}}^{2}{c_{1}}^{2}-2\,{c_{2}}^{4}-{%
c_{2}}^{2}{\eta _{0}}^{2}+2\,{c_{2}}^{2}{c_{4}}^{2}}{{\eta _{0}}^{2}\left( {%
c_{3}}^{2}-2\,{c_{4}}^{2}\right) \left( {c_{1}}^{2}-{c_{3}}^{2}\right) }}%
\text{,}C_{10,5}^{-1}\text{,}0\text{,}C_{10,7}^{-1}\text{,}0\text{,}  \notag
\\
&&1/2\,{\frac{\left( {c_{2}}^{2}-{c_{4}}^{2}\right) {c_{1}}^{2}}{{c_{4}}^{2}{%
\eta _{0}}^{2}c_{3}\,\left( {c_{1}}^{2}-{c_{3}}^{2}\right) }}\text{,}0\text{,%
}-1/2\,{\frac{{c_{1}}^{2}{c_{2}}^{2}-{c_{4}}^{2}{\eta _{0}}^{2}-{c_{4}}^{2}{%
c_{1}}^{2}+{c_{1}}^{2}{\eta _{0}}^{2}}{{c_{4}}^{2}{\eta _{0}}%
^{2}c_{3}\,\left( {c_{1}}^{2}-{c_{3}}^{2}\right) }}\text{,}  \notag \\
&&0\text{,}-1/2\,{\frac{{c_{1}}^{2}{c_{2}}^{2}-{c_{4}}^{2}{c_{1}}^{2}+{c_{4}}%
^{2}{\eta _{0}}^{2}}{{c_{4}}^{2}{\eta _{0}}^{2}c_{3}\,\left( {c_{1}}^{2}-{%
c_{3}}^{2}\right) }}\text{,}-1/2\,{\frac{{c_{4}}^{2}}{{c_{3}}^{2}\left( {%
c_{3}}^{2}-2\,{c_{4}}^{2}\right) \left( {c_{1}}^{2}-{c_{3}}^{2}\right) }}%
\text{,}  \notag \\
&&0\text{,}-1/2\,{\frac{-{c_{4}}^{2}+{c_{3}}^{2}}{{c_{3}}^{2}\left( {c_{3}}%
^{2}-2\,{c_{4}}^{2}\right) \left( {c_{1}}^{2}-{c_{3}}^{2}\right) }}]
\end{eqnarray}%
\begin{eqnarray}
\mathbf{C}_{11}^{-1} &=&[-1/2\,{\frac{{c_{4}}^{2}{c_{1}}^{2}}{\left( {c_{3}}%
^{2}-2\,{c_{4}}^{2}\right) \left( {c_{1}}^{2}-{c_{3}}^{2}\right) }}\text{,}%
-1/2\,{\frac{{c_{1}}^{2}}{c_{3}\,\left( {c_{1}}^{2}-{c_{3}}^{2}\right) }}%
\text{,}0\text{,}  \notag \\
&&-1/2\,{\frac{{c_{1}}^{2}{c_{2}}^{2}-{c_{4}}^{2}{c_{1}}^{2}-2\,{c_{2}}^{4}-{%
c_{2}}^{2}{\eta _{0}}^{2}+2\,{c_{2}}^{2}{c_{4}}^{2}}{{\eta _{0}}^{2}\left( {%
c_{3}}^{2}-2\,{c_{4}}^{2}\right) \left( {c_{1}}^{2}-{c_{3}}^{2}\right) }}%
\text{,}C_{11,5}^{-1}\text{,}0\text{,}C_{11,7}^{-1}\text{,}  \notag \\
&&-1/2\,{\frac{\left( {c_{2}}^{2}-{c_{4}}^{2}\right) {c_{1}}^{2}}{{c_{4}}^{2}%
{\eta _{0}}^{2}c_{3}\,\left( {c_{1}}^{2}-{c_{3}}^{2}\right) }}\text{,}0\text{%
,}1/2\,{\frac{{c_{1}}^{2}{c_{2}}^{2}-{c_{4}}^{2}{c_{1}}^{2}+{c_{4}}^{2}{\eta
_{0}}^{2}}{{c_{4}}^{2}{\eta _{0}}^{2}c_{3}\,\left( {c_{1}}^{2}-{c_{3}}%
^{2}\right) }}\text{,}  \notag \\
&&0\text{,}1/2\,{\frac{{c_{1}}^{2}{c_{2}}^{2}-{c_{4}}^{2}{\eta _{0}}^{2}-{%
c_{4}}^{2}{c_{1}}^{2}+{c_{1}}^{2}{\eta _{0}}^{2}}{{c_{4}}^{2}{\eta _{0}}%
^{2}c_{3}\,\left( {c_{1}}^{2}-{c_{3}}^{2}\right) }}\text{,}0\text{,}-1/2\,{%
\frac{-{c_{4}}^{2}+{c_{3}}^{2}}{{c_{3}}^{2}\left( {c_{3}}^{2}-2\,{c_{4}}%
^{2}\right) \left( {c_{1}}^{2}-{c_{3}}^{2}\right) }}\text{,}  \notag \\
&&0\text{,}-1/2\,{\frac{{c_{4}}^{2}}{{c_{3}}^{2}\left( {c_{3}}^{2}-2\,{c_{4}}%
^{2}\right) \left( {c_{1}}^{2}-{c_{3}}^{2}\right) }}]
\end{eqnarray}%
\begin{eqnarray}
\mathbf{C}_{12}^{-1} &=&[-1/2\,{\frac{{c_{4}}^{2}{c_{1}}^{2}}{\left( {c_{3}}%
^{2}-2\,{c_{4}}^{2}\right) \left( {c_{1}}^{2}-{c_{3}}^{2}\right) }}\text{,}0%
\text{,}-1/2\,{\frac{{c_{1}}^{2}}{c_{3}\,\left( {c_{1}}^{2}-{c_{3}}%
^{2}\right) }}\text{,}  \notag \\
&&-1/2\,{\frac{{c_{1}}^{2}{c_{2}}^{2}-{c_{4}}^{2}{c_{1}}^{2}-2\,{c_{2}}^{4}-{%
c_{2}}^{2}{\eta _{0}}^{2}+2\,{c_{2}}^{2}{c_{4}}^{2}}{{\eta _{0}}^{2}\left( {%
c_{3}}^{2}-2\,{c_{4}}^{2}\right) \left( {c_{1}}^{2}-{c_{3}}^{2}\right) }}%
\text{,}C_{12,5}^{-1}\text{,}0\text{,}C_{12,7}^{-1}\text{,}0\text{,}  \notag
\\
&&-1/2\,{\frac{\left( {c_{2}}^{2}-{c_{4}}^{2}\right) {c_{1}}^{2}}{{c_{4}}^{2}%
{\eta _{0}}^{2}c_{3}\,\left( {c_{1}}^{2}-{c_{3}}^{2}\right) }}\text{,}0\text{%
,}1/2\,{\frac{{c_{1}}^{2}{c_{2}}^{2}-{c_{4}}^{2}{\eta _{0}}^{2}-{c_{4}}^{2}{%
c_{1}}^{2}+{c_{1}}^{2}{\eta _{0}}^{2}}{{c_{4}}^{2}{\eta _{0}}%
^{2}c_{3}\,\left( {c_{1}}^{2}-{c_{3}}^{2}\right) }}\text{,}  \notag \\
&&0\text{,}1/2\,{\frac{{c_{1}}^{2}{c_{2}}^{2}-{c_{4}}^{2}{c_{1}}^{2}+{c_{4}}%
^{2}{\eta _{0}}^{2}}{{c_{4}}^{2}{\eta _{0}}^{2}c_{3}\,\left( {c_{1}}^{2}-{%
c_{3}}^{2}\right) }}\text{,}-1/2\,{\frac{{c_{4}}^{2}}{{c_{3}}^{2}\left( {%
c_{3}}^{2}-2\,{c_{4}}^{2}\right) \left( {c_{1}}^{2}-{c_{3}}^{2}\right) }}%
\text{,}0\text{,}  \notag \\
&&-1/2\,{\frac{-{c_{4}}^{2}+{c_{3}}^{2}}{{c_{3}}^{2}\left( {c_{3}}^{2}-2\,{%
c_{4}}^{2}\right) \left( {c_{1}}^{2}-{c_{3}}^{2}\right) }}]
\end{eqnarray}%
\begin{eqnarray}
\mathbf{C}_{13}^{-1} &=&[1/4\,{\frac{{c_{3}}^{2}{c_{1}}^{2}}{\left( {c_{3}}%
^{2}-2\,{c_{4}}^{2}\right) \left( {c_{1}}^{2}-2\,{c_{4}}^{2}\right) }}\text{,%
}0\text{,}0\text{,}  \notag \\
&&1/4\,{\frac{2\,{c_{1}}^{2}{c_{2}}^{2}-{c_{3}}^{2}{c_{1}}^{2}-4\,{c_{2}}%
^{4}-2\,{c_{2}}^{2}{\eta _{0}}^{2}+2\,{c_{2}}^{2}{c_{3}}^{2}}{{\eta _{0}}%
^{2}\left( {c_{3}}^{2}-2\,{c_{4}}^{2}\right) \left( {c_{1}}^{2}-2\,{c_{4}}%
^{2}\right) }}\text{,}  \notag \\
&&-1/4\,{\frac{2\,{c_{1}}^{2}{c_{2}}^{2}+{c_{1}}^{2}{\eta _{0}}^{2}-{c_{3}}%
^{2}{c_{1}}^{2}-4\,{c_{2}}^{4}-2\,{c_{2}}^{2}{\eta _{0}}^{2}+2\,{c_{2}}^{2}{%
c_{3}}^{2}+{\eta _{0}}^{2}{c_{3}}^{2}}{{\eta _{0}}^{2}\left( {c_{3}}^{2}-2\,{%
c_{4}}^{2}\right) \left( {c_{1}}^{2}-2\,{c_{4}}^{2}\right) }}\text{,}  \notag
\\
&&1/4\,{\frac{2\,{c_{2}}^{2}+{\eta _{0}}^{2}}{{c_{4}}^{2}\left( {\eta _{0}}%
^{2}+2\,{c_{2}}^{2}-2\,{c_{4}}^{2}\right) }}\text{,}  \notag \\
&&-1/4\,{\frac{2\,{c_{1}}^{2}{c_{2}}^{2}+{c_{1}}^{2}{\eta _{0}}^{2}-{c_{3}}%
^{2}{c_{1}}^{2}-4\,{c_{2}}^{4}-2\,{c_{2}}^{2}{\eta _{0}}^{2}+2\,{c_{2}}^{2}{%
c_{3}}^{2}+{\eta _{0}}^{2}{c_{3}}^{2}}{{\eta _{0}}^{2}\left( {c_{3}}^{2}-2\,{%
c_{4}}^{2}\right) \left( {c_{1}}^{2}-2\,{c_{4}}^{2}\right) }}\text{,}  \notag
\\
&&-1/4\,{\frac{{c_{2}}^{2}}{{\eta _{0}}^{2}{c_{4}}^{3}}}\text{,}-1/4\,{\frac{%
{c_{2}}^{2}}{{\eta _{0}}^{2}{c_{4}}^{3}}}\text{,}1/4\,{\frac{{c_{2}}^{2}}{{%
\eta _{0}}^{2}{c_{4}}^{3}}}\text{,}1/4\,{\frac{{c_{2}}^{2}+{\eta _{0}}^{2}}{{%
\eta _{0}}^{2}{c_{4}}^{3}}}\text{,}1/4\,{\frac{{c_{2}}^{2}+{\eta _{0}}^{2}}{{%
\eta _{0}}^{2}{c_{4}}^{3}}}\text{,}  \notag \\
&&1/4\,{\frac{{c_{2}}^{2}}{{\eta _{0}}^{2}{c_{4}}^{3}}}\text{,}1/4\,{\frac{1%
}{\left( {c_{3}}^{2}-2\,{c_{4}}^{2}\right) \left( {c_{1}}^{2}-2\,{c_{4}}%
^{2}\right) }}\text{,}-1/4\,{\frac{1}{{c_{4}}^{2}\left( {\eta _{0}}^{2}+2\,{%
c_{2}}^{2}-2\,{c_{4}}^{2}\right) }}\text{,}  \notag \\
&&1/4\,{\frac{1}{\left( {c_{3}}^{2}-2\,{c_{4}}^{2}\right) \left( {c_{1}}%
^{2}-2\,{c_{4}}^{2}\right) }}]
\end{eqnarray}%
\begin{eqnarray}
\mathbf{C}_{14}^{-1} &=&[1/4\,{\frac{{c_{3}}^{2}{c_{1}}^{2}}{\left( {c_{3}}%
^{2}-2\,{c_{4}}^{2}\right) \left( {c_{1}}^{2}-2\,{c_{4}}^{2}\right) }}\text{,%
}0\text{,}0\text{,}  \notag \\
&&1/4\,{\frac{2\,{c_{1}}^{2}{c_{2}}^{2}-{c_{3}}^{2}{c_{1}}^{2}-4\,{c_{2}}%
^{4}-2\,{c_{2}}^{2}{\eta _{0}}^{2}+2\,{c_{2}}^{2}{c_{3}}^{2}}{{\eta _{0}}%
^{2}\left( {c_{3}}^{2}-2\,{c_{4}}^{2}\right) \left( {c_{1}}^{2}-2\,{c_{4}}%
^{2}\right) }}\text{,}  \notag \\
&&-1/4\,{\frac{2\,{c_{1}}^{2}{c_{2}}^{2}+{c_{1}}^{2}{\eta _{0}}^{2}-{c_{3}}%
^{2}{c_{1}}^{2}-4\,{c_{2}}^{4}-2\,{c_{2}}^{2}{\eta _{0}}^{2}+2\,{c_{2}}^{2}{%
c_{3}}^{2}+{\eta _{0}}^{2}{c_{3}}^{2}}{{\eta _{0}}^{2}\left( {c_{3}}^{2}-2\,{%
c_{4}}^{2}\right) \left( {c_{1}}^{2}-2\,{c_{4}}^{2}\right) }}\text{,}  \notag
\\
&&-1/4\,{\frac{2\,{c_{2}}^{2}+{\eta _{0}}^{2}}{{c_{4}}^{2}\left( {\eta _{0}}%
^{2}+2\,{c_{2}}^{2}-2\,{c_{4}}^{2}\right) }}\text{,}  \notag \\
&&-1/4\,{\frac{2\,{c_{1}}^{2}{c_{2}}^{2}+{c_{1}}^{2}{\eta _{0}}^{2}-{c_{3}}%
^{2}{c_{1}}^{2}-4\,{c_{2}}^{4}-2\,{c_{2}}^{2}{\eta _{0}}^{2}+2\,{c_{2}}^{2}{%
c_{3}}^{2}+{\eta _{0}}^{2}{c_{3}}^{2}}{{\eta _{0}}^{2}\left( {c_{3}}^{2}-2\,{%
c_{4}}^{2}\right) \left( {c_{1}}^{2}-2\,{c_{4}}^{2}\right) }}\text{,}  \notag
\\
&&1/4\,{\frac{{c_{2}}^{2}}{{\eta _{0}}^{2}{c_{4}}^{3}}}\text{,}-1/4\,{\frac{{%
c_{2}}^{2}}{{\eta _{0}}^{2}{c_{4}}^{3}}}\text{,}-1/4\,{\frac{{c_{2}}^{2}}{{%
\eta _{0}}^{2}{c_{4}}^{3}}}\text{,}1/4\,{\frac{{c_{2}}^{2}+{\eta _{0}}^{2}}{{%
\eta _{0}}^{2}{c_{4}}^{3}}}\text{,}-1/4\,{\frac{{c_{2}}^{2}+{\eta _{0}}^{2}}{%
{\eta _{0}}^{2}{c_{4}}^{3}}}\text{,}  \notag \\
&&1/4\,{\frac{{c_{2}}^{2}}{{\eta _{0}}^{2}{c_{4}}^{3}}}\text{,}1/4\,{\frac{1%
}{\left( {c_{3}}^{2}-2\,{c_{4}}^{2}\right) \left( {c_{1}}^{2}-2\,{c_{4}}%
^{2}\right) }}\text{,}1/4\,{\frac{1}{{c_{4}}^{2}\left( {\eta _{0}}^{2}+2\,{%
c_{2}}^{2}-2\,{c_{4}}^{2}\right) }}\text{,}  \notag \\
&&1/4\,{\frac{1}{\left( {c_{3}}^{2}-2\,{c_{4}}^{2}\right) \left( {c_{1}}%
^{2}-2\,{c_{4}}^{2}\right) }}]
\end{eqnarray}%
\begin{eqnarray}
\mathbf{C}_{15}^{-1} &=&[1/4\,{\frac{{c_{3}}^{2}{c_{1}}^{2}}{\left( {c_{3}}%
^{2}-2\,{c_{4}}^{2}\right) \left( {c_{1}}^{2}-2\,{c_{4}}^{2}\right) }}\text{,%
}0\text{,}0\text{,}  \notag \\
&&1/4\,{\frac{2\,{c_{1}}^{2}{c_{2}}^{2}-{c_{3}}^{2}{c_{1}}^{2}-4\,{c_{2}}%
^{4}-2\,{c_{2}}^{2}{\eta _{0}}^{2}+2\,{c_{2}}^{2}{c_{3}}^{2}}{{\eta _{0}}%
^{2}\left( {c_{3}}^{2}-2\,{c_{4}}^{2}\right) \left( {c_{1}}^{2}-2\,{c_{4}}%
^{2}\right) }}\text{,}  \notag \\
&&-1/4\,{\frac{2\,{c_{1}}^{2}{c_{2}}^{2}+{c_{1}}^{2}{\eta _{0}}^{2}-{c_{3}}%
^{2}{c_{1}}^{2}-4\,{c_{2}}^{4}-2\,{c_{2}}^{2}{\eta _{0}}^{2}+2\,{c_{2}}^{2}{%
c_{3}}^{2}+{\eta _{0}}^{2}{c_{3}}^{2}}{{\eta _{0}}^{2}\left( {c_{3}}^{2}-2\,{%
c_{4}}^{2}\right) \left( {c_{1}}^{2}-2\,{c_{4}}^{2}\right) }}\text{,}  \notag
\\
&&1/4\,{\frac{2\,{c_{2}}^{2}+{\eta _{0}}^{2}}{{c_{4}}^{2}\left( {\eta _{0}}%
^{2}+2\,{c_{2}}^{2}-2\,{c_{4}}^{2}\right) }}\text{,}  \notag \\
&&-1/4\,{\frac{2\,{c_{1}}^{2}{c_{2}}^{2}+{c_{1}}^{2}{\eta _{0}}^{2}-{c_{3}}%
^{2}{c_{1}}^{2}-4\,{c_{2}}^{4}-2\,{c_{2}}^{2}{\eta _{0}}^{2}+2\,{c_{2}}^{2}{%
c_{3}}^{2}+{\eta _{0}}^{2}{c_{3}}^{2}}{{\eta _{0}}^{2}\left( {c_{3}}^{2}-2\,{%
c_{4}}^{2}\right) \left( {c_{1}}^{2}-2\,{c_{4}}^{2}\right) }}\text{,}  \notag
\\
&&1/4\,{\frac{{c_{2}}^{2}}{{\eta _{0}}^{2}{c_{4}}^{3}}}\text{,}1/4\,{\frac{{%
c_{2}}^{2}}{{\eta _{0}}^{2}{c_{4}}^{3}}}\text{,}-1/4\,{\frac{{c_{2}}^{2}}{{%
\eta _{0}}^{2}{c_{4}}^{3}}}\text{,}-1/4\,{\frac{{c_{2}}^{2}+{\eta _{0}}^{2}}{%
{\eta _{0}}^{2}{c_{4}}^{3}}}\text{,}-1/4\,{\frac{{c_{2}}^{2}+{\eta _{0}}^{2}%
}{{\eta _{0}}^{2}{c_{4}}^{3}}}\text{,}  \notag \\
&&-1/4\,{\frac{{c_{2}}^{2}}{{\eta _{0}}^{2}{c_{4}}^{3}}}\text{,}1/4\,{\frac{1%
}{\left( {c_{3}}^{2}-2\,{c_{4}}^{2}\right) \left( {c_{1}}^{2}-2\,{c_{4}}%
^{2}\right) }}\text{,}-1/4\,{\frac{1}{{c_{4}}^{2}\left( {\eta _{0}}^{2}+2\,{%
c_{2}}^{2}-2\,{c_{4}}^{2}\right) }}\text{,}  \notag \\
&&1/4\,{\frac{1}{\left( {c_{3}}^{2}-2\,{c_{4}}^{2}\right) \left( {c_{1}}%
^{2}-2\,{c_{4}}^{2}\right) }}]
\end{eqnarray}%
\begin{eqnarray}
\mathbf{C}_{16}^{-1} &=&[1/4\,{\frac{{c_{3}}^{2}{c_{1}}^{2}}{\left( {c_{3}}%
^{2}-2\,{c_{4}}^{2}\right) \left( {c_{1}}^{2}-2\,{c_{4}}^{2}\right) }}\text{,%
}0\text{,}0\text{,}  \notag \\
&&1/4\,{\frac{2\,{c_{1}}^{2}{c_{2}}^{2}-{c_{3}}^{2}{c_{1}}^{2}-4\,{c_{2}}%
^{4}-2\,{c_{2}}^{2}{\eta _{0}}^{2}+2\,{c_{2}}^{2}{c_{3}}^{2}}{{\eta _{0}}%
^{2}\left( {c_{3}}^{2}-2\,{c_{4}}^{2}\right) \left( {c_{1}}^{2}-2\,{c_{4}}%
^{2}\right) }}\text{,}  \notag \\
&&-1/4\,{\frac{2\,{c_{1}}^{2}{c_{2}}^{2}+{c_{1}}^{2}{\eta _{0}}^{2}-{c_{3}}%
^{2}{c_{1}}^{2}-4\,{c_{2}}^{4}-2\,{c_{2}}^{2}{\eta _{0}}^{2}+2\,{c_{2}}^{2}{%
c_{3}}^{2}+{\eta _{0}}^{2}{c_{3}}^{2}}{{\eta _{0}}^{2}\left( {c_{3}}^{2}-2\,{%
c_{4}}^{2}\right) \left( {c_{1}}^{2}-2\,{c_{4}}^{2}\right) }}\text{,}  \notag
\\
&&-1/4\,{\frac{2\,{c_{2}}^{2}+{\eta _{0}}^{2}}{{c_{4}}^{2}\left( {\eta _{0}}%
^{2}+2\,{c_{2}}^{2}-2\,{c_{4}}^{2}\right) }}\text{,}  \notag \\
&&-1/4\,{\frac{2\,{c_{1}}^{2}{c_{2}}^{2}+{c_{1}}^{2}{\eta _{0}}^{2}-{c_{3}}%
^{2}{c_{1}}^{2}-4\,{c_{2}}^{4}-2\,{c_{2}}^{2}{\eta _{0}}^{2}+2\,{c_{2}}^{2}{%
c_{3}}^{2}+{\eta _{0}}^{2}{c_{3}}^{2}}{{\eta _{0}}^{2}\left( {c_{3}}^{2}-2\,{%
c_{4}}^{2}\right) \left( {c_{1}}^{2}-2\,{c_{4}}^{2}\right) }}\text{,}  \notag
\\
&&-1/4\,{\frac{{c_{2}}^{2}}{{\eta _{0}}^{2}{c_{4}}^{3}}}\text{,}1/4\,{\frac{{%
c_{2}}^{2}}{{\eta _{0}}^{2}{c_{4}}^{3}}}\text{,}1/4\,{\frac{{c_{2}}^{2}}{{%
\eta _{0}}^{2}{c_{4}}^{3}}}\text{,}-1/4\,{\frac{{c_{2}}^{2}+{\eta _{0}}^{2}}{%
{\eta _{0}}^{2}{c_{4}}^{3}}}\text{,}1/4\,{\frac{{c_{2}}^{2}+{\eta _{0}}^{2}}{%
{\eta _{0}}^{2}{c_{4}}^{3}}}\text{,}  \notag \\
&&-1/4\,{\frac{{c_{2}}^{2}}{{\eta _{0}}^{2}{c_{4}}^{3}}}\text{,}1/4\,{\frac{1%
}{\left( {c_{3}}^{2}-2\,{c_{4}}^{2}\right) \left( {c_{1}}^{2}-2\,{c_{4}}%
^{2}\right) }}\text{,}  \notag \\
&&1/4\,{\frac{1}{{c_{4}}^{2}\left( {\eta _{0}}^{2}+2\,{c_{2}}^{2}-2\,{c_{4}}%
^{2}\right) }}\text{,}1/4\,{\frac{1}{\left( {c_{3}}^{2}-2\,{c_{4}}%
^{2}\right) \left( {c_{1}}^{2}-2\,{c_{4}}^{2}\right) }}]
\end{eqnarray}%
with
\begin{equation*}
C_{m\text{,}n}^{-1}=\frac{p_{m,n}}{q_{m,n}}
\end{equation*}%
where
\begin{eqnarray*}
p_{1,5} &=&2\,{c_{1}}^{2}{c_{2}}^{4}+{c_{1}}^{2}{c_{2}}^{2}{\eta _{0}}^{2}-{%
c_{1}}^{2}{c_{2}}^{2}{c_{3}}^{2}-2\,{c_{2}}^{2}{c_{4}}^{2}{c_{1}}^{2}-{c_{1}}%
^{2}{\eta _{0}}^{2}{c_{3}}^{2} \\
&&+{\eta _{0}}^{2}{c_{3}}^{2}{c_{4}}^{2}+{c_{1}}^{2}{c_{3}}^{2}{c_{4}}^{2}-{%
c_{1}}^{2}{c_{4}}^{2}{\eta _{0}}^{2}\text{,}
\end{eqnarray*}%
\begin{equation*}
q_{1,5}=2\left( {c_{1}}^{2}-2\,{c_{4}}^{2}\right) {c_{1}}^{2}{\eta _{0}}%
^{2}\left( {c_{1}}^{2}-{c_{3}}^{2}\right) \text{,}
\end{equation*}%
\begin{eqnarray*}
p_{1,7} &=&2\,{c_{1}}^{2}{c_{2}}^{4}-{c_{1}}^{2}{c_{2}}^{2}{c_{3}}^{2}+{c_{1}%
}^{2}{c_{2}}^{2}{\eta _{0}}^{2}-2\,{c_{2}}^{2}{c_{4}}^{2}{c_{1}}^{2} \\
&&+{c_{1}}^{2}{c_{3}}^{2}{c_{4}}^{2}-{c_{1}}^{2}{c_{4}}^{2}{\eta _{0}}^{2}-{%
\eta _{0}}^{2}{c_{3}}^{2}{c_{4}}^{2}\text{,}
\end{eqnarray*}%
\begin{equation*}
q_{1,7}=2\left( {c_{1}}^{2}-2\,{c_{4}}^{2}\right) {c_{1}}^{2}{\eta _{0}}%
^{2}\left( {c_{1}}^{2}-{c_{3}}^{2}\right) \text{,}
\end{equation*}

\begin{eqnarray*}
p_{2,5} &=&2\,{c_{1}}^{2}{c_{2}}^{4}-{c_{1}}^{2}{c_{2}}^{2}{c_{3}}^{2}+{c_{1}%
}^{2}{c_{2}}^{2}{\eta _{0}}^{2}-2\,{c_{2}}^{2}{c_{4}}^{2}{c_{1}}^{2} \\
&&+{c_{1}}^{2}{c_{3}}^{2}{c_{4}}^{2}-{c_{1}}^{2}{c_{4}}^{2}{\eta _{0}}^{2}-{%
\eta _{0}}^{2}{c_{3}}^{2}{c_{4}}^{2}\text{,}
\end{eqnarray*}%
\begin{equation*}
q_{2,5}=2\left( {c_{1}}^{2}-2\,{c_{4}}^{2}\right) {c_{1}}^{2}{\eta _{0}}%
^{2}\left( {c_{1}}^{2}-{c_{3}}^{2}\right) \text{,}
\end{equation*}

\begin{eqnarray*}
p_{2,7} &=&2\,{c_{1}}^{2}{c_{2}}^{4}+{c_{1}}^{2}{c_{2}}^{2}{\eta _{0}}^{2}-{%
c_{1}}^{2}{c_{2}}^{2}{c_{3}}^{2}-2\,{c_{2}}^{2}{c_{4}}^{2}{c_{1}}^{2}-{c_{1}}%
^{2}{\eta _{0}}^{2}{c_{3}}^{2} \\
&&+{\eta _{0}}^{2}{c_{3}}^{2}{c_{4}}^{2}+{c_{1}}^{2}{c_{3}}^{2}{c_{4}}^{2}-{%
c_{1}}^{2}{c_{4}}^{2}{\eta _{0}}^{2}\text{,}
\end{eqnarray*}%
\begin{equation*}
q_{2,7}=2\left( {c_{1}}^{2}-2\,{c_{4}}^{2}\right) {c_{1}}^{2}{\eta _{0}}%
^{2}\left( {c_{1}}^{2}-{c_{3}}^{2}\right) \text{,}
\end{equation*}%
\begin{eqnarray*}
p_{3,5} &=&2\,{c_{1}}^{2}{c_{2}}^{4}+{c_{1}}^{2}{c_{2}}^{2}{\eta _{0}}^{2}-{%
c_{1}}^{2}{c_{2}}^{2}{c_{3}}^{2}-2\,{c_{2}}^{2}{c_{4}}^{2}{c_{1}}^{2}-{c_{1}}%
^{2}{\eta _{0}}^{2}{c_{3}}^{2} \\
&&+{\eta _{0}}^{2}{c_{3}}^{2}{c_{4}}^{2}+{c_{1}}^{2}{c_{3}}^{2}{c_{4}}^{2}-{%
c_{1}}^{2}{c_{4}}^{2}{\eta _{0}}^{2}\text{,}
\end{eqnarray*}%
\begin{equation*}
q_{3,5}=2\left( {c_{1}}^{2}-2\,{c_{4}}^{2}\right) {c_{1}}^{2}{\eta _{0}}%
^{2}\left( {c_{1}}^{2}-{c_{3}}^{2}\right) \text{,}
\end{equation*}%
\begin{eqnarray*}
p_{3,7} &=&2\,{c_{1}}^{2}{c_{2}}^{4}-{c_{1}}^{2}{c_{2}}^{2}{c_{3}}^{2}+{c_{1}%
}^{2}{c_{2}}^{2}{\eta _{0}}^{2}-2\,{c_{2}}^{2}{c_{4}}^{2}{c_{1}}^{2}+{c_{1}}%
^{2}{c_{3}}^{2}{c_{4}}^{2} \\
&&-{c_{1}}^{2}{c_{4}}^{2}{\eta _{0}}^{2}-{\eta _{0}}^{2}{c_{3}}^{2}{c_{4}}%
^{2}\text{,}
\end{eqnarray*}%
\begin{equation*}
q_{3,7}=2\left( {c_{1}}^{2}-2\,{c_{4}}^{2}\right) {c_{1}}^{2}{\eta _{0}}%
^{2}\left( {c_{1}}^{2}-{c_{3}}^{2}\right) \text{,}
\end{equation*}%
\begin{eqnarray*}
p_{4,5} &=&2\,{c_{1}}^{2}{c_{2}}^{4}-{c_{1}}^{2}{c_{2}}^{2}{c_{3}}^{2}+{c_{1}%
}^{2}{c_{2}}^{2}{\eta _{0}}^{2}-2\,{c_{2}}^{2}{c_{4}}^{2}{c_{1}}^{2} \\
&&+{c_{1}}^{2}{c_{3}}^{2}{c_{4}}^{2}-{c_{1}}^{2}{c_{4}}^{2}{\eta _{0}}^{2}-{%
\eta _{0}}^{2}{c_{3}}^{2}{c_{4}}^{2}\text{,}
\end{eqnarray*}%
\begin{equation*}
q_{4,5}=2\left( {c_{1}}^{2}-2\,{c_{4}}^{2}\right) {c_{1}}^{2}{\eta _{0}}%
^{2}\left( {c_{1}}^{2}-{c_{3}}^{2}\right) \text{,}
\end{equation*}%
\begin{eqnarray*}
p_{4,7} &=&2\,{c_{1}}^{2}{c_{2}}^{4}+{c_{1}}^{2}{c_{2}}^{2}{\eta _{0}}^{2}-{%
c_{1}}^{2}{c_{2}}^{2}{c_{3}}^{2}-2\,{c_{2}}^{2}{c_{4}}^{2}{c_{1}}^{2}-{c_{1}}%
^{2}{\eta _{0}}^{2}{c_{3}}^{2} \\
&&+{\eta _{0}}^{2}{c_{3}}^{2}{c_{4}}^{2}+{c_{1}}^{2}{c_{3}}^{2}{c_{4}}^{2}-{%
c_{1}}^{2}{c_{4}}^{2}{\eta _{0}}^{2}\text{,}
\end{eqnarray*}%
\begin{equation*}
q_{4,7}=2\left( {c_{1}}^{2}-2\,{c_{4}}^{2}\right) {c_{1}}^{2}{\eta _{0}}%
^{2}\left( {c_{1}}^{2}-{c_{3}}^{2}\right) \text{,}
\end{equation*}%
\begin{eqnarray*}
p_{9,5} &=&{c_{1}}^{2}{c_{2}}^{2}{c_{3}}^{2}+{c_{1}}^{2}{\eta _{0}}^{2}{c_{3}%
}^{2}-{c_{1}}^{2}{c_{3}}^{2}{c_{4}}^{2}-{c_{1}}^{2}{c_{4}}^{2}{\eta _{0}}%
^{2}-2\,{c_{3}}^{2}{c_{2}}^{4} \\
&&-{c_{3}}^{2}{c_{2}}^{2}{\eta _{0}}^{2}+2\,{c_{2}}^{2}{c_{4}}^{2}{c_{3}}%
^{2}+{\eta _{0}}^{2}{c_{3}}^{2}{c_{4}}^{2}\text{,}
\end{eqnarray*}%
\begin{equation*}
q_{9,5}=2{\eta _{0}}^{2}{c_{3}}^{2}\left( {c_{3}}^{2}-2\,{c_{4}}^{2}\right)
\left( {c_{1}}^{2}-{c_{3}}^{2}\right) \text{,}
\end{equation*}%
\begin{eqnarray*}
p_{9,7} &=&{c_{1}}^{2}{c_{2}}^{2}{c_{3}}^{2}-{c_{1}}^{2}{c_{3}}^{2}{c_{4}}%
^{2}+{c_{1}}^{2}{c_{4}}^{2}{\eta _{0}}^{2}-2\,{c_{3}}^{2}{c_{2}}^{4}-{c_{3}}%
^{2}{c_{2}}^{2}{\eta _{0}}^{2} \\
&&+2\,{c_{2}}^{2}{c_{4}}^{2}{c_{3}}^{2}+{\eta _{0}}^{2}{c_{3}}^{2}{c_{4}}^{2}%
\text{,}
\end{eqnarray*}%
\begin{equation*}
q_{9,7}=2{\eta _{0}}^{2}{c_{3}}^{2}\left( {c_{3}}^{2}-2\,{c_{4}}^{2}\right)
\left( {c_{1}}^{2}-{c_{3}}^{2}\right) \text{,}
\end{equation*}%
\begin{eqnarray*}
p_{10,5} &=&{c_{1}}^{2}{c_{2}}^{2}{c_{3}}^{2}-{c_{1}}^{2}{c_{3}}^{2}{c_{4}}%
^{2}+{c_{1}}^{2}{c_{4}}^{2}{\eta _{0}}^{2}-2\,{c_{3}}^{2}{c_{2}}^{4}-{c_{3}}%
^{2}{c_{2}}^{2}{\eta _{0}}^{2} \\
&&+2\,{c_{2}}^{2}{c_{4}}^{2}{c_{3}}^{2}+{\eta _{0}}^{2}{c_{3}}^{2}{c_{4}}^{2}%
\text{,}
\end{eqnarray*}%
\begin{equation*}
q_{10,5}=2{\eta _{0}}^{2}{c_{3}}^{2}\left( {c_{3}}^{2}-2\,{c_{4}}^{2}\right)
\left( {c_{1}}^{2}-{c_{3}}^{2}\right) \text{,}
\end{equation*}%
\begin{eqnarray*}
p_{10,7} &=&{c_{1}}^{2}{c_{2}}^{2}{c_{3}}^{2}+{c_{1}}^{2}{\eta _{0}}^{2}{%
c_{3}}^{2}-{c_{1}}^{2}{c_{3}}^{2}{c_{4}}^{2}-{c_{1}}^{2}{c_{4}}^{2}{\eta _{0}%
}^{2}-2\,{c_{3}}^{2}{c_{2}}^{4} \\
&&-{c_{3}}^{2}{c_{2}}^{2}{\eta _{0}}^{2}+2\,{c_{2}}^{2}{c_{4}}^{2}{c_{3}}%
^{2}+{\eta _{0}}^{2}{c_{3}}^{2}{c_{4}}^{2}\text{,}
\end{eqnarray*}%
\begin{equation*}
q_{10,7}=2{\eta _{0}}^{2}{c_{3}}^{2}\left( {c_{3}}^{2}-2\,{c_{4}}^{2}\right)
\left( {c_{1}}^{2}-{c_{3}}^{2}\right) \text{,}
\end{equation*}

\begin{eqnarray*}
p_{11,5} &=&{c_{1}}^{2}{c_{2}}^{2}{c_{3}}^{2}+{c_{1}}^{2}{\eta _{0}}^{2}{%
c_{3}}^{2}-{c_{1}}^{2}{c_{3}}^{2}{c_{4}}^{2}-{c_{1}}^{2}{c_{4}}^{2}{\eta _{0}%
}^{2}-2\,{c_{3}}^{2}{c_{2}}^{4} \\
&&-{c_{3}}^{2}{c_{2}}^{2}{\eta _{0}}^{2}+2\,{c_{2}}^{2}{c_{4}}^{2}{c_{3}}%
^{2}+{\eta _{0}}^{2}{c_{3}}^{2}{c_{4}}^{2}\text{,}
\end{eqnarray*}%
\begin{equation*}
q_{11,5}=2{\eta _{0}}^{2}{c_{3}}^{2}\left( {c_{3}}^{2}-2\,{c_{4}}^{2}\right)
\left( {c_{1}}^{2}-{c_{3}}^{2}\right) \text{,}
\end{equation*}%
\begin{eqnarray*}
p_{11,7} &=&{c_{1}}^{2}{c_{2}}^{2}{c_{3}}^{2}-{c_{1}}^{2}{c_{3}}^{2}{c_{4}}%
^{2}+{c_{1}}^{2}{c_{4}}^{2}{\eta _{0}}^{2}-2\,{c_{3}}^{2}{c_{2}}^{4}-{c_{3}}%
^{2}{c_{2}}^{2}{\eta _{0}}^{2} \\
&&+2\,{c_{2}}^{2}{c_{4}}^{2}{c_{3}}^{2}+{\eta _{0}}^{2}{c_{3}}^{2}{c_{4}}^{2}%
\text{,}
\end{eqnarray*}%
\begin{equation*}
q_{11,7}=2{\eta _{0}}^{2}{c_{3}}^{2}\left( {c_{3}}^{2}-2\,{c_{4}}^{2}\right)
\left( {c_{1}}^{2}-{c_{3}}^{2}\right) \text{,}
\end{equation*}%
\begin{eqnarray*}
p_{12,5} &=&{c_{1}}^{2}{c_{2}}^{2}{c_{3}}^{2}-{c_{1}}^{2}{c_{3}}^{2}{c_{4}}%
^{2}+{c_{1}}^{2}{c_{4}}^{2}{\eta _{0}}^{2}-2\,{c_{3}}^{2}{c_{2}}^{4}-{c_{3}}%
^{2}{c_{2}}^{2}{\eta _{0}}^{2} \\
&&+2\,{c_{2}}^{2}{c_{4}}^{2}{c_{3}}^{2}+{\eta _{0}}^{2}{c_{3}}^{2}{c_{4}}^{2}%
\text{,}
\end{eqnarray*}%
\begin{equation*}
q_{12,5}=2{\eta _{0}}^{2}{c_{3}}^{2}\left( {c_{3}}^{2}-2\,{c_{4}}^{2}\right)
\left( {c_{1}}^{2}-{c_{3}}^{2}\right) \text{,}
\end{equation*}%
\begin{eqnarray*}
p_{12,7} &=&{c_{1}}^{2}{c_{2}}^{2}{c_{3}}^{2}+{c_{1}}^{2}{\eta _{0}}^{2}{%
c_{3}}^{2}-{c_{1}}^{2}{c_{3}}^{2}{c_{4}}^{2}-{c_{1}}^{2}{c_{4}}^{2}{\eta _{0}%
}^{2}-2\,{c_{3}}^{2}{c_{2}}^{4} \\
&&-{c_{3}}^{2}{c_{2}}^{2}{\eta _{0}}^{2}+2\,{c_{2}}^{2}{c_{4}}^{2}{c_{3}}%
^{2}+{\eta _{0}}^{2}{c_{3}}^{2}{c_{4}}^{2}\text{,}
\end{eqnarray*}%
\begin{equation*}
q_{12,7}=2{\eta _{0}}^{2}{c_{3}}^{2}\left( {c_{3}}^{2}-2\,{c_{4}}^{2}\right)
\left( {c_{1}}^{2}-{c_{3}}^{2}\right) \text{,}
\end{equation*}


\end{subequations}

\end{document}